\documentclass[12pt,epsf]{article}
\usepackage{amsfonts}
\usepackage{amssymb,amsmath}
\usepackage{graphicx}
\usepackage[dvips]{color}
\usepackage[vcentermath]{youngtab}
\usepackage[margin=1in]{geometry}
\usepackage{amsmath,amssymb,amsfonts,amsthm,mathtools}
\usepackage{hyperref}
\usepackage{bm}
\usepackage{braket}
\usepackage{amssymb}
\usepackage{bbm}
\usepackage{gnuindex}

\newcommand{\be}{\begin{equation}}
\newcommand{\ee}{\end{equation}}
\newcommand{\bea}{\begin{eqnarray}}
\newcommand{\eea}{\end{eqnarray}}
\newcommand{\beas}{\begin{eqnarray}}
\newcommand{\eeas}{\end{eqnarray}}
\newcommand{\ba}{\begin{array}}
\newcommand{\ea}{\end{array}}
\newtheorem{lemm}{Lemma}
\newtheorem{prop}{Proposition}
\newtheorem{protocol}{Protocol}
\newtheorem{coro}{Corollary}

\newcommand{\nbox}{{\,\lower0.9pt\vbox{\hrule \hbox{\vrule height 0.2 cm \hskip 0.19 cm \vrule height 0.2 cm}\hrule}\,}}

\def\href#1#2{#2}
\input{tcilatex}
\begin{document}


\newpage


\begin{titlepage}
\hfill
\vbox{
    \halign{#\hfil         \cr
           } 
      }  

\hbox to \hsize{{}\hss \vtop{ \hbox{}

}}

%

\vspace*{20mm}

\begin{center}

{\large \textbf{Logical-Class Diagnostic Layer for Stabilizer Codes and Decoders via Supercharge Structures \vspace{0.4cm}}}

{\normalsize \vspace{10mm} }

{\normalsize {Hai Lin}}

{\normalsize \vspace{10mm} }

{\small \emph{\textit{Shing-Tung Yau Center of Southeast University and School of Mathematics, Southeast University,\\
Nanjing 210096, China
}} }

{\normalsize \vspace{0.4cm} }

\end{center}


\begin{abstract}

{\normalsize \vspace{0.3cm} }

We develop a decoder-agnostic logical-class diagnostic layer for abelian Pauli stabilizer codes, motivated by fault-tolerant quantum computing. An ancilla fermionic register defines a nilpotent supercharge whose clean degree-zero harmonic sector is the code space. Projected-return blocks derived from graded commutators organize the recovery test of whether a decoder-selected correction preserves every encoded input for every error in a declared same-syndrome family; the passing syndromes form the certified faithful window. Demoting the supercharge to an exact binary diagnostic formulation classifies residuals modulo stabilizers, separating a check violation, a harmless stabilizer residual, and a syndrome-silent nontrivial logical action, with a two-sector form for CSS codes. Because the unknown fault cancels, the relative logical class of two record-compatible candidates is computable online. We verify the diagnostic formalisms on the Steane code and specify a static--circuit bridge audit protocol for the Gross bivariate-bicycle quantum low-density parity-check (qLDPC) code. The method supplements existing decoders without additional physical qubits for software use.

\end{abstract}

\end{titlepage}

\vskip 1cm \newpage \newpage \newpage \newpage

\tableofcontents

\section{Introduction}

\renewcommand{\theequation}{1.\arabic{equation}} \setcounter{equation}{0} %
\renewcommand{\thethm}{1.\arabic{thm}} \setcounter{thm}{0} %
\renewcommand{\theprop}{1.\arabic{prop}} \setcounter{prop}{0} %
\renewcommand{\thelemm}{1.\arabic{lemm}} \setcounter{lemm}{0}

\label{sec:introduction}

Consider a quantum processor storing one or more encoded qubits in a
low-density parity-check memory during repeated syndrome extraction. Every
reported check can be quiet while the encoded information has moved to a
different logical state. This is not a defect of the check measurement. A
syndrome reports local stabilizer violations; it does not reveal a logical
Pauli that commutes with all the checks. Let $\mathcal{S}$ be an abelian
Pauli stabilizer group and $N(\mathcal{S})$ its normalizer. A
post-correction residual in $\mathcal{S}$ is harmless, whereas a residual in
$N(\mathcal{S})\backslash \mathcal{S}$ changes the encoded state without
producing a syndrome. The local--global distinction is familiar from the
stabilizer formalism and the Knill--Laflamme criterion \cite%
{Gottesman1997,KnillLaflamme1997,NielsenChuang2010,Terhal2015}. Our question
is how to preserve that distinction when auditing a decoder's proposed
recovery.

In this paper, we propose and develop, to supply quantum decoders with a
logical-address diagnostic and audit layer: an exact, decoder-agnostic
apparatus that certifies not merely that a correction has returned the state
to the code space, but that it has returned the state to the correct logical
fibre inside it. It does not replace a base decoder, and it does not attempt
to infer an unobserved fault from a silent record. It answers two sharper
questions: what does a proposed correction do on a declared error family,
and do two record-compatible candidates differ logically.

The need is practical, as well as conceptual. High-rate quantum low-density
parity-check codes have matured from asymptotic constructions into a
developing field with end-to-end memory protocols and real-time decoder
stacks \cite{Bravyi2024,Breuckmann2021,Muller:2025rue,Maurer:2025tbw}. These
codes are strongly degenerate \cite{Panteleev2021}: one detector record is
compatible with a structurally large family of fault histories, and repeated
noisy syndrome extraction enlarges that family further \cite{Dennis2002}, in
a way shaped by the chosen extraction circuit \cite{Shaw2025}. Belief
propagation with ordered-statistics post-processing \cite{Panteleev2021} and
localized-statistics decoding \cite{Hillmann2024} are built to decode such
degenerate codes, and post-processing strategies already attack decoder
degeneracy directly \cite{Tsubouchi:2025hgy}. Degeneracy is thus the normal
operating condition of a real decoder, and it is the mechanism by which two
detector-compatible accounts of one observation can differ by a protected
logical action. What is still missing is an exact, per-record logical
verdict.

We organize this audit with a nilpotent supercharge. Section~\ref%
{sec:susy-stabilizer} associates one fermionic ancilla-register mode with
each stabilizer generator $S_{a}$. The violated-check projector $\Pi
_{a}^{1}=(I-S_{a})/2$ controls creation in that mode: $Q=\sum_{a=1}^{r}\Pi
_{a}^{1}\otimes c_{a}^{\dagger }$, $Q^{2}=0$. The commuting projectors and
anticommuting register operators make the square vanish. Acting on an empty
register, $Q$ creates a coherent one-excitation superposition supported on
the violated checks; it does not itself perform a stabilizer measurement.
The auxiliary operator $H_{\mathrm{SUSY}}=\{Q,Q^{\dagger }\}$ counts those
violations; Its \textquotedblleft energy\textquotedblright\ is syndrome
weight. Bright, syndrome-bearing directions are paired by $Q$. A code state
is dark, but a code state carrying an old register excitation is dark too.
The clean protected manifold therefore needs the ancilla-degree restriction $%
\mathcal{K}=\ker Q\cap \ker Q^{\dagger }\cap \ker N_{\mathrm{anc}}=\mathcal{C%
}_{\mathrm{QEC}}\otimes \lvert 0\rangle _{\mathrm{anc}}$. This cut removes
stale flags and identifies the landing space that a recovery must actually
reach. It also gives the code space a degree-zero harmonic interpretation
familiar from supersymmetric complexes \cite{Witten1982}. The value of this
language is not a new physical protection mechanism: it makes the
syndrome-active directions and the clean encoded sector visible in one
algebra.

Three perspectives organize the paper: we see the departure, because $Q$
writes the syndrome record into the ancilla register exactly once; we
certify the return, because a syndrome-resolved block tests the landing of a
whole sector; we scale the audit, because the same test survives demotion to
finite-field arithmetic and then circuit-level detector histories.

Four cooperating objects carry these promises, and naming their roles once
makes the formal development easier to follow. The destination is $\mathcal{K%
}$: the arena every certified recovery must hit. The route is the
projected-return block, displayed here in its decoder-facing form $B_{s}=P_{%
\mathcal{K}}\,C_{s}\,\widetilde{P}_{s}$, in which $\widetilde{P}_{s}$
selects a clean bright syndrome sector, the decoder-selected factor $C_{s}$
proposes the return, and $P_{\mathcal{K}}$ tests the landing; Sec.~\ref%
{sec:syndrome-fibres-KL-window} derives this block from a graded commutator
internal to the $Q$-complex, so the apparatus is intrinsic rather than
appended. The warrant is the certified faithful window $\mathcal{S}_{\mathrm{%
faith}}$, the syndrome addresses on which the selected correction is proved,
not merely scored, to carry one declared family to one logical fibre \cite%
{KnillLaflamme1997,NielsenChuang2010}. The diagnosis is the software
residual class $H^{1}(\tilde{Q}^{\mathrm{full}})$, obtained by demoting $Q$
to an exact finite-field differential in Sec.~\ref%
{sec:software-supercharge-cohomology} using the binary chain complex
associated with the CSS code construction \cite{Calderbank1996,Steane1996}.
Route and diagnosis answer different questions: one is a statement about
transport of an operator block, the other a quotient-theoretic statement
about a committed residual. Keeping both visible is what dinstiguishes a
scalar cost, a soft score, or a symmetry label, from a certificate of
logical faithfulness.

The construction is checked at two scales, and the two audits test different
interfaces. The Steane $[\![7,1,3]\!]$ analysis \cite{Steane1996} of Subsec.~%
\ref{subsec:steane-static-audit} exhausts the complete weight-one catalogue
by hand, an executability proof of the conventions rather than a benchmark
of large-code performance. The Steane code also supports experimentally
demonstrated fault-tolerant logical gate operations, including magic-state
injection \cite{Postler:2021ddz}. The gross $[\![144,12,12]\!]$
bivariate-bicycle analysis of Sec.~\ref{sec:bb-circuit-reranking} moves to a
fixed extraction circuit in the setting of Ref.~\cite{Bravyi2024}. There a
spacetime fault history $f$ produces a detector record $d=Df$, and its
logical action is recorded by the linear map $A$. Two candidates $f_{a}$ and
$f_{b}$ sharing a detector record share a logical address if and only if $%
A(f_{a}+f_{b})=0$, the exact criterion established in Subsec.~\ref%
{ssec:bb-fibre-test}. The unknown true fault cancels from this comparison,
so the test needs no online oracle. The audit is specified for the
static--circuit bridge of Subsec.~\ref{subsec:circuit-static-bridge} on a
declared low-weight domain.

Two downstream refinements preserve the same logical order. The
noise-calibrated weighted lift of Sec.~\ref{sec:weighted-lift} may reorder
candidates without changing the exact binary logical quotient. The
character-resolved compression of Subsec.~\ref{subsec:compress-Bs} may
reorganize an already-certified ledger whenever a finite Abelian action on
syndrome labels is verified. Finally, the static Witten index of Sec.~\ref%
{sec:witten-static-decoder} counts the clean protected states \cite%
{Witten1982}. The order is strict: first certify, and only then weight,
rerank, resolve symmetry, and compress.

The main new results are four. First, an ancilla-register supercharge for an
arbitrary abelian Pauli stabilizer code, whose clean degree-zero harmonic
sector is exactly $\mathcal{C}_{\mathrm{QEC}}$. Second, the
graded-commutator return block, its sectorwise collapse, and the certified
faithful window built on it. Third, the demotion $Q\mapsto \tilde{Q}$ used
as an accept--reject certificate on $p_{\mathrm{fs}}$, together with an
online ambiguity flag computed from relative classes of candidates, in
Subsec.~\ref{subsec:ambiguity-flag-rate}. Fourth, the static--circuit bridge
with its mismatch ledger, and a matched ablation that separates the value of
logical information from the value of its implementation, in Subsec.~\ref%
{ssec:bb-metrics}. Three supporting results keep the architecture
consistent: the invariance of the logical address under noise-calibrated
weighting, a character-resolved symmetry reorganization of the certified
ledger, and a projected Witten index that fixes the size of $\mathcal{K}$.
All these statements concern logical classification.

The paper follows the order of dependence. Section~\ref{sec:susy-stabilizer}
builds $Q$, $\mathcal{K}$, and the return blocks, and records the
compatibility of an external non-Clifford resource with $\mathcal{K}$ in
Subsec.~\ref{subsec:logical-resource}. Section~\ref%
{sec:syndrome-fibres-KL-window} introduces the decoder map, the faithfulness
test, and $\mathcal{S}_{\mathrm{faith}}$, and ends with the certified
symmetry reorganization. Section~\ref{sec:software-supercharge-cohomology}
develops the finite-field certificate and checks it by hand on the Steane $%
[\![7,1,3]\!]$ code in Subsec.~\ref{subsec:steane-static-audit}. Section~\ref%
{sec:bb-circuit-reranking} carries the construction to circuit level on the
gross code. Section~\ref{sec:weighted-lift} adds bias-calibrated weighting,
and Sec.~\ref{sec:witten-static-decoder} fixes the size of the protected
arena. Section~\ref{sec:discussion} concludes and discusses related
perspectives.

\section{Ancilla-Register Supercharge and Clean Protected Manifold for
Stabilizer Codes}

\label{sec:susy-stabilizer}

\renewcommand{\theequation}{2.\arabic{equation}} \setcounter{equation}{0} %
\renewcommand{\thethm}{2.\arabic{thm}} \setcounter{thm}{0} %
\renewcommand{\theprop}{2.\arabic{prop}} \setcounter{prop}{0} %
\renewcommand{\thecoro}{2.\arabic{coro}} \setcounter{coro}{0} %
\renewcommand{\thelemm}{2.\arabic{lemm}} \setcounter{lemm}{0}

This section develops, in a model-independent and intrinsic way, a
construction that embeds an arbitrary stabilizer code, of
Calderbank--Shor--Steane type or of general abelian Pauli type, into a
graded supersymmetric complex, by adjoining a fermionic ancilla-register
sector and by defining a single nilpotent supercharge $Q$. The output is the
static operator-theoretic structure used throughout the paper: a nilpotent $%
Q $, a non-negative operator $H_{\mathrm{SUSY}}=\{Q,Q^{\dagger}\}$, an
ancilla-degree grading of the enlarged Hilbert space, a Hodge decomposition
adapted to that grading, and a clean protected manifold $\mathcal{K}$ whose
degree-zero harmonic representatives are precisely the encoded code states.

The construction in this section is kinematic. Here, the identification of $%
\mathcal{K}$ follows from the algebra of $Q$ alone together with the
ancilla-degree grading that $Q$ already respects, in the standard sense in
which a supersymmetric complex is read degree by degree \cite{Witten1982}.
The scope is deliberately narrow, and stating it once prevents later
over-reading. What transfers from the general supersymmetric formalism to a
stabilizer code is the static algebraic structure alone: nilpotency, the
fermion-number grading, the Hodge decomposition, and the cohomological
identification of the code subspace. The operator $H_{\mathrm{SUSY}}$ is
used throughout as a non-negative auxiliary grading operator that resolves
syndrome sectors. The objects fixed here are the intrinsic input used later
in Sec.~\ref{sec:syndrome-fibres-KL-window} for the syndrome-resolved
projected-return blocks and the certified faithful window, in Sec.~\ref%
{sec:software-supercharge-cohomology} for the exact finite-field residual
classifier, in Sec.~\ref{sec:witten-static-decoder} for the static index
count of protected logical fibres, in Sec.~\ref{subsec:logical-resource} for
the logical non-Clifford interface, and in Sec.~\ref%
{sec:bb-circuit-reranking} for the circuit-level detector-fibre audit on the
gross code. The separation between local check information and global
logical action that motivates the whole architecture is the essential
insight from the stabilizer formalism and the Knill--Laflamme criterion \cite%
{Gottesman1997,KnillLaflamme1997,NielsenChuang2010}, and the present section
is its operator-level realization.

\subsection{Setup: Stabilizer code and ancilla-register Hilbert space}

\label{subsec:sg-setup}

Let $\mathcal{S}=\langle S_{1},\dots ,S_{r}\rangle \subset \mathcal{P}_{n}$
be an abelian Pauli stabilizer group on $n$ data qubits, with $\mathcal{H}_{%
\mathrm{data}}=(\mathbb{C}^{2})^{\otimes n}$. Each generator $S_{a}$, with $%
a=1,\dots ,r$, is Hermitian, unitary, and squares to the identity,
\begin{equation}
S_{a}^{\dagger }=S_{a},\qquad S_{a}^{2}=I,\qquad \left[ \,S_{a},S_{b}\,%
\right] =0.  \label{eq:sg-stab-algebra}
\end{equation}%
The single-stabilizer syndrome projectors are
\begin{equation}
\Pi _{a}^{0}=\tfrac{1}{2}\left( I+S_{a}\right) ,\qquad \Pi _{a}^{1}=\tfrac{1%
}{2}\left( I-S_{a}\right) ,  \label{eq:sg-syndrome-proj}
\end{equation}%
which are mutually orthogonal projectors obeying $\Pi _{a}^{1}=I-\Pi
_{a}^{0} $, $\Pi _{a}^{0}\Pi _{a}^{1}=0$, and $\left[ \,\Pi _{a}^{s},\Pi
_{b}^{s^{\prime }}\,\right] =0$ for all $a$, $b$, $s$, $s^{\prime }$. The
code subspace is the simultaneous $+1$ eigenspace of all generators,
\begin{equation}
\mathcal{C}_{\mathrm{QEC}}=\left\{ \left\vert \psi \right\rangle \in
\mathcal{H}_{\mathrm{data}}\;\middle|\;S_{a}\left\vert \psi \right\rangle
=\left\vert \psi \right\rangle \;\;\text{for all }a\right\}
=\prod_{a=1}^{r}\Pi _{a}^{0}\,\mathcal{H}_{\mathrm{data}},
\label{eq:sg-code-def}
\end{equation}%
with $\dim \mathcal{C}_{\mathrm{QEC}}=2^{k}$ and $k=n-r$ the number of
logical qubits for an independent generating set. This is the standard
stabilizer-code starting point \cite{Gottesman1997,NielsenChuang2010}. Its
complement is resolved by the violation projectors $\Pi _{a}^{1}$, and that
is the property the supercharge structure of Subsec.~\ref{subsec:sg-Q} is
closely related to.

To each stabilizer generator $S_{a}$ we adjoin one fermionic
ancilla-register mode, with creation and annihilation operators $%
c_{a}^{\dagger }$ and $c_{a}$ satisfying the canonical anticommutation
relations
\begin{equation}
\left\{ \,c_{a},c_{b}^{\dagger }\,\right\} =\delta _{ab},\qquad \left\{
\,c_{a},c_{b}\,\right\} =0,\qquad \left\{ \,c_{a}^{\dagger },c_{b}^{\dagger
}\,\right\} =0.  \label{eq:sg-cccr}
\end{equation}%
The ancilla-register Fock space is $\mathcal{H}_{\mathrm{anc}}=\bigwedge
{}^{\bullet }\mathbb{C}^{r}$, with fermion-number operator and parity
\begin{equation}
N_{\mathrm{anc}}=\sum_{a=1}^{r}c_{a}^{\dagger }c_{a},\qquad
(-1)^{F}=(-1)^{N_{\mathrm{anc}}}.  \label{eq:sg-nu}
\end{equation}%
The full Hilbert space of the construction is the graded tensor product
\begin{equation}
\mathcal{H}=\mathcal{H}_{\mathrm{data}}\otimes \mathcal{H}_{\mathrm{anc}%
},\qquad \dim \mathcal{H}=2^{n}\,2^{r}.  \label{eq:sg-Htot}
\end{equation}%
The grading $(-1)^{F}$ defined by fermion number plays the role of the
fermion parity throughout the paper.

It is convenient, and it will turn out to be indispensable in Subsec.~\ref%
{subsec:sg-hodge}, to resolve $\mathcal{H}$ into its ancilla-degree slices
from the very beginning,
\begin{equation}
\mathcal{H}=\bigoplus_{p=0}^{r}\mathcal{H}^{(p)},\qquad \mathcal{H}%
^{(p)}=\ker \left( N_{\mathrm{anc}}-p\right) =\mathcal{H}_{\mathrm{data}%
}\otimes \textstyle\bigwedge^{p}\mathbb{C}^{r},
\label{eq:sg-degree-decomposition}
\end{equation}%
\begin{equation}
\mathcal{H}^{(0)}=\mathcal{H}_{\mathrm{data}}\otimes \left\vert
0\right\rangle _{\mathrm{anc}},\qquad \dim \mathcal{H}^{(0)}=2^{n}.
\label{eq:sg-degree-zero}
\end{equation}%
The integer $p$ is the number of ancilla-register excitations, that is, the
number of check flags currently written into the register. It is the degree
of the syndrome record. The slice $p=0$ is the clean register, in which no
check has been flagged, while $p>0$ records an occupied subset of violated
checks. The operator $N_{\mathrm{anc}}$ is the fermion-number, equivalently
form-degree, operator of a standard supersymmetric complex, in which the
graded structure and its degreewise cohomology are part of the original
formulation rather than an addition to it \cite{Witten1982}. Recording the
grading explicitly at this point is what allows every cohomological
statement in Subsecs.~\ref{subsec:sg-hodge} and \ref{subsec:sg-KL} to be
degree-resolved, and it is what prevents the full kernel of the supercharge
from being confused with the degree-zero protected sector required by the
projected-return blocks of Subsec.~\ref{subsec:projected-return-faithfulness}
and by the fibre tests of Sec.~\ref{sec:syndrome-fibres-KL-window}.

\subsection{The Supercharge $Q$}

\label{subsec:sg-Q}

The supercharge that promotes the stabilizer code to the ancilla-register
construction is the nilpotent differential
\begin{equation}
Q=\sum_{a=1}^{r}\Pi _{a}^{1}\otimes c_{a}^{\dagger }=\sum_{a=1}^{r}\tfrac{1}{%
2}\left( I-S_{a}\right) \otimes c_{a}^{\dagger },  \label{eq:sg-Q-def}
\end{equation}%
with adjoint
\begin{equation}
Q^{\dagger }=\sum_{a=1}^{r}\Pi _{a}^{1}\otimes c_{a}.  \label{eq:sg-Qdag-def}
\end{equation}%
By construction $Q$ raises the ancilla-register number by one and $%
Q^{\dagger }$ lowers it by one,
\begin{equation}
\left[ \,N_{\mathrm{anc}},Q\,\right] =+Q,\qquad \left[ \,N_{\mathrm{anc}%
},Q^{\dagger }\,\right] =-Q^{\dagger }.  \label{eq:sg-degree-raising}
\end{equation}%
The grading also implies $\left[ \,(-1)^{F},Q\,\right] \neq 0$ and $\left\{
\,(-1)^{F},Q\,\right\} =0$, so that $Q$ and $Q^{\dagger }$ are odd operators
with respect to the fermion parity, exactly as a supercharge must be \cite%
{Witten1982}.

The convention in Eq.~\eqref{eq:sg-Q-def} is physical and is fixed in this
paper: the ancilla-register vacuum $\left\vert 0\right\rangle _{\mathrm{anc}%
} $ denotes the clean, no-syndrome register, and $c_{a}^{\dagger }$ creates
the excitation associated with a violation of check $a$. The opposite
particle-hole convention, obtained by replacing $c_{a}^{\dagger }$ with $%
c_{a}$ in Eq.~\eqref{eq:sg-Q-def}, is a genuine particle-hole conjugation
only after a simultaneous redefinition of both the reference vacuum and the
fermion number, that is, after the unitary exchanging $\left\vert
0\right\rangle _{a}$ and $\left\vert 1\right\rangle _{a}$ on every mode. It
reverses the signs in Eq.~\eqref{eq:sg-degree-raising} and is not used
anywhere below.

The choice of $\Pi _{a}^{1}$, and not $\Pi _{a}^{0}$, in Eq.~%
\eqref{eq:sg-Q-def} is in such a way that, the supercharge must act only on
bright directions carrying nonzero syndrome weight, and must annihilate the
protected sector. Since $\Pi _{a}^{1}$ annihilates $\mathcal{C}_{\mathrm{QEC}%
}$ for every $a$ by Eq.~\eqref{eq:sg-code-def}, the operator $Q$ attaches
one ancilla-register excitation precisely where a stabilizer violation is
present, and acts as zero on every state of $\mathcal{C}_{\mathrm{QEC}%
}\otimes \left\vert 0\right\rangle _{\mathrm{anc}}$.

Since the projectors $\Pi _{a}^{1}$ commute while the creation operators
anticommute,
\begin{equation}
Q^{2}=\sum_{a,b=1}^{r}\Pi _{a}^{1}\Pi _{b}^{1}\otimes
c_{a}^{\dagger}c_{b}^{\dagger} =\frac{1}{2}\sum_{a,b=1}^{r}\Pi _{a}^{1}\Pi
_{b}^{1}\otimes \left\{ \,c_{a}^{\dagger},c_{b}^{\dagger}\,\right\} =0 ,
\qquad \left( Q^{\dagger}\right) ^{2}=0 .  \label{eq:sg-Qsq-zero}
\end{equation}
Nilpotency is what permits the Hodge-type decomposition of Subsec.~\ref%
{subsec:sg-hodge}, exactly as in the differential-form setting on which this
construction is modelled \cite{Witten1982}. It is also the structural
reason, exploited in Sec.~\ref{sec:syndrome-fibres-KL-window}, that the
coherent syndrome record is created once and can never be overcounted.

Using $\Pi _{a}^{1\dagger }=\Pi _{a}^{1}$, $\left( \Pi _{a}^{1}\right)
^{2}=\Pi _{a}^{1}$, we give the two orderings separately,
\begin{equation}
QQ^{\dagger }=\sum_{a,b=1}^{r}\Pi _{a}^{1}\Pi _{b}^{1}\otimes c_{a}^{\dagger
}c_{b},\qquad Q^{\dagger }Q=\sum_{a,b=1}^{r}\Pi _{a}^{1}\Pi _{b}^{1}\otimes
c_{a}c_{b}^{\dagger },  \label{eq:sg-QQdag}
\end{equation}%
so that, on summing, the terms with $a\neq b$ cancel in pairs because $\Pi
_{a}^{1}\Pi _{b}^{1}$ is symmetric under $a\leftrightarrow b$ while $%
c_{a}^{\dagger }c_{b}-c_{b}c_{a}^{\dagger }$ is antisymmetric, and the terms
with $a=b$ sum by $\left\{ \,c_{a},c_{a}^{\dagger }\,\right\} =1$. The
result is
\begin{equation}
H_{\mathrm{SUSY}}=\left\{ \,Q,Q^{\dagger }\,\right\} =\sum_{a=1}^{r}\Pi
_{a}^{1}\otimes I_{\mathrm{anc}}=\frac{1}{2}\left( r\,I_{\mathrm{data}%
}-\sum_{a=1}^{r}S_{a}\right) \otimes I_{\mathrm{anc}},
\label{eq:sg-HSUSY-clean}
\end{equation}%
which is manifestly non-negative, $H_{\mathrm{SUSY}}\geq 0$. Here $I_{%
\mathrm{anc}}$ denotes the identity operator on the entire fermionic
ancilla-register Fock space $\mathcal{H}_{\mathrm{anc}}$. When a data-space
operator $W$ is lifted to the composite Hilbert space it acts as $W\otimes
I_{\mathrm{anc}}$, which must not be confused with the vacuum state $%
\left\vert 0\right\rangle _{\mathrm{anc}}$ or with the vacuum projector $%
P_{0,\mathrm{anc}}=\left\vert 0\right\rangle _{\mathrm{anc}}\left\langle
0\right\vert _{\mathrm{anc}}$ appearing in Eq.~\eqref{eq:sg-PK} below. The
distinction is that: $I_{\mathrm{anc}}$ selects nothing and reads nothing,
whereas $P_{0,\mathrm{anc}}$ enforces an empty ledger.

On any joint eigenstate $\left\vert \psi _{s}\right\rangle$ with stabilizer
eigenvalues $S_{a}\left\vert \psi _{s}\right\rangle =(-1)^{s_{a}}\left\vert
\psi _{s}\right\rangle$ and $s_{a}\in \{0,1\}$, one has
\begin{equation}
\Pi _{a}^{1}\left\vert \psi _{s}\right\rangle =s_{a}\left\vert \psi
_{s}\right\rangle , \qquad H_{\mathrm{SUSY}}\left\vert \psi
_{s}\right\rangle =\left\vert s\right\vert \left\vert \psi _{s}\right\rangle
, \qquad \left\vert s\right\vert =\sum_{a=1}^{r}s_{a} ,
\label{eq:sg-syndrome-weight}
\end{equation}
where $\left\vert s\right\vert$ is the syndrome weight and the spectrum of $%
H_{\mathrm{SUSY}}$ is the integer set $\{0,1,\dots ,r\}$. Here $s_{a}=0$
means that constraint $a$ is satisfied, while $s_{a}=1$ means that a Pauli
error has anticommuted with $S_{a}$ and flipped its eigenvalue. Thus $H_{%
\mathrm{SUSY}}$ literally counts the number of violated stabilizers and
diagnoses data-register syndrome weight \cite{Gottesman1997}. It does not,
by itself, distinguish the different ancilla-register occupations carried by
a zero-syndrome data state, because Eq.~\eqref{eq:sg-HSUSY-clean} acts as
the identity on the register factor. That blindness is the physical origin
of the raw harmonic cylinder recorded in the next subsection, and it is the
reason the clean degree-zero cut must be imposed before $\mathcal{K}$ is
handed to the projected-return blocks of Subsec.~\ref%
{subsec:projected-return-faithfulness}.

\subsection{Hodge decomposition and the clean protected manifold}

\label{subsec:sg-hodge}

The finite-dimensional Hodge decomposition of the enlarged Hilbert space
reads
\begin{equation}
\mathcal{H}=\mathcal{K}_{\mathrm{raw}}\oplus \func{im}Q\oplus \func{im}%
Q^{\dagger },\qquad \mathcal{K}_{\mathrm{raw}}=\ker Q\cap \ker Q^{\dagger
}=\ker H_{\mathrm{SUSY}},  \label{eq:sg-hodge}
\end{equation}%
the three summands being mutually orthogonal. Since $Q$ and $Q^{\dagger }$
act on the ancilla-register factor only through $c_{a}^{\dagger }$ and $%
c_{a} $, and on the data factor only through the violation projectors $\Pi
_{a}^{1} $, a product state $\left\vert \psi \right\rangle \otimes
\left\vert \eta \right\rangle _{\mathrm{anc}}$ is annihilated by both $Q$
and $Q^{\dagger }$ if and only if $\Pi _{a}^{1}\left\vert \psi \right\rangle
=0$ for every $a$, that is, if and only if $\left\vert \psi \right\rangle
\in \mathcal{C}_{\mathrm{QEC}}$. Consequently
\begin{equation}
\mathcal{K}_{\mathrm{raw}}=\ker Q\cap \ker Q^{\dagger }=\mathcal{C}_{\mathrm{%
QEC}}\otimes \mathcal{H}_{\mathrm{anc}},\qquad \dim \mathcal{K}_{\mathrm{raw}%
}=2^{k}\,2^{r}=2^{n}.  \label{eq:sg-Kraw}
\end{equation}

Equation~\eqref{eq:sg-Kraw} states plainly a fact that must not be blurred.
The raw harmonic space is a full cylinder over the code space, carrying
every possible ancilla-register record, and it is therefore too large, by
the factor $2^{r}$, to serve by itself as the protected code subspace of the
decoder. The physical reason is transparent. The supercharge writes a
violated-check pattern into the register whenever it acts nontrivially, but
nothing in $Q$ or $Q^{\dagger }$ ever erases an ancilla-register excitation
that is already present on a state for which every $\Pi _{a}^{1}$ vanishes.
A leftover, previously written flag sitting on top of a code state is
invisible to $H_{\mathrm{SUSY}}$, since Eq.~\eqref{eq:sg-HSUSY-clean} reads
only the data register, and yet such a state still lies in $\mathcal{K}_{%
\mathrm{raw}}$.

The physically relevant protected manifold is obtained by intersecting the
raw harmonic space with the degree-zero sector $\mathcal{H}^{(0)}=\ker N_{%
\mathrm{anc}}$ of Eq.~\eqref{eq:sg-degree-decomposition},
\begin{equation}
\mathcal{K}=\mathcal{K}_{\mathrm{raw}}\cap \ker N_{\mathrm{anc}}=\ker Q\cap
\ker Q^{\dagger }\cap \ker N_{\mathrm{anc}}=\mathcal{C}_{\mathrm{QEC}%
}\otimes \left\vert 0\right\rangle _{\mathrm{anc}}\cong \mathcal{C}_{\mathrm{%
QEC}},  \label{eq:sg-K-equals-CQEC}
\end{equation}%
with $\dim \mathcal{K}=2^{k}$. In other words, $\mathcal{K}=\ker H_{\mathrm{%
SUSY}}\ |_{p=0}$. This is not an addition external to the supersymmetric
complex. The operator $N_{\mathrm{anc}}$ is exactly the fermion-number
grading operator already fixed in Eq.~\eqref{eq:sg-nu}, and restricting a
graded cohomological statement to a single homogeneous degree before quoting
it is standard practice in any graded complex of Witten type \cite%
{Witten1982}, where the cohomology of definite degree $p$ is the object
identified with the harmonic representatives at that degree, and never with
the union of harmonic representatives at all degrees simultaneously. The
associated orthogonal projector is
\begin{equation}
P_{\mathcal{K}}=P_{\mathcal{C}_{\mathrm{QEC}}}\otimes P_{0,\mathrm{anc}%
},\qquad P_{0,\mathrm{anc}}=\left\vert 0\right\rangle _{\mathrm{anc}%
}\left\langle 0\right\vert _{\mathrm{anc}},\qquad P_{\mathcal{K}}^{2}=P_{%
\mathcal{K}}=P_{\mathcal{K}}^{\dagger },  \label{eq:sg-PK}
\end{equation}%
so that $P_{\mathcal{K}}\mathcal{H}=\mathcal{K}$. Both tensor factors are
essential: the first requires that no check is violated, the second that no
stale flag remains in the register. This is the projector that lands on the
clean encoded manifold in the projected-return blocks of Subsec.~\ref%
{subsec:projected-return-faithfulness} and in the faithfulness tests of Sec.~%
\ref{sec:syndrome-fibres-KL-window}, and it is the projector used again in
the index count of Sec.~\ref{sec:witten-static-decoder}.

The supercharge and its adjoint annihilate the protected manifold, giving
\begin{equation}
QP_{\mathcal{K}}=Q^{\dagger }P_{\mathcal{K}}=0,\qquad P_{\mathcal{K}}Q=P_{%
\mathcal{K}}Q^{\dagger }=0.  \label{eqsg-PK-darkness}
\end{equation}%
This operator-level identity certifies that $\mathcal{K}$ is dark and
prevents leakage into bright ancilla-excited sectors.

Because $Q$ is degree-raising rather than degree-preserving, no additional
condition survives at the bottom of the complex. On $\mathcal{H}^{(0)}$ the
operator $Q^{\dagger }$ vanishes automatically, since it would lower the
degree below zero and $\mathcal{H}^{(-1)}=0$ by Eq.~%
\eqref{eq:sg-degree-decomposition}, while $Q$ maps $\mathcal{H}^{(0)}$ into $%
\mathcal{H}^{(1)}$. Hence $\mathcal{K}=\ker \left( Q\big\vert_{\mathcal{H}%
^{(0)}}\right) =H^{0}(Q)$, with no quotient appearing, since $\mathcal{H}%
^{(-1)}=0$ and there is no incoming image at degree zero. One therefore has
the cohomological identification
\begin{equation}
\mathcal{K}=H^{0}(Q)\cong \mathcal{C}_{\mathrm{QEC}}.
\label{eq:sg-K-cohomology}
\end{equation}%
The harmonic representatives of $H^{0}(Q)$, read in the degree-zero
clean-register sense of Eq.~\eqref{eq:sg-K-cohomology}, are precisely the
encoded code states.

The graded trace familiar from supersymmetric quantum mechanics \cite%
{Witten1982} is available here. The Witten index
\begin{equation}
I_{W}=\mathrm{Tr}_{\mathcal{H}}\left[ (-1)^{F}P_{\mathcal{K}}e^{-\beta H_{%
\mathrm{SUSY}}}\right]  \label{eq:sg-witten-fullspace}
\end{equation}%
hence counts the logical dimension. The count of protected logical fibres in
the present section is therefore,
\begin{equation}
\dim H^{0}(Q)=\dim \mathcal{K}=2^{k},  \label{eq:sg-degree-zero-count}
\end{equation}%
and it must never be replaced by an unrestricted supertrace. The definition
of a deformation-pinned static index that isolates this protected count from
the vanishing full-space trace, its stabilizer-code specialization, and its
reading as a topological certificate of the number of protected logical
fibres are developed in Sec.~\ref{sec:witten-static-decoder}. No index
statement is used here to establish Eq.~\eqref{eq:sg-K-equals-CQEC}; that
identity follows directly from the stabilizer projectors and the grading.

\subsection{SUSY doublets from Pauli errors}

\label{subsec:sg-doublets}

Let $W\in \mathcal{P}_{n}$ be a Pauli error and define its anticommuting set
\begin{equation}
V(W)=\left\{ a\in \{1,\dots ,r\}\;\middle\vert\; S_{a}W=-WS_{a}\right\} .
\label{eq:sg-V-def}
\end{equation}
For $\left\vert \psi \right\rangle \in \mathcal{C}_{\mathrm{QEC}}$ the
corrupted state $W\left\vert \psi \right\rangle$ has stabilizer eigenvalues $%
s_{a}=1$ precisely for $a\in V(W)$ and $s_{a}=0$ otherwise. The graded
commutator supplies the analytic error-to-syndrome map
\begin{equation}
\left[ \,Q,\,W\otimes I_{\mathrm{anc}}\,\right] =\sum_{a\in
V(W)}WS_{a}\otimes c_{a}^{\dagger} ,  \label{eq:sg-error-to-syndrome}
\end{equation}
which vanishes identically if and only if $W$ commutes with every generator.
Equation~\eqref{eq:sg-error-to-syndrome} realizes, inside the
ancilla-register complex, the standard correspondence between a Pauli error
and the syndrome it produces \cite{Gottesman1997}, and it is the intrinsic
forward syndrome oracle whose decoder-facing reading is used in Sec.~\ref%
{sec:syndrome-fibres-KL-window}. The content of the identity is that the
syndrome is not extracted by an appended primitive but is generated by the
differential of the complex itself.

For $V(W)\neq \varnothing $ the pair
\begin{equation}
\left\vert b_{W}\right\rangle =W\left\vert \psi \right\rangle \otimes
\left\vert 0\right\rangle _{\mathrm{anc}},\qquad \left\vert
f_{W}\right\rangle =\frac{1}{\sqrt{\left\vert V(W)\right\vert }}\sum_{a\in
V(W)}W\left\vert \psi \right\rangle \otimes c_{a}^{\dagger }\left\vert
0\right\rangle _{\mathrm{anc}}  \label{eq:sg-doublet}
\end{equation}%
forms a supersymmetric doublet of energy $E=\left\vert V(W)\right\vert $,
exchanged by the canonical Hodge pairing
\begin{equation}
Q\left\vert b_{W}\right\rangle =\sqrt{\left\vert V(W)\right\vert }\left\vert
f_{W}\right\rangle ,\qquad Q^{\dagger }\left\vert f_{W}\right\rangle =\sqrt{%
\left\vert V(W)\right\vert }\left\vert b_{W}\right\rangle ,
\label{eq:sg-Hodge-pairing}
\end{equation}%
with $H_{\mathrm{SUSY}}\left\vert b_{W}\right\rangle =\left\vert
V(W)\right\vert \left\vert b_{W}\right\rangle $ and $H_{\mathrm{SUSY}%
}\left\vert f_{W}\right\rangle =\left\vert V(W)\right\vert \left\vert
f_{W}\right\rangle $. The degree bookkeeping is explicit: a syndrome-bearing
error starts in the clean degree-zero sector and is carried by $Q$ into
degree one, exactly as required by Eq.~\eqref{eq:sg-degree-raising}. This
bright-state pairing is the finite-dimensional stabilizer-code counterpart
of the standard supersymmetric pairing of nonzero-energy states \cite%
{Witten1982}, and it is the sharpest available statement of the structural
claim of the paper: the energy of the enlarged system is the syndrome
weight, and a detectable error is a bright, doublet-paired direction of the
complex rather than a mere classical label.

The spectral gap above the clean protected sector is
\begin{equation}
\Delta E=\min_{W\notin N(\mathcal{S})}\left\vert V(W)\right\vert =1,
\label{eq:sg-Emin}
\end{equation}%
whenever a nonzero single-check syndrome sector exists, the minimum being
taken over Pauli operators outside the normalizer $N(\mathcal{S})$. The
qualification is essential rather than cosmetic. Operators lying inside $N(%
\mathcal{S})$ have $V(W)=\varnothing $, are syndrome-silent, remain inside $%
\mathcal{K}$, and are therefore entirely invisible to Eq.~%
\eqref{eq:sg-error-to-syndrome} by construction rather than by accident,
since $Q$ is assembled from the very check algebra with which such an
operator commutes. No refinement of the supercharge and no deformation
preserving nilpotency can extract information that the check algebra does
not contain. Diagnosing that silent sector is the task of the logical-fibre
test of Sec.~\ref{sec:syndrome-fibres-KL-window}, of the exact finite-field
residual classifier of Sec.~\ref{sec:software-supercharge-cohomology}, and
it is the precise reason a clean syndrome is not by itself a certificate of
logical recovery.

\paragraph{The Ising-chain illustration}

The smallest example that displays both the features and usefulness of the
ancilla-register construction would be the three-qubit repetition code. Its
commuting Pauli checks and violated-check projectors are
\begin{equation}
S_{1}=Z_{1}Z_{2},\qquad S_{2}=Z_{2}Z_{3},\qquad \Pi _{a}^{(1)}=\frac{1}{2}%
\bigl(I-S_{a}\bigr).  \label{eq:ising-checks}
\end{equation}%
The protected data space is $\mathcal{C}_{\mathrm{QEC}}=\func{span}%
\{|000\rangle ,|111\rangle \}$, and one ancilla-register mode is assigned to
each bond. The supercharge therefore specializes to
\begin{equation}
Q=\Pi _{1}^{(1)}c_{1}^{\dagger }+\Pi _{2}^{(1)}c_{2}^{\dagger }=\frac{1}{2}%
(I-Z_{1}Z_{2})c_{1}^{\dagger }+\frac{1}{2}(I-Z_{2}Z_{3})c_{2}^{\dagger }.
\label{eq:ising-supercharge}
\end{equation}%
Because the two projectors commute whereas the two creation operators
anticommute, the cross terms cancel:
\begin{equation}
Q^{2}=\Pi _{1}^{(1)}\Pi _{2}^{(1)}\bigl(c_{1}^{\dagger }c_{2}^{\dagger
}+c_{2}^{\dagger }c_{1}^{\dagger }\bigr)=0.  \label{eq:ising-nilpotency}
\end{equation}%
This two-bond calculation makes the general nilpotency mechanism visible
without introducing additional code-specific structure. It is the finite
stabilizer realization of the graded pairing familiar from supersymmetric
complexes, along the line of~\cite{Witten1982}.

Let $|\psi _{L}\rangle \in \mathcal{C}_{\mathrm{QEC}}$ and apply the
middle-qubit error $W=X_{2}$. Since $X_{2}$ anticommutes with both checks,
both violation projectors act as the identity on $X_{2}|\psi _{L}\rangle $.
Hence
\begin{equation}
QX_{2}|\psi _{L}\rangle |0\rangle _{\mathrm{anc}}=X_{2}|\psi _{L}\rangle %
\bigl(c_{1}^{\dagger }+c_{2}^{\dagger }\bigr)|0\rangle _{\mathrm{anc}}.
\label{eq:ising-register-write}
\end{equation}%
Thus one action of $Q$ writes a coherent one-excitation superposition over
the two violated bonds; it does not create the doubly occupied register
state. With
\begin{equation}
|b_{2}\rangle =X_{2}|\psi _{L}\rangle |0\rangle _{\mathrm{anc}},\qquad
|f_{2}\rangle =\frac{1}{\sqrt{2}}X_{2}|\psi _{L}\rangle \bigl(c_{1}^{\dagger
}+c_{2}^{\dagger }\bigr)|0\rangle _{\mathrm{anc}},
\label{eq:ising-doublet-states}
\end{equation}%
one obtains $Q|b_{2}\rangle =\sqrt{2}|f_{2}\rangle $ and $Q^{\dagger
}|f_{2}\rangle =\sqrt{2}|b_{2}\rangle $. Consequently,
\begin{equation}
H_{\mathrm{SUSY}}|b_{2}\rangle =2|b_{2}\rangle ,\qquad H_{\mathrm{SUSY}%
}|f_{2}\rangle =2|f_{2}\rangle ,  \label{eq:ising-doublet-energy}
\end{equation}%
which reads the two violated checks as syndrome weight. This is precisely
the Pauli error--syndrome correspondence of the stabilizer formalism~\cite%
{Gottesman1997}, now represented as motion from ancilla degree zero to
degree one. The example separates two tasks cleanly: $Q$ exposes local check
violations, whereas the later logical-fibre test must decide whether a
return inside $\mathcal{K}$ is logically faithful.

\subsection{Cohomological syndrome-detectability criterion and
Knill--Laflamme Conditions}

\label{subsec:sg-KL}

The Knill--Laflamme conditions acquire a cohomological reading once the
protected manifold has been correctly identified on the degree-zero slice
through Eq.~\eqref{eq:sg-K-equals-CQEC}. Let $\mathcal{E}=\{E_{\alpha}\}$ be
a declared correctable error set and lift each element to the enlarged
Hilbert space by
\begin{equation}
\widetilde{E}_{\alpha}=E_{\alpha}\otimes I_{\mathrm{anc}} .
\label{eq:sg-Etilde}
\end{equation}
With the clean projector of Eq.~\eqref{eq:sg-PK}, the standard conditions
\cite{KnillLaflamme1997,NielsenChuang2010} read
\begin{equation}
P_{\mathcal{K}}\,\widetilde{E}_{\alpha}^{\dagger}\widetilde{E}_{\beta}\, P_{%
\mathcal{K}}=c_{\alpha \beta}\,P_{\mathcal{K}} , \qquad c_{\alpha
\beta}=c_{\beta \alpha}^{\ast} , \qquad c\geq 0 ,  \label{eq:sg-KL}
\end{equation}
with $c$ a Hermitian positive-semidefinite matrix independent of the encoded
state. Two features deserve emphasis. First, the projector is the clean one:
it is this $P_{\mathcal{K}}$, and not the projector onto the raw harmonic
cylinder $\mathcal{K}_{\mathrm{raw}}$ of Eq.~\eqref{eq:sg-Kraw}, that is
inherited by the syndrome-labelled projected-return blocks of Subsec.~\ref%
{subsec:projected-return-faithfulness} and by the certified faithful window $%
\mathcal{S}_{\mathrm{faith}}$ of Sec.~\ref{sec:syndrome-fibres-KL-window}.
Since both outer projectors contain $P_{0,\mathrm{anc}}$, the coefficients $%
c_{\alpha \beta}$ are exactly the data-code Gram coefficients, and
ancilla-register occupations cannot enter them; replacing $P_{\mathcal{K}}$
by a projector onto $\mathcal{K}_{\mathrm{raw}}$ would admit configurations
that are harmonic only because $H_{\mathrm{SUSY}}$ does not read the
register. Second, it is the off-diagonal entries with $\alpha \neq \beta$
that carry the decisive content, since they are the amplitudes for coherent
leakage between logical fibres inside $\mathcal{K}$, and a family of errors
can be individually harmless yet jointly non-correctable.

On the enlarged complex the equivalent statement is
\begin{equation}
\lbrack Q,\,\widetilde{E}_{\alpha }]_{\mathrm{gr}}\,P_{\mathcal{K}}\neq
0\iff V(E_{\alpha })\neq \varnothing ,  \label{eq:sg-error-image}
\end{equation}%
and here the subscript gr denotes graded commutator but for most relavent
situations the error operators are bosonic; so that detectable errors are
precisely those with nontrivial $Q$-image, moving a degree-zero,
clean-register reference state into a bright higher-degree direction,
whereas undetectable, that is, logical, errors commute with every $S_{a}$,
act inside $\func{End}(\mathcal{K})$, and are represented by the degree-zero
harmonic elements of $H^{0}(Q)$ identified in Eq.~\eqref{eq:sg-K-cohomology}%
. This is the precise cohomological incarnation, internal to the
ancilla-register complex, of the classical algebraic separation between
detectable and undetectable errors familiar from the stabilizer and
Calderbank--Shor--Steane literature \cite{Calderbank1996,Steane1996}, and
for Calderbank--Shor--Steane presentations the same firewall reappears as
the middle cohomology of the binary complex in Sec.~\ref%
{sec:software-supercharge-cohomology}.

Finally, the grading fixes the source projector as well as the target. For a
syndrome word $s=(s_{1},\dots ,s_{r})\in \mathbb{F}_{2}^{r}$ the data-space
syndrome projector and its clean-source lift $\widetilde{P}_{s}$ are
\begin{equation}
P_{s}=\prod_{a=1}^{r}\Pi _{a}^{s_{a}},\qquad \widetilde{P}_{s}=P_{s}\otimes
P_{0,\mathrm{anc}}.  \label{eq:sg-clean-source}
\end{equation}%
The tilde is load-bearing and must not be dropped: $P_{s}$ resolves only the
data-register syndrome sector, whereas $\widetilde{P}_{s}$ additionally
holds the ancilla register in its clean reference vacuum at the source. The
elementary block geometry
\begin{equation}
\widetilde{P}_{s}\widetilde{P}_{s^{\prime }}=\delta _{ss^{\prime }}%
\widetilde{P}_{s},\qquad \widetilde{P}_{s}P_{\mathcal{K}}=0\quad \text{for }%
s\neq 0,  \label{eq:sg-block-geometry}
\end{equation}%
expresses the orthogonality of every nonzero, that is, bright, syndrome
block to the dark manifold $\mathcal{K}$, and it is the geometry on which
the graded-commutator return blocks of Subsec.~\ref%
{subsec:projected-return-faithfulness} are built. Every projected-return
block of Sec.~\ref{sec:syndrome-fibres-KL-window}, in the collapsed
decoder-facing form $B_{s}=P_{\mathcal{K}}\,C_{s}\,\widetilde{P}_{s}$, is
therefore to be read with the degree-zero target projector of Eq.~%
\eqref{eq:sg-PK} and the clean-source lift of Eq.~\eqref{eq:sg-clean-source}%
, and the same convention governs the character-resolved compression of Sec.~%
\ref{subsec:compress-Bs} and the hand-executed audit of the Steane $%
[[7,1,3]] $ code in Sec.~\ref{subsec:steane-static-audit}.

No decoder map $s\mapsto E_{s}$ is defined in the present section, and none
is needed. The cohomological distinction that closes the section is
transparent: a detectable error has nonzero syndrome and leaves the
protected degree-zero sector; a syndrome-silent stabilizer dressing acts
trivially on the encoded fibre; and a syndrome-silent nontrivial logical
operator acts inside $\mathcal{K}$ but changes that fibre. The last
possibility is exactly why the syndrome-resolved projected-return test and
the certified faithful window of Sec.~\ref{sec:syndrome-fibres-KL-window}
are needed in addition to syndrome clearance, why the residual classifier of
Sec.~\ref{sec:software-supercharge-cohomology} is its exact finite-field
counterpart, and why the detector-fibre audit of Sec.~\ref%
{sec:bb-circuit-reranking} must repeat the same test at circuit level. The
supercharge fixes the geometry of protected, bright and syndrome-silent
directions; the decoder-facing sections act on that fixed geometry without
redefining it, e.g. without redefining $Q$, $\mathcal{K}$,$~P_{\mathcal{K}}$.

\subsection{Graded-commutator projected return blocks}

\label{subsec:graded-commutator-projected-return-blocks}

\paragraph{Purpose and scope.}

Subsections~2.1--2.5 identified the protected manifold $\mathcal{K}=\ker
Q\cap \ker Q^{\dagger }\cap \ker N_{\mathrm{anc}}$ and the clean syndrome
projectors $\widetilde{P}_{s}$. The purpose of this subsection is to
construct the return route from the persepctive of supercharge structure: A
syndrome-labelled operator that carries amplitude from a bright sector back
toward $\mathcal{K}$, defined before any decoder is introduced. The single
structural idea is to separate two tasks that must not be confused. The
differential $Q$ is universal: it records which checks are violated, but it
cannot choose an inverse. A data-space factor $C_{s}$ supplies the
syndrome-labelled direction of return, and the supercharge then tests that
direction. The subsection defines this object, proves that it is intrinsic
to the supercharge structure, and specifies the notation used everywhere
below. No decoder map $s\mapsto E_{s}$, no coset language, and no correction
operator specialty enter here: $C_{s}$ is an unspecified even data-space
operator until Subsecs.~\ref{subsec:decoder-map} and \ref%
{subsec:projected-return-faithfulness}. Every statement follows from
nilpotency of $Q$ and from orthogonality of projectors.

\paragraph{The $\mathbb{Z}_{2}$-graded commutator.}

For homogeneous operators we use
\begin{equation}  \label{eq:gr-commutator}
[A,B]_{\mathrm{gr}} = AB - (-1)^{|A||B|} BA , \qquad |A|\in\{0,1\} ,
\end{equation}
where $|A|$ is the fermion parity fixed in Subsec.~\ref{subsec:sg-setup}.
Two odd operators give an anticommutator; one even and one odd operator give
an ordinary commutator. This one symbol lets every identity below be written
once rather than case by case.

\paragraph{The syndrome-resolved odd seed.}

For each nonzero syndrome word $s$, with violated-check set $V_s = \{a : s_a
= 1\}$ and weight $w_s = |V_s|$, we introduce an odd seed $M_s$, $|M_s| = 1$%
. The structural statements below need only this oddness; for the stabilizer
realization used in the collapse we take the factorized form
\begin{equation}  \label{eq:odd-seed}
M_s = \frac{1}{\sqrt{w_s}}\, C_s\, c_s , \qquad c_s = \frac{1}{\sqrt{w_s}}
\sum_{a \in V_s} c_a , \qquad \{c_s, c_s^{\dagger}\} = 1 ,
\end{equation}
where $C_s$ acts on the data register, lifted as $C_s \otimes I_{\mathrm{anc}%
}$, and $c_s$ is the normalized collective ancilla-register annihilator
supported exactly on the violated checks. Two features of Eq.~%
\eqref{eq:odd-seed} are physical design, not ornament. First, the collective
operator is a sum, not a product. A sum of $w_s$ annihilators is odd for
every $w_s$, whereas a product $\prod_{a \in V_s} c_a$ has parity $%
(-1)^{w_s} $ and would violate $|M_s| = 1$ whenever $w_s$ is even; it would
also fail the contraction identity below for every $w_s > 1$, confining the
sectorwise collapse to the weight-one window. Second, the two factors of $%
w_s^{-1/2}$ are not a duplication. One canonically normalizes the collective
ancilla-register direction; the other cancels the amplitude $\sqrt{w_s}$
with which $Q$ writes that direction, so the projected block comes out with
unit coefficient. In words, the seed is routing data inserted \emph{into}
the nilpotent complex, not an operation appended to it: $c_s$ reads the
record the supercharge writes, and $C_s$ chooses where to send the amplitude.

\paragraph{Carrier and projected return block.}

The carrier attached to the label $s$ is the graded response of the complex
to the seed,
\begin{equation}  \label{eq:carrier}
\Xi_s \;\equiv\; [Q, M_s]_{\mathrm{gr}} \;=\; Q M_s + M_s Q ,
\end{equation}
so that $\Xi_s$ is even, $|\Xi_s| = 0$. For a general odd seed this is all
that can be said; for the factorized seed one can say more. Since $[N_{%
\mathrm{anc}}, Q] = Q$ and $[N_{\mathrm{anc}}, c_s] = -c_s$, each term of
Eq.~\eqref{eq:carrier} individually preserves the ancilla degree, giving $%
[N_{\mathrm{anc}}, \Xi_s] = 0$: the carrier is manufactured internally by
the graded algebra, in the standard spirit of supersymmetric complexes \cite%
{Witten1982}, and it respects the degree slicing exactly. The fundamental
object of this subsection is
\begin{equation}  \label{eq:return-block}
B_s \;\equiv\; P_K\, \Xi_s\, \widetilde{P}_s \;=\; P_K\, [Q, M_s]_{\mathrm{gr%
}}\, \widetilde{P}_s , \qquad s \neq 0 ,
\end{equation}
built by the chain
\begin{equation}  \label{eq:return-chain}
M_s \;\longrightarrow\; \Xi_s = [Q, M_s]_{\mathrm{gr}} \;\longrightarrow\;
B_s = P_K\, \Xi_s\, \widetilde{P}_s .
\end{equation}
Read from right to left: $\widetilde{P}_s$ selects a definite bright source
with a clean ancilla register, $\Xi_s$ transports, and $P_K$ deposits the
harmonic component. The good effect of Eq.~\eqref{eq:return-block} is what
it isolates. The block $B_s$ is the point where code physics ends and
decoder choice begins: it is a syndrome-resolved probe of return from the
bright sector into $K$, defined before any representative is selected, and
it is the common algebraic input on which every later faithfulness test acts.

\paragraph{Closure of the carrier.}

Transport along Eq.~\eqref{eq:return-chain} never leaves the nilpotent
complex.

\begin{lemm}[$Q$-closure of the carrier]
\label{lem:Q-closure} For every odd seed $M_s$,
\begin{equation}  \label{eq:carrier-closure}
[Q, \Xi_s]_{\mathrm{gr}} = 0 .
\end{equation}
\end{lemm}

\noindent {\textit{Proof}}. Since $\Xi_s$ is even, the outer bracket is an
ordinary commutator, and with $\Xi_s = Q M_s + M_s Q$, $[Q, \Xi_s]_{\mathrm{%
gr}} = Q(QM_s + M_sQ) - (QM_s + M_sQ)Q = Q^{2}M_s - M_s Q^{2} = 0$ by
nilpotency. {\hfill $\square $}\newline

The carrier is $Q$-closed, indeed $Q$-exact by construction, and one can
sharpen this: on any $Q$-closed vector $|\phi \rangle $, $\Xi _{s}|\phi
\rangle =Q\,(M_{s}|\phi \rangle )$ is $Q$-exact and hence orthogonal to the
harmonic manifold, since the Hodge decomposition makes im$\,Q$ orthogonal to
$K_{\mathrm{raw}}$; equivalently, $K\subseteq \ker Q^{\dagger }$ implies $%
P_{K}Q=0$, so the protected component of the carrier vanishes on $Q$-closed
input. The carrier therefore cannot counterfeit a harmonic class, nor mix
one into another: it interrogates only how a \emph{bright} (non-closed)
sector couples to the manifold that already exists. One more useful reading
follows immediately: on the clean source,
\begin{equation}
B_{s}=P_{K}\,(QM_{s}+M_{s}Q)\,\widetilde{P}_{s}=P_{K}\,M_{s}Q\,\widetilde{P}%
_{s},
\end{equation}%
because $P_{K}Q=0$. The return is accordingly always \textquotedblleft
write, then read\textquotedblright : the supercharge first writes the
syndrome record into the ancilla register, and the seed then reads it out.

\paragraph{The two clean-source contraction identities.}

The behaviour of the collective annihilator on the clean source window is
the algebraic heart of the construction.

\begin{lemm}[Clean-source contraction identities]
\label{lem:drop} For the factorized seed \eqref{eq:odd-seed} and every $s
\neq 0$,
\begin{equation}  \label{eq:drop-identities}
c_s\, \widetilde{P}_s = 0 , \qquad [Q, c_s]_{\mathrm{gr}}\, \widetilde{P}_s
= \sqrt{w_s}\, \widetilde{P}_s .
\end{equation}
\end{lemm}

\noindent {\textit{Proof}}. The first identity is immediate: $\widetilde{P}%
_s $ carries the factor $P_{0,\mathrm{anc}}$, and every $c_a$ annihilates
the ancilla-register vacuum, hence so does the sum $c_s$. For the second,
both factors are odd, so the bracket is the anticommutator, and only the
diagonal contractions survive:
\begin{equation*}
[Q, c_s]_{\mathrm{gr}} = \frac{1}{\sqrt{w_s}} \sum_{b=1}^{r} \sum_{a \in
V_s} \Pi^{1}_{b}\, \{c^{\dagger}_{b}, c_{a}\} = \frac{1}{\sqrt{w_s}} \sum_{a
\in V_s} \Pi^{1}_{a} .
\end{equation*}
On the source block every check in $V_s$ is violated, so $\Pi^{1}_{a}
\widetilde{P}_s = \widetilde{P}_s$ for $a \in V_s$, and the $w_s$ equal
terms give $\sqrt{w_s}\, \widetilde{P}_s$. {\hfill $\square $}\newline

Read physically, the first identity says that a clean incoming register
contains no pre-existing record for the seed to erase, so the seed cannot
act before the differential has written one. The second says that, once the
differential has acted, contracting the universal syndrome-reading operator
against the collective drain of exactly the declared modes returns the
syndrome weight as a pure number. Both hold for arbitrary $w_s$: that
freedom is the dividend of the sum form in Eq.~\eqref{eq:odd-seed}.

\paragraph{The sectorwise collapse.}

We now combine the two lemmas. Since $C_s$ is even and acts only on the data
register, the graded Leibniz rule applied to the seed gives
\begin{equation}  \label{eq:carrier-leibniz}
\Xi_s = \frac{1}{\sqrt{w_s}}\, [Q, C_s]_{\mathrm{gr}}\, c_s + \frac{1}{\sqrt{%
w_s}}\, C_s\, [Q, c_s]_{\mathrm{gr}} .
\end{equation}
Right multiplication by $\widetilde{P}_s$ kills the first term by the first
identity of Lemma~\ref{lem:drop} and converts the second by the second
identity, so
\begin{equation}  \label{eq:collapse-source}
\Xi_s\, \widetilde{P}_s = \frac{1}{\sqrt{w_s}}\, C_s \sqrt{w_s}\, \widetilde{%
P}_s = C_s\, \widetilde{P}_s ,
\end{equation}
the two normalizations cancelling exactly against $\sqrt{w_s}$. Left
multiplication by $P_K$ then gives the central result.

\begin{prop}[Sectorwise collapse of the projected return block]
\label{prop:collapse} For the factorized seed \eqref{eq:odd-seed} and every
nonzero syndrome word $s$, on the clean ancilla-register source window,
\begin{equation}  \label{eq:sectorwise-collapse}
B_s \;=\; P_K\, \Xi_s\, \widetilde{P}_s \;=\; P_K\, C_s\, \widetilde{P}_s .
\end{equation}
\end{prop}

The physical meaning is the payoff of the whole construction. Between the
two projectors, the graded transport and the bare operator $C_s$ are
indistinguishable: the ancilla register has served as a coherent
intermediary that is written and then erased exactly, leaving no register
structure in the block, with no measurement and no classical processing. A
syndrome-resolved transport problem has been compressed to an ordinary
data-space block that decoding software can test directly; this is what
makes the faithfulness test of Subsec.~\ref%
{subsec:projected-return-faithfulness} a question on the data register
alone. Two scope restrictions are firm. First, the identity is sectorwise:
it holds between $P_K$ on the left and $\widetilde{P}_s$ on the right, not
as an operator identity on all of $\mathcal{H}$. Second, it is an algebraic
fact about an arbitrary even $C_s$: it asserts neither that $C_s$ is the
correct correction nor even that $B_s \neq 0$.

\paragraph{State-level check on the supersymmetric doublet.}

The collapse is an operator statement between projectors; its sharpest check
is on the explicit doublet of Subsec.~\ref{subsec:sg-doublets}. Take the
matching fibre $V_s = V(W)$, so that $w_s = |V(W)|$. Every annihilator $c_a$
kills the register vacuum, while $c_s$ removes exactly the modes declared by
the syndrome of $W$, giving
\begin{equation}  \label{eq:seed-action-doublet}
M_s |b_W\rangle = 0 , \qquad M_s |f_W\rangle = \frac{1}{\sqrt{w_s}}\, C_s W
|\psi\rangle \otimes |0\rangle_{\mathrm{anc}} ,
\end{equation}
and hence the carrier action with unit coefficient,
\begin{equation}  \label{eq:carrier-action-doublet}
\Xi_s |b_W\rangle = C_s W |\psi\rangle \otimes |0\rangle_{\mathrm{anc}} ,
\qquad \Xi_s |f_W\rangle = \frac{1}{\sqrt{w_s}}\, Q\bigl( C_s W |\psi\rangle
\otimes |0\rangle_{\mathrm{anc}} \bigr) .
\end{equation}
The first line is Lemma~\ref{lem:drop} read on a single vector, and it also
fixes the annihilator convention: $c_s^{\dagger}$ would create on the vacuum
and send the seed the wrong way along the complex. The second line of Eq.~%
\eqref{eq:carrier-action-doublet} vanishes, on any given encoded input,
precisely when $C_s W |\psi\rangle \in C_{\mathrm{QEC}}$, that is, when the
residual preserves the code space on that state; for a Pauli residual and
every encoded input together this is the condition $C_s W \in N(S)$, the
syndrome matching imposed once $C_s$ is specialized to a correction.
Commutation with every generator is sufficient but is framed here as a
statement about states, which is the exact one. Projecting between encoded
states $|\psi\rangle, |\psi^{\prime }\rangle$ gives
\begin{equation}  \label{eq:protected-fibre-element}
\bigl( \langle\psi^{\prime }| \otimes \langle 0|_{\mathrm{anc}} \bigr)\,
P_K\, \Xi_s\, |b_W\rangle = \langle\psi^{\prime }|\, C_s W\, |\psi\rangle ,
\end{equation}
the Knill--Laflamme matrix element of the residual $C_sW$ read on one fibre
\cite{KnillLaflamme1997}. The doublet therefore separates the two conditions
a decoder must meet: the odd leg tests syndrome restoration, the projected
even leg tests restoration of encoded information. These are state-level
identities; the fibrewise audit belongs to Subsec.~\ref%
{subsec:projected-return-faithfulness}.

\paragraph{Exact darkness of the harmonic manifold.}

The protected manifold is never a source.

\begin{prop}[Exact darkness]
\label{prop:darkness} For every nonzero syndrome word $s$,
\begin{equation}  \label{eq:exact-darkness}
B_s\, P_K = 0 .
\end{equation}
\end{prop}

\noindent {\textit{Proof}}. $B_s P_K = P_K \Xi_s \widetilde{P}_s P_K$, and $%
\widetilde{P}_s P_K = 0$ for every $s \neq 0$. {\hfill $\square $}\newline

The statement follows from source--target orthogonality alone, so it is
exact and independent of the seed, of $C_{s}$, and of any later reweighting
of the family. A state already at rest in $K$ has no bright component for $%
\widetilde{P}_{s}$ to read: the return blocks are invisible on the protected
manifold.

\paragraph{Equivariance under the normalizer of the supercharge.}

\begin{coro}[$Q$-normalizer covariance]
\label{cor:equivariance} Let $U$ be an even unitary on $\mathcal{H}$
satisfying
\begin{equation}  \label{eq:covariance-fixed}
U Q U^{\dagger} = Q , \qquad U P_K U^{\dagger} = P_K ,
\end{equation}
\begin{equation}  \label{eq:covariance-moved}
U \widetilde{P}_s U^{\dagger} = \widetilde{P}_{\pi(s)} , \qquad U M_s
U^{\dagger} = M_{\pi(s)} ,
\end{equation}
where $\pi$ denotes the induced permutation of syndrome labels. Then
\begin{equation}  \label{eq:covariance-result}
U \Xi_s U^{\dagger} = \Xi_{\pi(s)} , \qquad U B_s U^{\dagger} = B_{\pi(s)} .
\end{equation}
\end{coro}

\noindent {\textit{Proof}}. Evenness of $U$ preserves the graded commutator,
hence $U\Xi _{s}U^{\dagger }=[UQU^{\dagger },UM_{s}U^{\dagger }]_{\mathrm{gr}%
}$ $=[Q,M_{\pi (s)}]_{\mathrm{gr}}=\Xi _{\pi (s)}$; conjugating the sandwich %
\eqref{eq:return-block} with the projectors' covariance then gives $%
UB_{s}U^{\dagger }=B_{\pi (s)}$. {\hfill $\square $}\newline

Equivariance says that the family $\{B_s\}$ belongs to the intrinsic
symmetry structure of the complex, not to a presentation of the labels. The
hypothesis on the seeds is a condition on the chosen returns, $C_{\pi(s)} =
U C_s U^{\dagger}$, and is not automatic: $Q$ alone never selects $C_s$. Its
practical use is that the character-resolved compression of Subsec.~\ref%
{subsec:compress-Bs}, applied through the projectors $P_{\chi}$ of a finite
Abelian symmetry group, may reorganize already certified blocks only subject
to the covariant action of Eq.~\eqref{eq:covariance-result}, and hence
remains downstream of the present definition.

\paragraph{Boundary of the statements.}

Established here, unconditionally and by algebra alone, are the definition
of the family $\{B_{s}\}$ through the chain \eqref{eq:return-chain}, the
closure of Lemma~\ref{lem:Q-closure}, the two clean-source contraction
identities of Lemma~\ref{lem:drop}, the sectorwise collapse of Proposition~%
\ref{prop:collapse}, the exact darkness of Proposition~\ref{prop:darkness},
and the covariance of Corollary~\ref{cor:equivariance}. Not established
here, and deliberately not, are the decoder map $s\mapsto E_{s}$, the
identification $C_{s}=E_{s}^{-1}$, the statement that an entire syndrome
fibre returns to one logical fibre of $K$, the certified faithful window $%
\mathcal{S}_{\mathrm{faith}}$, and the associated Knill--Laflamme Gram
condition. Those data refine the present algebraic block, in Subsecs.~\ref%
{subsec:decoder-map} and \ref{subsec:projected-return-faithfulness}, into
the certified family
\begin{equation}
B_{s}=P_{K}\,C_{s}\,\widetilde{P}_{s},\qquad s\in \mathcal{S}_{\mathrm{faith}%
},  \label{eq:certified-family}
\end{equation}%
and only there does the collapse \eqref{eq:sectorwise-collapse} acquire its
operational reading. The compression layer of Subsec.~\ref%
{subsec:compress-Bs} takes the family $\{B_{s}\}$ as already defined and
studies only its certified reorganisation; it reopens nothing established
above.

\subsection{CSS codes and optional weighting}

\label{subsec:css-codes-weighting}

\paragraph{CSS codes.}

The constructions of Subsecs.~2.1--2.6 are written for an arbitrary abelian
Pauli stabilizer group. The present subsection records the
Calderbank--Shor--Steane specialization that later supplies the two
check-family labels of Subsecs.~4.1 and~6.1 \cite{Calderbank1996,Steane1996}%
, and it reopens neither the undeformed supercharge of Subsec.~2.2, nor the
degree-zero cut of Subsec.~2.3, nor the cohomological reading of the
Knill--Laflamme conditions in Subsec.~2.5 \cite{KnillLaflamme1997}. For a
CSS code with $X$-type stabilizers $\{S_{a}^{X}\}_{a=1}^{r_{X}}$ and $Z$%
-type stabilizers $\{S_{b}^{Z}\}_{b=1}^{r_{Z}}$, with $r_{X}+r_{Z}=r$, the
supercharge of Eq.~(2.9) splits by check species,
\begin{equation}
Q\;=\;Q_{X}\;+\;Q_{Z},  \label{eq:css-split}
\end{equation}%
\begin{equation}
Q\;=\;\sum_{a=1}^{r_{X}}\Pi _{a}^{X,(1)}\otimes c_{a,X}^{\dagger
}+\sum_{b=1}^{r_{Z}}\Pi _{b}^{Z,(1)}\otimes c_{b,Z}^{\dagger },
\end{equation}%
\begin{equation}
Q_{X}\;=\;\sum_{a=1}^{r_{X}}\Pi _{a}^{X,(1)}\otimes c_{a,X}^{\dagger
},\qquad Q_{Z}\;=\;\sum_{b=1}^{r_{Z}}\Pi _{b}^{Z,(1)}\otimes
c_{b,Z}^{\dagger },  \label{eq:css-pieces}
\end{equation}%
with separate ancilla-register sectors for the $X$- and the $Z$-syndrome,
and with the violated-check projectors $\Pi _{a}^{X,(1)}=\tfrac{1}{2}%
(I-S_{a}^{X})$ and $\Pi _{b}^{Z,(1)}=\tfrac{1}{2}(I-S_{b}^{Z})$ of
Subsec.~2.1, each of which answers the yes-or-no question whether its check
has fired and here the letters $X$ and $Z$ name check species only. They
must not be merged with the barred logical species $H_{\overline{X}}^{1}$
and $H_{\overline{Z}}^{1}$ classified in Subsec.~\ref%
{subsec:full-logical-address}, and they must not be read as the species of
the Pauli that fires a given check, since an $X$-type check responds to the $%
Z$-type component of a residual and therefore fires on $Y$-type support as
well.

\paragraph{Bicomplex structure and common darkness.}

The binary shadow of the CSS presentation is the orthogonality of the two
incidence matrices whose rows are the supports of the two families,
\begin{equation}
H_{X}H_{Z}^{\mathsf{T}}=0,\qquad H_{Z}H_{X}^{\mathsf{T}}=0,
\label{eq:css-orthogonality-phys}
\end{equation}%
over $\mathbb{F}_{2}$,$~$which is the finite-field statement used again in
Subsec.~\ref{subsec:full-logical-address}. At the operator level the
corresponding fact is $[\Pi _{a}^{X,(1)},\Pi _{b}^{Z,(1)}]=0$, valid in any
CSS presentation \cite{Gottesman1997}, and together with the anticommutation
of distinct ancilla-register creators it yields the two-family identities
\begin{equation}
Q_{X}^{2}=0,\qquad Q_{Z}^{2}=0,\qquad \left\{ Q_{X},Q_{Z}\right\} =0,\qquad
Q^{2}=0.  \label{eq:css-bicomplex}
\end{equation}%
Anticommutation in Eq.~\eqref{eq:css-bicomplex} makes the physical complex a
bicomplex, and nothing more: the two differentials act on one and the same
data register, so no tensor-product factorization of the cohomology is
available without further hypotheses, and a dimension count already refutes
one, since for the Steane code the two one-family dark spaces have dimension
sixteen each while the protected manifold has dimension two. The
load-bearing degree-zero statement is instead the common-darkness condition,
namely that a clean-register state is dark precisely when both check
families are satisfied, $\mathcal{K}=\ker Q\cap \ker Q^{\dagger }\cap \ker
N_{\mathrm{anc}}=\mathcal{C}_{\mathrm{QEC}}\otimes |0\rangle _{\mathrm{anc}}$%
, where the second equality uses the fact, established in Subsec.~2.3, that
the adjoint annihilates the empty-register sector because it would lower the
degree below zero; Eq.~\eqref{eq:css-weighted-darkness} is therefore the
manifold already identified in Eq.~(2.20). After demotion the two species
recombine differently, as a direct sum rather than a product: the completed
classifier of Subsec.~\ref{subsec:full-logical-address} is the direct sum of
the two conjugate sectors, of total dimension $2k$ over $\mathbb{F}_{2}$,
and it is that sum, not any product, which the weighted construction of
Subsec.~7.1 rescales. Physical darkness is an intersection of kernels, the
software address is a sum of conjugate logical sectors, and the two
combination rules are compatible and intended.

\paragraph{Concrete examples.}

The Steane code $[[7,1,3]]$ has $r_{X}=r_{Z}=3$, and its certification
ledger of Subsec.~2.7 closes already on the weight-one window \cite%
{Steane1996,Calderbank1996}; the corresponding static audit is carried out
in Subsec.~\ref{subsec:steane-static-audit}. The bivariate-bicycle gross
code $[[144,12,12]]$ measures one check per data qubit, that is seventy-two
per species, of which $r=132$ are independent because $k=12$ \cite%
{Bravyi2024}; it is the setting of the circuit-level bridge of Sec.~\ref%
{sec:bb-circuit-reranking}. For both codes Eq.~%
\eqref{eq:css-weighted-darkness} states that the protected manifold is read
by requiring every check of both species to be satisfied.

\paragraph{Optional weighted deformation for the CSS case.}

The two family amplitudes of Subsec.~6.1 can be discussed already at the
physical layer. For two real positive numbers, define
\begin{equation}
Q(t_{X},t_{Z})\;=\;t_{X}\,Q_{X}\;+\;t_{Z}\,Q_{Z},\qquad t_{X}>0,\qquad
t_{Z}>0,\qquad Q(1,1)=Q.  \label{eq:css-weighted-supercharge}
\end{equation}%
This pair is the family-uniform CSS restriction, chosen because one physical
bias parameter must not be split across arrow directions. Equation~%
\eqref{eq:css-weighted-supercharge} defines neither a new code nor a new
protected sector, and it is not offered as a parent formula for later
non-commuting presentations.

The cancellation that proves $Q^{2}=0$ in Subsec.~2.2 uses only the symmetry
of the projector product under exchange of the two check labels against the
antisymmetry of the corresponding creator product, so it is blind to the
coefficients, and Eq.~\eqref{eq:css-bicomplex} gives at once
\begin{equation}
Q(t_{X},t_{Z})^{2}\;=\;t_{X}^{2}\,Q_{X}^{2}+t_{X}t_{Z}\left\{
Q_{X},Q_{Z}\right\} +t_{Z}^{2}\,Q_{Z}^{2}\;=\;0.
\label{eq:css-weighted-nilpotency}
\end{equation}%
Since the anticommutator is bilinear in the differential, the associated
non-negative operator of Subsec.~2.2 carries the amplitudes quadratically,
\begin{equation}
H_{\mathrm{SUSY}}(t_{X},t_{Z})\;=\;\left\{
Q(t_{X},t_{Z}),\,Q(t_{X},t_{Z})^{\dagger }\right\}
\;=\;t_{X}^{2}\sum_{a=1}^{r_{X}}\Pi _{a}^{X,(1)}\otimes I_{\mathrm{anc}%
}\;+\;t_{Z}^{2}\sum_{b=1}^{r_{Z}}\Pi _{b}^{Z,(1)}\otimes I_{\mathrm{anc}}.
\label{eq:css-weighted-hamiltonian}
\end{equation}%
Let $\lvert \psi _{m_{X},m_{Z}}\rangle $ be a normalized simultaneous
eigenstate of all commuting violated-check projectors. Applying Eq.~%
\eqref{eq:css-weighted-hamiltonian} therefore gives
\begin{equation}
H_{\mathrm{SUSY}}(t_{X},t_{Z})\lvert \psi _{m_{X},m_{Z}}\rangle =\bigl(%
t_{X}^{2}m_{X}+t_{Z}^{2}m_{Z}\bigr)\lvert \psi _{m_{X},m_{Z}}\rangle .
\end{equation}%
So a simultaneous syndrome eigenstate with $m_{X}$ violated $X$-type checks
and $m_{Z}$ violated $Z$-type checks carries the syndrome energy, and the
first excitation above the dark manifold costs,
\begin{equation}
E(m_{X},m_{Z})\;=\;t_{X}^{2}\,m_{X}+t_{Z}^{2}\,m_{Z},\qquad \Delta _{\min
}(t_{X},t_{Z})\;=\;\min \!\left\{ t_{X}^{2},\,t_{Z}^{2}\right\} .
\label{eq:css-weighted-energy}
\end{equation}%
The two amplitudes are thus read physically as excitation per violated check
of each species. Their action is confined to the bright sectors, since they
rescale energies and never enter the dark condition $\Pi _{a}^{X,(1)}|\psi
\rangle =\Pi _{b}^{Z,(1)}|\psi \rangle =0$ of Subsec.~2.3, and consequently
the protected manifold of Eq.~\eqref{eq:css-weighted-darkness} is rigid
throughout the open positive quadrant,
\begin{equation}
\mathcal{K}(t_{X},t_{Z})\;=\;\mathcal{K}\;\cong \;\mathcal{C}_{\mathrm{QEC}%
},\qquad t_{X}>0,\qquad t_{Z}>0.  \label{eq:css-weighted-darkness}
\end{equation}%
Neither the exact-darkness statement of Subsec.~2.6, nor the projected
return blocks built from $\mathcal{K}$, acquires any amplitude dependence;
only the relative illumination of the two bright species moves.

\paragraph{Conventions and scope}

Two conventions are declared here because neither is forced by the algebra.
Reality of the amplitudes is one of them: Eq.~%
\eqref{eq:css-weighted-nilpotency} would survive complex coefficients, only
their moduli reach Eq.~\eqref{eq:css-weighted-hamiltonian}, and a phase is
removable by the ancilla-register redefinitions $c_{a,X}^{\dagger }\mapsto
e^{\mathrm{i}\varphi _{X}}c_{a,X}^{\dagger }$ and $c_{b,Z}^{\dagger }\mapsto
e^{\mathrm{i}\varphi _{Z}}c_{b,Z}^{\dagger }$, which preserve the relations
of Subsec.~2.1, so real positive amplitudes are a gauge-fixed and physically
complete choice matching the positive likelihood ratios used in Subsec.~6.2.
The second convention concerns layers: the amplitudes multiply the
differential linearly in Eq.~\eqref{eq:css-weighted-supercharge} while the
penalties of Eq.~\eqref{eq:css-weighted-hamiltonian} are quadratic, so if
one prefers to specify linear penalties $w_{X}$ and $w_{Z}$ directly, then $%
t_{X}=\sqrt{w_{X}}$ and $t_{Z}=\sqrt{w_{Z}}$, a choice of layer between the
present subsection and Subsec.~6.1. We note that the overall amplitude scale
is physical here, since it fixes the absolute penalties and the first
excitation penalty of Eq.~\eqref{eq:css-weighted-energy}, in this convention.

\subsection{Protected logical sector and external non-Clifford resource}

\label{subsec:logical-resource}

The ancilla-register construction of Subsecs.~2.1--2.3 has fixed the clean
protected manifold of the nilpotent complex, $K=\ker Q\,\cap \,\ker
Q^{\dagger }\,\cap \,\ker N_{\mathrm{anc}}\;=\;\mathcal{C}_{\mathrm{QEC}%
}\otimes |0\rangle _{\mathrm{anc}}$, together with its orthogonal projector $%
P_{K}$. This is the unique clean protected manifold the construction admits,
and it is the landing arena of every certified recovery in this paper. The
present subsection records one kinematic fact about that same manifold, a
fact that is not part of the identification of $K$ itself: the dark sector
is not memory-only, and an ideal logical non-Clifford resource, supplied
from outside the complex, may be chosen so as to act strictly inside it. The
statement is static and operator-theoretic throughout. No decoder-selected
correction, no syndrome fibre, no residual classifier, no finite-field
differential and no extraction or gadget circuit is introduced here; the
decoder-facing audit of faults associated with such a resource is developed
in Secs.~3, 5 and~6 with the objects defined there.

Fix once and for all a two-dimensional logical subspace of the protected
manifold, equivalently one designated logical tensor factor of the encoded
space,
\begin{equation}
K_{L}\;=\;\mathrm{span}\bigl\{\,|0_{L}\rangle \,,\,|1_{L}\rangle \,\bigr\}%
\;\subseteq \;K,\qquad \langle i_{L}|j_{L}\rangle \;=\;\delta _{ij},\qquad
i,j\in \{0,1\}.  \label{eq:logical-factor}
\end{equation}%
Since $K_{L}$ is a subspace of the manifold already established in Sec. 2.3,
its basis vectors inherit exact darkness,
\begin{equation}
Q\,|j_{L}\rangle \;=\;0,\qquad Q^{\dagger }\,|j_{L}\rangle \;=\;0,\qquad H_{%
\mathrm{SUSY}}\,|j_{L}\rangle \;=\;0,\qquad j\in \{0,1\},
\label{eq:inherited-darkness}
\end{equation}%
the last relation following from $H_{\mathrm{SUSY}}=\{Q,Q^{\dagger }\}$.
This is inherited darkness rather than a new protection theorem. Equation~%
\eqref{eq:inherited-darkness} is the restriction of the harmonic
identification of Subsec.~2.3 to a chosen logical frame inside $\mathcal{C}_{%
\mathrm{QEC}}$, in the standard stabilizer formalism \cite{Gottesman1997},
and the protection it expresses is exactly the one already established by
the supercharge complex: the chosen factor is a clean, zero-syndrome sector
of the same complex that carries the encoded memory, so no stabilizer check
is violated on it and no flag is raised in the ancilla register. Selecting $%
K_{L}$ therefore introduces no additional ancilla-register occupation, no
enlargement of the encoded space, and no second protected-manifold
definition.

An encoded Clifford operation is required only to preserve the declared
sector. Writing $C_{L}$ for such a Clifford operation, understood throughout
as an effective operation on the encoded qubit rather than as a physical
operator on the enlarged Hilbert space, and denoting by $\mathrm{Cliff}_{L}$
the encoded Clifford group on the designated logical factor, the requirement
is
\begin{equation}
C_{L}\,K_{L}\;\subseteq \;K_{L},\qquad C_{L}\in \mathrm{Cliff}_{L}.
\label{eq:clifford-invariance}
\end{equation}%
The Sec.~2.3 establishes the harmonic identification for the encoded sector.

On the ordered basis of $K_{L}$ the resource is the encoded $\pi / 8$ phase
operator,
\begin{equation}  \label{eq:resource-operator}
T_{L} \, | 0_{L} \rangle \;=\; | 0_{L} \rangle , \qquad T_{L} \, | 1_{L}
\rangle \;=\; e^{ \, i \pi / 4 } \, | 1_{L} \rangle , \qquad T_{L} \;=\;
\mathrm{diag} \bigl( 1 , \, e^{ \, i \pi / 4 } \bigr) \quad \text{on} \quad
K_{L} ,
\end{equation}
and the encoded state it prepares from $| + _{L} \rangle = ( | 0_{L} \rangle
+ | 1_{L} \rangle ) / \sqrt{2}$ is
\begin{equation}  \label{eq:resource-state}
| T_{L} \rangle \;=\; T_{L} \, | + _{L} \rangle \;=\; \frac{1}{\sqrt{2}} %
\Bigl( \, | 0_{L} \rangle \; + \; e^{ \, i \pi / 4 } \, | 1_{L} \rangle \, %
\Bigr) , \qquad T_{L} \, K_{L} \;\subseteq\; K_{L} , \qquad | T_{L} \rangle
\in K_{L} \subseteq K .
\end{equation}
The compatibility demand placed on the resource is exhausted by the two
inclusions displayed in Eq.~\eqref{eq:resource-state}. Because $T_{L}$ is
diagonal in a frame annihilated by both $Q$ and its adjoint $Q^{\dagger}$,
linearity gives at once
\begin{equation}  \label{eq:resource-darkness}
Q \, T_{L} \, | \psi_{L} \rangle \;=\; 0 , \qquad Q^{\dagger} \, T_{L} \, |
\psi_{L} \rangle \;=\; 0 \qquad \text{for every} \qquad | \psi_{L} \rangle
\in K_{L} ,
\end{equation}
and in particular $Q \, | T_{L} \rangle = Q^{\dagger} \, | T_{L} \rangle = 0$%
. Linearity applies to superpositions no less than to basis vectors, so the
encoded coherences of a state on the declared frame, and not merely its
encoded populations, lie in the dark sector.

The supercharge of Subsec.~2.2 is a detector rather than an actuator: it
attaches one ancilla-register excitation exactly where a stabilizer
constraint is violated, and it annihilates the protected sector. It can
therefore certify that a state or an operator is dark, that is, carries no
violated check, but it does not prepare $|T_{L}\rangle $ or implement $T_{L}$%
. The eqs.~\eqref{eq:resource-state} and \eqref{eq:resource-darkness} show
that the encoded computation and the encoded memory thus live in one and the
same dark sector, so a single protected-manifold unified description
suffices for both, and the bright and dark distinction fixed in Subsec.~2.3
survives unchanged when one passes from storage language to encoded-gate
language.

The resource is therefore supplied from outside the complex, which is also
the standard completion route around the obstruction to a transversal
universal encoded gate set \cite{BravyiKitaev2005},\cite{Postler:2021ddz}.
The result stated in this subsection here is the compatibility of a declared
external resource with the harmonic protected manifold: no resource factory,
preparation schedule, distillation yield, is presupposed or implied. The
supercharge audits injection without changing $\mathcal{K}$.

\section{Syndrome Fibres, Projected Return, and the Certified Faithful Window%
}

\label{sec:syndrome-fibres-KL-window}

\renewcommand{\theequation}{3.\arabic{equation}} \setcounter{equation}{0} %
\renewcommand{\thethm}{3.\arabic{thm}} \setcounter{thm}{0} %
\renewcommand{\theprop}{3.\arabic{prop}} \setcounter{prop}{0} %
\renewcommand{\thelemm}{3.\arabic{lemm}} \setcounter{lemm}{0}

This section establishes the operational bridge between the microscopic
supercharge operator and classical decoding software \cite%
{KnillLaflamme1997,Gottesman1997,NielsenChuang2010}. The kinematics
established in Sec.~2 adjoined a fermionic ancilla register to the data
Hilbert space, constructing a nilpotent supercharge $Q$ with $Q^2 = 0$, its
non-negative Hamiltonian $H_{\mathrm{SUSY}} = \{Q, Q^\dagger\}$, and the
clean BPS harmonic manifold
\begin{equation}
\mathcal{K} = \ker Q \cap \ker Q^\dagger \cap \ker N_{\mathrm{anc}} =
\mathcal{C}_{\mathrm{QEC}} \otimes |0\rangle_{\mathrm{anc}}, \qquad \dim
\mathcal{K} = 2^k,  \label{eq:sec3-K}
\end{equation}
which is intrinsically identified with the protected codespace. The upstream
complex determines which stabilizer constraints have been violated, but it
does not specify which classical recovery operator to apply. A decoder may
clear every measured syndrome while inadvertently applying a non-trivial
logical operator that destroys the encoded quantum information \cite%
{Dennis2002,Terhal2015}.

The mandate of this section is to close this gap by formulating an exact,
address-resolved certification layer. The purpose of this layer is to
guarantee that a decoder-selected correction returns every error in a
declared same-syndrome family to the correct logical fibre within $\mathcal{C%
}_{\mathrm{QEC}}$, rather than merely returning amplitude to the codespace.
The organizational architecture follows the ordered recovery workflow
\begin{equation}
Q\;\longrightarrow \;s\;\longrightarrow \;\mathcal{F}(s)\;\longrightarrow
\;E_{s}=\mathcal{R}_{\mathrm{dec}}(s)\;\longrightarrow
\;C_{s}\;\longrightarrow \;B_{s}\;\longrightarrow \;R_{s},
\label{eq:sec3-pipeline}
\end{equation}%
where the last arrow denotes a diagnostic relation. The passing addresses
form the certified faithful window $\mathcal{S}_{\mathrm{faith}}$, used
downstream in Subsection~\ref{subsec:compress-Bs} and Section~\ref%
{sec:bb-circuit-reranking}.

\subsection{Syndrome-silent logical failure and the target event}

\label{subsec:fs-target}

Let $\mathcal{S} = \langle S_1, \dots, S_r \rangle \subset \mathcal{P}_n$ be
an abelian Pauli stabilizer group on $n$ physical data qubits, with
normalizer $N(\mathcal{S})$ and protected codespace $\mathcal{C}_{\mathrm{QEC%
}} \cong \mathcal{K}$. Overall Pauli phases are suppressed throughout. The
extraction apparatus evaluates the syndrome homomorphism $\sigma: \mathcal{P}%
_n \to \mathbb{F}_2^r$, whose binary components are given by
\begin{equation}
[\sigma(W)]_a = 0 \quad \text{if } [S_a, W] = 0, \qquad [\sigma(W)]_a = 1
\quad \text{if } \{S_a, W\} = 0.  \label{eq:sec3-sigma}
\end{equation}
The kernel of this map is precisely $N(\mathcal{S})$. In gauge-theoretic
terms, stabilizer checks act as local constraints, and an observed syndrome
bitstring $s = \sigma(E_{\mathrm{true}})$ marks the localized violations
produced by a physical fault $E_{\mathrm{true}} \in \mathcal{P}_n$. Because $%
N(\mathcal{S})$ contains both harmless stabilizer dressing operators and
non-trivial logical Paulis, $\sigma$ is non-injective and identifies an
entire coset of possible Pauli explanations.

The classical decoder observes the word $s$, selects an explanatory Pauli
candidate $E_s$, and applies the unitary recovery operator $C_s = E_s^{-1}$.
The net physical transformation on the protected memory is governed by the
static residual operator
\begin{equation}
R_s = C_s E_{\mathrm{true}} \in \mathcal{P}_n.  \label{eq:sec3-residual}
\end{equation}
The operator $R_s$ is an offline audit variable that depends on the
unobserved physical history $E_{\mathrm{true}}$. A syndrome-matched
correction ensures that $\sigma(R_s) = 0$. Recovery succeeds when $R_s \in
\mathcal{S}$, acting trivially on the codespace. The dangerous alternative
is the syndrome-silent logical failure
\begin{equation}
R_s \in N(\mathcal{S}) \setminus \mathcal{S}, \qquad \sigma(R_s) = 0,
\label{eq:sec3-silent-event}
\end{equation}
in which every check is satisfied, yet an undetected logical Pauli operation
has acted. The target quantity of this section is its probability,
\begin{equation}
p_{\mathrm{fs}} = \mathrm{Prob}\big(R_s \in N(\mathcal{S}) \setminus
\mathcal{S} \;\text{and}\; \sigma(R_s) = 0\big).  \label{eq:sec3-pfs}
\end{equation}
The second clause emphasizes invisibility and follows mathematically from
the first. All local checks remain quiet while the encoded state has changed
\cite{Dennis2002,Terhal2015}.

\subsection{Pauli fibres, decoder maps, and online relative logical address}

\label{subsec:decoder-map}

The collection of all physical Pauli errors consistent with an observed
syndrome $s$ constitutes the Pauli fibre
\begin{equation}
\mathcal{F}(s) = \sigma^{-1}(s) = E_s N(\mathcal{S}) \quad (\text{modulo
overall phases}).  \label{eq:sec3-fibre}
\end{equation}
Two distinct physical degeneracies coexist within each fibre. Harmless
stabilizer dressing occurs when an alternative explanation takes the form $%
E_s g$ with $g \in \mathcal{S}$, which acts trivially on the codespace $%
\mathcal{K}$. Harmful logical degeneracy occurs when an alternative
explanation takes the form $E_s L$ with $L \in N(\mathcal{S}) \setminus
\mathcal{S}$, which changes the logical state while presenting the same
syndrome. A classical decoding algorithm is formalized as an explanation
section on a declared working domain $\Sigma_{\mathrm{corr}} \subseteq
\mathbb{F}_2^r$,
\begin{equation}
\mathcal{R}_{\mathrm{dec}}: \Sigma_{\mathrm{corr}} \to \mathcal{P}_n, \qquad
s \mapsto E_s \in \mathcal{F}(s), \qquad \mathcal{R}_{\mathrm{dec}}(0) = I.
\label{eq:sec3-decoder-map}
\end{equation}
The corresponding recovery operation and its invariant stabilizer coset are
given by
\begin{equation}
C_s = E_s^{-1}, \qquad C_0 = I, \qquad C_s \mathcal{S} = \{C_s g : g \in
\mathcal{S}\}.  \label{eq:sec3-correction-coset}
\end{equation}
For nonzero $s$, the correction must satisfy $\sigma(C_s) = s$; requiring $%
C_s \in N(\mathcal{S})$ would prevent it from removing the check violations.
The inversion applies specifically to the selected Pauli representative and
does not invert a general noisy quantum channel.

To quantify the representational economy of a candidate explanation, one
defines the stabilizer-invariant coset cost
\begin{equation}
J_s = \min_{g \in \mathcal{S}} \mathrm{wt}(E_s g).  \label{eq:sec3-cost}
\end{equation}
The cost $J_s$ serves as a practical likelihood surrogate under independent
error models. However, minimizing $J_s$ does not guarantee logical
correctness in degenerate quantum LDPC codes, because a lower-weight
operator can belong to an incorrect logical coset \cite{Bravyi2024}.

When a base decoder produces several explanations of one word, write their
corrections as
\begin{equation}
\mathcal{C}_{s}=\{C_{s,j}=E_{s,j}^{-1}\}_{j=1}^{M_{s}},\qquad E_{s,j}\in
\mathcal{F}(s),\qquad \sigma (C_{s,j})=s.  \label{eq:sec3-candidate-list}
\end{equation}%
For offline audit, each candidate induces an absolute residual $%
R_{s,j}=C_{s,j}E_{\mathrm{true}}\in N(\mathcal{S})$. Writing $\ell $ for the
logical-class map of the stabilizer formalism \cite{Gottesman1997},
\begin{equation}
\ell :N(\mathcal{S})\rightarrow N(\mathcal{S})/\mathcal{S}\cong \mathbb{F}%
_{2}^{2k},\qquad \ell (W)=[W]_{\mathcal{S}},
\label{eq:sec3-logical-class-map}
\end{equation}%
the absolute logical address $\ell (R_{s,j})$ is an offline audit variable
requiring $E_{\mathrm{true}}$. What a running engine can compute online from
candidate data alone is the relative logical address between two candidate
explanations,
\begin{equation}
\Delta _{jj^{\prime }}=\ell \big(C_{s,j}C_{s,j^{\prime }}^{-1}\big)=\ell %
\big(R_{s,j}R_{s,j^{\prime }}^{-1}\big),\qquad \Delta _{jj^{\prime }}=0\
\Longleftrightarrow \ C_{s,j}\mathcal{S}=C_{s,j^{\prime }}\mathcal{S}.
\label{eq:sec3-relative-address}
\end{equation}%
Because $(C_{s,j}E_{\mathrm{true}})(C_{s,j^{\prime }}E_{\mathrm{true}%
})^{-1}=C_{s,j}C_{s,j^{\prime }}^{-1}$, the unknown physical fault $E_{%
\mathrm{true}}$ cancels identically in this quotient. If $\ell
(C_{s,j}C_{s,j^{\prime }}^{-1})=0$, the two candidates differ only by a
stabilizer and are logically equivalent; if $\ell (C_{s,j}C_{s,j^{\prime
}}^{-1})\neq 0$, they represent rival logical actions. This relative
quotient partitions the list into distinct logical classes without requiring
knowledge of the true error.

Section 4 represents these relative logical classes in $H_{\mathrm{full}%
}^{1} $, enabling online candidate partitioning and ambiguity detection e.g.
$I_{\mathrm{amb}}(s)$ in \eqref{eq:ambflag}. Section 5 audits their
correspondence with circuit-level logical addresses, via the static--circuit
bridge, before having class-based decoder reranking.

\subsection{Projected return and Knill--Laflamme faithfulness}

\label{subsec:projected-return-faithfulness}

The projected block tests the route into $\mathcal{K}$. We now ask whether
that route reaches the correct logical address. Let $\Sigma_{\mathrm{adm}}$
be a declared address set. For $s \in \Sigma_{\mathrm{adm}}$, fix the
audited Pauli family
\begin{equation}
\mathcal{F}_s = \{E_{s,a}\}_{a \in I_s} \subseteq \mathcal{F}(s), \qquad
\widetilde{P}_s E_{s,a} P_K = E_{s,a} P_K.  \label{eq:audited-family}
\end{equation}
At an address considered for certification, it includes the decoder's
selected explanation $E_{s,j^{\star}(s)}$. The restriction to a declared
family is essential: if $k > 0$, the full fibre contains logically
inequivalent errors, which one fixed $C_s$ cannot all correct.

The sectorwise collapse of Subsection~\ref%
{subsec:graded-commutator-projected-return-blocks} yields the decoder-facing
projected-return block
\begin{equation}
B_s = P_K C_s \widetilde{P}_s, \qquad s \neq 0.
\label{eq:projected-return-block}
\end{equation}
The source selects the clean syndrome sector, $C_s$ supplies the proposed
return, and $P_K$ keeps the protected component. Arrival is not a test of
logical identity,
\begin{equation}
B_s \neq 0 \;\not\Longrightarrow\; \text{the return at } s \text{ is a
logical identity on } \mathcal{K}.  \label{eq:return-not-faithfulness}
\end{equation}
Multiplying a valid correction by a non-trivial logical Pauli preserves
syndrome clearance while rotating the encoded logical state.

The chosen correction must act as a scalar on the selected branch for every
protected input, so that $P_{K}C_{s}E_{s,j^{\star }(s)}P_{K}=\lambda
_{s}P_{K}$,$~\lambda _{s}\in \mathbb{C}$. A syndrome does not specify which
declared branch occurred. The same correction must therefore satisfy
\begin{equation}
P_{K}C_{s}E_{s,a}P_{K}=\lambda _{s,a}P_{K},\qquad \lambda _{s,a}\in \mathbb{C%
},\qquad \forall a\in I_{s}.  \label{eq:all-branch-scalar-return}
\end{equation}%
For syndrome-matched Paulis, $C_{s}E_{s,a}\in N(\mathcal{S})$. Its action on
$\mathcal{K}$ is scalar precisely when it is a stabilizer up to phase. Thus
Eq.~(\ref{eq:all-branch-scalar-return}) says that the entire audited family
lies in the selected stabilizer coset $E_{s}\mathcal{S}$, up to phases, and $%
|\lambda _{s,a}|=1$. This is logical homogeneity: errors that the declared
record cannot distinguish reach the same encoded address under this
decoder's return.

The fibrewise Knill--Laflamme condition \cite%
{KnillLaflamme1997,NielsenChuang2010} is
\begin{equation}
P_K E_{s,a}^\dagger E_{s,b} P_K = h_{ab}^{(s)} P_K, \qquad \forall a, b \in
I_s.  \label{eq:fibrewise-kl-condition}
\end{equation}
In the present fixed-Pauli setting it follows directly from Eq.~(\ref%
{eq:all-branch-scalar-return}). Inserting $C_s^\dagger C_s = I$ between the
two errors yields
\begin{equation}
P_K E_{s,a}^\dagger E_{s,b} P_K = P_K E_{s,a}^\dagger C_s^\dagger C_s
E_{s,b} P_K = \bar{\lambda}_{s,a} \lambda_{s,b} P_K = h_{ab}^{(s)} P_K.
\label{eq:kl-derivation}
\end{equation}
This establishes that $h_{ab}^{(s)} = \bar{\lambda}_{s,a} \lambda_{s,b}$ is
a positive-semidefinite matrix of rank one, ensuring that no state-dependent
information leaks across different error branches.

\begin{definition}[Fibrewise Faithfulness Predicate]
\label{def:faith-predicate} For fixed data $(\widetilde{P}%
_s,C_s,j^{\star}(s),\mathcal{F}_s)$, define
\begin{equation}
\mathrm{Faith}_{\mathrm{KL}}(s; C_s, \mathcal{F}_s) = 1
\label{eq:faith-predicate}
\end{equation}
when Eq.~(\ref{eq:all-branch-scalar-return}) holds for every declared
branch, with Eq.~(\ref{eq:fibrewise-kl-condition}) recorded as its
Pauli-family consequence; otherwise, set the predicate to zero.
\end{definition}

This predicate uses fixed code, decoder, and family data and does not
inspect the unobserved fault $E_{\mathrm{true}}$.

\subsection{$Q$ as forward syndrome oracle}

\label{subsec:forward-oracle}

The operator supercharge $Q$ acts as an intrinsic forward syndrome oracle
through the graded commutator algebra of the supersymmetric complex \cite%
{Witten1982}. For an arbitrary Pauli operator $W$ with violated check set $%
V(W)=\{a:\{S_{a},W\}=0\}$, the graded bracket generates the violated check
excitations in the register:
\begin{equation}
\lbrack Q,W\otimes I_{\mathrm{anc}}]_{\mathrm{gr}}=\sum_{a\in
V(W)}WS_{a}\otimes c_{a}^{\dagger }.  \label{eq:sec3-oracle}
\end{equation}%
This commutator vanishes identically if and only if $W$ commutes with all
stabilizer generators. When acting on a clean input state, the supercharge
writes a superposition of excitations marking precisely the violated
constraints.

The forward map has a direct energetic reading. By the spectral
decomposition of $H_{\mathrm{SUSY}}=\{Q,Q^{\dagger }\}$, a clean-register
state in syndrome sector $s$ carries $H_{\mathrm{SUSY}}$ eigenvalue $w_{s}$,
representing one unit of energy per violated check. Three structural
properties govern this map. First, nilpotency, $Q^{2}=0$, ensures that
applying the differential a second time yields zero, so the syndrome record
is written without double-counting. Second, because the protected manifold
satisfies $Q\mathcal{K}=0$, the codespace is identically dark to the oracle.
Third, $[Q,W\otimes I_{\mathrm{anc}}]_{\mathrm{gr}}=0$ for all $W\in N(%
\mathcal{S})$. Consequently, non-trivial logical operators in $N(\mathcal{S}%
)\setminus \mathcal{S}$ create no excitations and cost zero energy. They are
invisible to the check algebra, confirming that the forward oracle $Q$
diagnoses violated constraints, the decoder proposes the recovery operator $%
C_{s}$, and the projected block $B_{s}$ records the return to $\mathcal{K}$,
and the fibrewise scalar-return test certifies its logical action.

\subsection{Software demotion and the residual class}

\label{subsec:software-demotion}

For Calderbank--Shor--Steane (CSS) codes defined by binary check matrices $%
H_{X}\in \mathbb{F}_{2}^{m_{X}\times n}$ and $H_{Z}\in \mathbb{F}%
_{2}^{m_{Z}\times n}$ satisfying the commutativity condition
\begin{equation}
H_{X}H_{Z}^{\mathsf{T}}=0\qquad \text{over }\mathbb{F}_{2},
\label{eq:sec3-css-orth}
\end{equation}%
the physical supercharge demotes to a binary complex of vector spaces over $%
\mathbb{F}_{2}$ \cite{Calderbank1996,Steane1996}:
\begin{equation}
C_{2}\xrightarrow{H_Z^{\mathsf{T}}}C_{1}\xrightarrow{H_X}C_{0},\qquad C_{2}=%
\mathbb{F}_{2}^{m_{Z}},\quad C_{1}=\mathbb{F}_{2}^{n},\quad C_{0}=\mathbb{F}%
_{2}^{m_{X}}.  \label{eq:sec3-css-complex}
\end{equation}%
The nilpotent software differential $\tilde{Q}^{\overline{Z}}$ is defined on
these chain spaces by
\begin{equation}
\tilde{Q}^{\overline{Z}}|_{C_{2}}=H_{Z}^{\mathsf{T}},\qquad \tilde{Q}^{%
\overline{Z}}|_{C_{1}}=H_{X},\qquad \tilde{Q}^{\overline{Z}%
}|_{C_{0}}=0,\qquad (\tilde{Q}^{\overline{Z}})^{2}=0.  \label{eq:sec3-Qsw}
\end{equation}%
The software differential $\tilde{Q}^{\overline{Z}}$ discards the Hilbert
space inner product while retaining the exact incidence algebra of check
violations and stabilizer boundaries.

For a designated CSS sector, the exact residual classifier is the middle
cohomology group
\begin{equation}
H^{1}\bigl(\tilde{Q}^{\overline{Z}}\bigr)=H_{\overline{Z}}^{1}=\frac{\ker
H_{X}}{\func{im}\,H_{Z}^{\mathsf{T}}}.  \label{eq:sec3-H1}
\end{equation}

For a binary residual vector $z_{s}^{\mathsf{T}}\in C_{1}$ in this $%
\overline{Z}$ sector, its cohomology class in$~H^{1}(\tilde{Q}^{\overline{Z}%
})$ is defined only when the sector closure condition $H_{X}z_{s}^{\mathsf{T}%
}=0$ holds. Because this single CSS sector reads only $Z$-type support and
does not by itself classify every logical Pauli, Section~\ref%
{sec:software-supercharge-cohomology} completes the software complex across
both conjugate sectors in $H_{\mathrm{full}}^{1}$ (Subsec.~\ref%
{subsec:full-logical-address}) and establishes the exact three-tier residual
certificate in Subsec.~\ref{ssec:tier-taxonomy-certificate}~\cite%
{Calderbank1996,Steane1996,Gottesman1997}.

\subsection{Faithful window}

\label{subsec:certified-faithful-window}

The certified faithful window collects the addresses at which the
decoder-selected correction passes the declared-family faithfulness
predicate:

\begin{definition}[Certified Faithful Window]
\label{def:certified-faithful-window} Let $\Sigma_{\mathrm{corr}%
}\subseteq\Sigma_{\mathrm{adm}}$ denote the domain on which the decoder
selects a correction. The certified faithful window is defined by
\begin{equation}
\mathcal{S}_{\mathrm{faith}} = \{s \in \Sigma_{\mathrm{corr}} : \mathrm{Faith%
}_{\mathrm{KL}}(s; C_s, \mathcal{F}_s) = 1\}.
\label{eq:certified-faithful-window}
\end{equation}
\end{definition}

Membership in $\mathcal{S}_{\mathrm{faith}}$ certifies the selected
correction on every error of the declared family $\mathcal{F}_{s}$ and every
protected input, up to a scalar phase. It does not certify an undeclared
member of $\mathcal{F}(s)$. In practical decoding implementations, one
retains an explicit subset of certified addresses,
\begin{equation}
\mathcal{S}_{\mathrm{impl}}\subseteq \mathcal{S}_{\mathrm{faith}}.
\label{eq:implemented-faithful-window}
\end{equation}%
The residual diagnostic sequence
\begin{equation}
B_{s}\;\longrightarrow \;R_{s}=C_{s}E_{\mathrm{true}}\;\longrightarrow
\;u_{s}=(x_{s},z_{s})~\longrightarrow \;[u_{s}]\in H^{1}(\tilde{Q}^{\mathrm{%
full}})  \label{eq:residual-dictionary}
\end{equation}%
is an audit workflow. The last class requires a closed residual $u_{s}$ in
the full CSS sectors, which is analyzed in detail in Sec.~\ref%
{sec:software-supercharge-cohomology}. Subsection~\ref{subsec:compress-Bs}
may reorganize the certified ledger without enlarging $\mathcal{S}_{\mathrm{%
faith}}$. Sec.~\ref{sec:bb-circuit-reranking} may rank candidates outside
this static window, but such ranking is not an enlarging of its certificate.

\subsection{Certified ledger: orthogonality and symmetry reconstruction}

\label{subsec:compress-Bs}

\paragraph{Purpose, scope, and the certification-first order.}

This subsection is the unique home of the static organisation of the
projected-return ledger. Its placement is part of the logical content. The
decoder map and correction cosets have been fixed in Subsec.~\ref%
{subsec:decoder-map}. The branchwise scalar-return and fibrewise
Knill--Laflamme conditions have been imposed in Subsec.~\ref%
{subsec:projected-return-faithfulness}. The passing addresses have been
collected into the certified faithful window $\mathcal{S}_{\mathrm{faith}}$
of Definition~\ref{def:certified-faithful-window}. The distinction between
returning to the protected manifold and returning to the same logical fibre
is the state-independent core of quantum error correction~\cite%
{KnillLaflamme1997,Gottesman1997}. Ledger organisation sits downstream of
that distinction and cannot be used to establish it.

The input is the already-certified family
\begin{equation}
B_{s}=P_{\mathcal{K}}\,C_{s}\,\widetilde{P}_{s},\qquad s\in \mathcal{S}_{%
\mathrm{impl}}\subseteq \mathcal{S}_{\mathrm{faith}},  \label{eq:sec27-block}
\end{equation}%
where $P_{\mathcal{K}}$ is the clean protected-manifold projector, $%
\widetilde{P}_{s}$ is the clean ancilla-register source projector, and $%
C_{s} $ is the decoder-selected Pauli correction. The sectorwise collapse,
exact darkness, and covariance of Subsec.~\ref%
{subsec:graded-commutator-projected-return-blocks} are used here as already
proved.

Two questions are settled below, and naming them first makes the purpose of
the subsection clear. The first question is how many independent certified
return actions the ledger contains. The answer is exactly the number of
certified addresses: distinct syndrome records are mutually exclusive check
outcomes, and no exact algebraic step can make one block stand for two of
them. The second question is how many of these blocks must be stored
explicitly. The answer is one per orbit of a verified symmetry, with every
other block reconstructed exactly by conjugation. The first answer protects
the certificate. The second reduces storage. No $B_{s}$ is approximated or
redefined, $\mathcal{S}_{\mathrm{faith}}$ is not enlarged, and no address
outside the implemented window is certified.

The mandatory order of operations is
\begin{eqnarray}
\text{fibrewise certification}\; &\longrightarrow &\;\mathcal{S}_{\mathrm{%
impl}}\;\longrightarrow \;\text{per-syndrome coset bookkeeping}\;  \notag \\
&\longrightarrow &\;\text{optional symmetry reconstruction}.
\label{eq:sec27-pipeline}
\end{eqnarray}%
No arrow may be reversed. If no symmetry has been verified, the last arrow
is omitted and the ledger is stored in full. Organisation can expose
repeated storage, but it can never convert an uncertified return into a
certified one.

\paragraph{Raw-ledger separation and organisation-stable invariants.}

The source geometry inherited from Subsec.~\ref%
{subsec:graded-commutator-projected-return-blocks} is
\begin{equation}
\widetilde{P}_{s}\widetilde{P}_{s^{\prime }}=\delta_{s,s^{\prime }}%
\widetilde{P}_{s}, \qquad \widetilde{P}_{s}P_{\mathcal{K}}=P_{\mathcal{K}}%
\widetilde{P}_{s}=0 \quad (s\neq 0).  \label{eq:sec27-source-geometry}
\end{equation}
Its meaning is direct. Two different syndrome words are mutually exclusive
outcomes of one commuting check measurement, so their clean source sectors
share no state. A bright source has no overlap with the dark manifold. These
two facts decide in advance what any later organisation can and cannot do.

\begin{lemm}[Hilbert--Schmidt separation and isometry form of the raw ledger]

\label{lem:sec27-HS} For nonzero $s,s^{\prime }\in\mathcal{S}_{\mathrm{impl}%
} $,
\begin{equation}
B_{s}^{\dagger}B_{s^{\prime }} =\widetilde{P}_{s}\,C_{s}^{\dagger}P_{%
\mathcal{K}}C_{s^{\prime }}\,\widetilde{P}_{s^{\prime }},
\label{eq:sec27-product}
\end{equation}
and
\begin{equation}
\func{Tr}\!\left(B_{s}^{\dagger}B_{s^{\prime }}\right) =\delta_{s,s^{\prime
}}\, \func{Tr}\!\left(\widetilde{P}_{s}\,C_{s}^{\dagger}P_{\mathcal{K}%
}C_{s}\,\widetilde{P}_{s}\right).  \label{eq:sec27-HS}
\end{equation}
If, in addition, $C_{s}$ is a syndrome-matched Pauli correction, $%
\sigma(C_{s})=s$, and $s$ is a realized address, then
\begin{equation}
B_{s}^{\dagger}B_{s}=\widetilde{P}_{s}, \qquad B_{s}B_{s}^{\dagger}=P_{%
\mathcal{K}}, \qquad \lVert B_{s}\rVert_{\mathrm{HS}}^{2}=\func{Tr}P_{%
\mathcal{K}}=2^{k}.  \label{eq:sec27-isometry}
\end{equation}
Consequently the nonzero raw blocks are mutually orthogonal in the
Hilbert--Schmidt inner product, all have the same norm, and each is an
isometry from the clean source sector $\widetilde{P}_{s}\mathcal{H}$ onto $%
\mathcal{K}$.
\end{lemm}

\noindent {\textit{Proof}}. Equation~\eqref{eq:sec27-product} follows from
Eq.~\eqref{eq:sec27-block} and $P_{\mathcal{K}}^{2}=P_{\mathcal{K}}$.
Cyclicity of the trace moves the two outer source projectors next to one
another, and their product is evaluated by Eq.~%
\eqref{eq:sec27-source-geometry}; this gives Eq.~\eqref{eq:sec27-HS}. For
Eq.~\eqref{eq:sec27-isometry}, a Pauli operator maps the data syndrome
sector $s$ onto the sector $s+\sigma(C_{s})$ and commutes with the register
vacuum projector, so a syndrome-matched correction obeys $C_{s}\widetilde{P}%
_{s}C_{s}^{\dagger}=\widetilde{P}_{0}$. The clean sector $\widetilde{P}_{0}%
\mathcal{H}$ is the code space with an empty register, so $\widetilde{P}%
_{0}=P_{\mathcal{K}}$ in any presentation, before any conjugation. The
unitary $C_{s}$ carries it onto the realized sector $\widetilde{P}_{s}%
\mathcal{H}$, which therefore also has dimension $2^{k}$; hence $%
C_{s}^{\dagger}P_{\mathcal{K}}C_{s}=\widetilde{P}_{s}$ for every realized
address, whether the presentation is independent or redundant. Substituting
into Eq.~\eqref{eq:sec27-product} with $s=s^{\prime }$ gives $%
B_{s}^{\dagger}B_{s}=\widetilde{P}_{s}$, and $B_{s}B_{s}^{\dagger}=P_{%
\mathcal{K}}\widetilde{P}_{0}P_{\mathcal{K}}=P_{\mathcal{K}}$. The trace is $%
\dim\mathcal{K}=2^{k}$. {\hfill $\square $}\newline

The trace-level qualification is essential. In general $B_{s}^{%
\dagger}B_{s^{\prime }}$ need not vanish as an operator for $s\neq s^{\prime
}$, because $C_{s}^{\dagger}P_{\mathcal{K}}C_{s^{\prime }}$ lies between the
source projectors. What Lemma~\ref{lem:sec27-HS} does establish is that the
Gram matrix of the raw family,
\begin{equation}
\left(\mathcal{G}_{\mathrm{Gram}}\right)_{ss^{\prime }} =\func{Tr}%
\!\left(B_{s}^{\dagger}B_{s^{\prime }}\right) =2^{k}\,\delta_{s,s^{\prime }},
\label{eq:sec27-Gram-positive}
\end{equation}
is already diagonal with strictly positive entries. No invertible linear
recombination, unitary rotation, or Fourier transform of the family can
lower its rank; a non-invertible recombination merely discards information
and certifies nothing. The number of independent certified return directions
is therefore
\begin{equation}
N_{\mathrm{cert}}=\func{rank}\mathcal{G}_{\mathrm{Gram}}=\lvert\mathcal{S}_{%
\mathrm{impl}}\rvert .  \label{eq:sec27-certified-rank}
\end{equation}
The same exclusivity holds at the level of correction cosets. A stabilizer
coset is syndrome-pure, since $\sigma(C_{s}g)=\sigma(C_{s})=s$ for every $%
g\in\mathcal{S}$, so
\begin{equation}
C_{s}\mathcal{S}=C_{s^{\prime }}\mathcal{S} \;\Longrightarrow\; s=s^{\prime
},  \label{eq:sec27-no-merge}
\end{equation}
and no correction coset can serve two addresses. Within one address,
candidates that differ by a stabilizer $g$ give $P_{\mathcal{K}}C_{s}g%
\widetilde{P}_{s}=\pm B_{s}$, because $g$ acts as a sign on the syndrome
eigenspace; this degeneracy is already absorbed by the coset $C_{s}\mathcal{S%
}$ of Eq.~\eqref{eq:sec3-correction-coset} in the standard stabilizer
accounting~\cite{Gottesman1997}.

For each realized nonzero address $s$, a syndrome-matched Pauli block $B_{s}$
restricts to an isometry from the clean source sector $\widetilde{P}_{s}%
\mathcal{H}$ onto $\mathcal{K}$, by Eq. \eqref{eq:sec27-isometry}. Logical
faithfulness additionally requires the declared-family scalar-return test of
Eq.~\eqref{eq:all-branch-scalar-return}. Blocks at distinct syndrome
addresses are Hilbert--Schmidt orthogonal; verified symmetry can reduce
explicit storage without reducing their linear-span dimension.

Linear recombination preserves exact darkness. Under the stated covariance,
symmetry conjugation reconstructs the corresponding return blocks.

\begin{prop}[Organisation stability of the inherited certificates]
\label{prop:sec27-stability} Let $B=\sum_{a=1}^{m}u_{a}B_{a}$, with $u_{a}\in%
\mathbb{C}$, be a finite linear combination of certified blocks. Then
\begin{equation}
B\,P_{\mathcal{K}}=0.  \label{eq:sec27-linear-darkness}
\end{equation}
If an even unitary $U_{g}$ preserves $Q$, $P_{\mathcal{K}}$, and the
clean-source decomposition while inducing $s\mapsto g\cdot s$, and if the
seed family is covariant, which for the collapsed blocks means $%
U_{g}C_{s}U_{g}^{\dagger}=C_{g\cdot s}$, then
\begin{equation}
U_{g}B_{s}U_{g}^{\dagger}=B_{g\cdot s}.  \label{eq:sec27-covariance}
\end{equation}
Every complete retained covariant family inherits these two statements.
\end{prop}

\noindent {\textit{Proof}}. Exact darkness is linear and holds term by term.
Covariance follows by conjugating the inherited projected-return identity of
Subsec.~\ref{subsec:graded-commutator-projected-return-blocks} and using the
covariance of $Q$, $P_{\mathcal{K}}$, the seeds, and the source projectors.
\hfill {$\square $}

\paragraph{Implemented window and honest coverage.}

Only realized addresses enter the ledger. For an independent generating set
every word $s\in \mathbb{F}_{2}^{r}$ is realized. For a redundant
presentation some formal words are inconsistent and give $\widetilde{P}%
_{s}=0 $; they are not ledger entries,
\begin{equation}
\mathcal{S}_{\mathrm{impl}}\subseteq \{\,s\in \mathbb{F}_{2}^{r}:\widetilde{P%
}_{s}\neq 0\,\}.  \label{eq:sec27-realized}
\end{equation}%
The smallest local base is the set of nonzero syndrome addresses generated
by single-qubit Pauli operators. Writing $s(W)=\sigma (W)$ for the syndrome
word of $W$, define
\begin{equation}
\mathcal{S}_{1}=\left\{ s(X_{i}),\,s(Y_{i}),\,s(Z_{i})\;\middle|\;i=1,\ldots
,n\right\} \setminus \{0\},\qquad N_{w1}=\lvert \mathcal{S}_{1}\rvert \leq
3n.  \label{eq:sec27-weight-one-count}
\end{equation}%
The inequality is the general statement because distinct one-qubit Pauli
operators can share a syndrome address. Equality holds exactly when no
nonidentity element of $N(\mathcal{S})$ has weight at most two, since two
single-qubit Paulis share a syndrome if and only if their product, of weight
at most two, commutes with every check. For a distance-three code this base
reaches the designed radius. For a higher-distance code it is only the first
audited shell. In particular, the Gross $[[144,12,12]]$ bivariate-bicycle
benchmark requires an explicit account of higher-weight coverage rather than
an inference from distance alone~\cite{Bravyi2024}.

A larger implemented window may be declared as
\begin{equation}
\mathcal{S}_{\mathrm{impl}}=\mathcal{S}_{1}\cup \mathcal{S}_{\mathrm{cert}%
}^{(2)}\cup \cdots \cup \mathcal{S}_{\mathrm{cert}}^{(w_{\max })}\subseteq
\mathcal{S}_{\mathrm{faith}},  \label{eq:sec27-higher-shells}
\end{equation}%
where $\mathcal{S}_{\mathrm{cert}}^{(w)}$ collects the addresses first
reached at weight $w$ that have independently passed the fibrewise predicate
of Definition~\ref{def:faith-predicate}, so that the union is disjoint and
every address is counted once. First occurrence classifies addresses, not
errors: when higher-weight errors add branches at an address already in the
window, the enlarged declared family must pass the same fixed-correction
test of Definition~\ref{def:faith-predicate} again. Equation~%
\eqref{eq:sec27-higher-shells} is a declaration of certified coverage, not a
consequence of symmetry. Nothing below is restricted to low weight: the
orthogonality of Lemma~\ref{lem:sec27-HS} and the reconstruction of
Proposition~\ref{prop:sec27-reconstruction} hold for every certified
address, whatever the weight of the errors that reach it. Weight enters only
through the declared coverage and the cost of certifying each shell, not
through a different identity. The large reduction from the full inventory of
$2^{r}-1$ nonzero syndrome words to an auditable ledger is secured by this
chosen window and by nothing else in the subsection. Every later operation
acts inside the window and cannot repair an omission from it.

\paragraph{Optional symmetry reconstruction.}

What a symmetry can do is different from merging, and it is useful. Suppose
a finite Abelian group $G$ acts through even unitaries $U_{g}$ and preserves
the complete certified geometry,
\begin{equation}
U_{g}QU_{g}^{\dagger}=Q, \qquad U_{g}P_{\mathcal{K}}U_{g}^{\dagger}=P_{%
\mathcal{K}}, \qquad U_{g}\widetilde{P}_{s}U_{g}^{\dagger}=\widetilde{P}%
_{g\cdot s}, \qquad g\cdot\mathcal{S}_{\mathrm{impl}}=\mathcal{S}_{\mathrm{%
impl}}.  \label{eq:sec27-character-covariance}
\end{equation}
Suppose in addition that the decoder is equivariant on the implemented
window,
\begin{equation}
C_{g\cdot s}=U_{g}\,C_{s}\,U_{g}^{\dagger} \qquad\text{for all } g\in G,\
s\in\mathcal{S}_{\mathrm{impl}},  \label{eq:sec27-decoder-equivariance}
\end{equation}
with consistent Pauli phases. The first line is a property of the code and
its presentation, for instance the translation group of a bivariate-bicycle
code acting on qubits and checks together~\cite{Bravyi2024}. The second line
is a property of the decoder and is not automatic, because $Q$ alone never
selects $C_{s}$. A minimum-weight rule with a translation-invariant
tie-break satisfies it, while a rule that breaks ties by qubit index
generally does not. A boundary, a decoder tie rule, or a change of
presentation can break either condition, and each must be verified rather
than assumed. Invariance of noise weights, of the extraction schedule, or of
the logical-frame convention is required in addition only when scores or
circuit-level certificates are reused; it is not a hypothesis of the static
block identity below. Symmetry-aware decoding has been used elsewhere to
generate candidate corrections~\cite{Koutsioumpas2025}. Here the symmetry is
used only after certification, and only to reconstruct already-certified
records.

\begin{prop}[Exact orbit reconstruction of the certified ledger]
\label{prop:sec27-reconstruction} Assume Eqs.~%
\eqref{eq:sec27-character-covariance} and~%
\eqref{eq:sec27-decoder-equivariance}. Choose one representative $s_{O}$ in
each $G$-orbit $O\subseteq\mathcal{S}_{\mathrm{impl}}$. Then for every $g\in
G$,
\begin{equation}
B_{g\cdot s_{O}}=U_{g}\,B_{s_{O}}\,U_{g}^{\dagger},
\label{eq:sec27-orbit-reconstruction}
\end{equation}
and the prescription is independent of the choice of $g$ within a coset of
the isotropy subgroup $G_{s_{O}}=\{h\in G:h\cdot s_{O}=s_{O}\}$ of the group
action, not to be confused with the code stabilizer $\mathcal{S}$. Every $%
g\cdot s_{O}$ lies in $\mathcal{S}_{\mathrm{faith}}$ with the
symmetry-transformed family $U_{g}\mathcal{F}_{s_{O}}U_{g}^{\dagger}$; the
symmetry-transformed certificate covers that family, or a declared subfamily
of it, and no newly chosen branch at $g\cdot s_{O}$. The full ledger is
reproduced exactly from the stored pairs $\bigl(B_{s_{O}},\{U_{g}\}_{g\in G}%
\bigr)$, and the two counts are
\begin{equation}
N_{\mathrm{store}}=\lvert\mathcal{S}_{\mathrm{impl}}/G\rvert, \qquad N_{%
\mathrm{cert}}=\lvert\mathcal{S}_{\mathrm{impl}}\rvert\ \text{unchanged},
\label{eq:sec27-storage-count}
\end{equation}
where $\mathcal{S}_{\mathrm{impl}}/G$ denotes the set of $G$-orbits in $%
\mathcal{S}_{\mathrm{impl}}$ and not a quotient group. The counts obey the
orbit-stabilizer bound
\begin{equation}
\frac{\lvert\mathcal{S}_{\mathrm{impl}}\rvert}{\lvert G\rvert} \;\leq\; N_{%
\mathrm{store}}\;\leq\;\lvert\mathcal{S}_{\mathrm{impl}}\rvert ,
\label{eq:sec27-store-bound}
\end{equation}
the lower bound being attained when $G$ acts freely on $\mathcal{S}_{\mathrm{%
impl}}$.
\end{prop}

\noindent {\textit{Proof}}. Insert $U_{g}^{\dagger}U_{g}$ between the
factors of Eq.~\eqref{eq:sec27-block} and use Eqs.~%
\eqref{eq:sec27-character-covariance} and~%
\eqref{eq:sec27-decoder-equivariance}; this gives $U_{g}B_{s_{O}}U_{g}^{%
\dagger}=P_{\mathcal{K}}C_{g\cdot s_{O}}\widetilde{P}_{g\cdot
s_{O}}=B_{g\cdot s_{O}}$. If $h\cdot s_{O}=s_{O}$, the same identity gives $%
U_{h}B_{s_{O}}U_{h}^{\dagger}=B_{s_{O}}$, so the reconstruction does not
depend on the representative $g$ of the coset. Conjugating Eq.~%
\eqref{eq:all-branch-scalar-return} by $U_{g}$ shows that $C_{g\cdot s_{O}}$
acts as the same scalar $\lambda_{s_{O},a}$ on each symmetry-transformed
branch $U_{g}E_{s_{O},a}U_{g}^{\dagger}$, so the predicate of Definition~\ref%
{def:faith-predicate} is preserved. The counts are the definition of an
orbit decomposition, and Eq.~\eqref{eq:sec27-store-bound} is the
orbit-stabilizer theorem. {\hfill $\square $}\newline

The proposition is a storage statement, not a rank statement, and the two
must be kept apart. By Lemma~\ref{lem:sec27-HS}, the blocks $B_{s_{O}}$ and $%
U_{g}B_{s_{O}}U_{g}^{\dagger}$ are orthogonal and both nonzero whenever $%
g\cdot s_{O}\neq s_{O}$; for $g\in G_{s_{O}}$ they coincide. The symmetry
tells the stack how to write the second from the first; it does not say that
the second is redundant. Physically, a code with a translation symmetry has
the same local physics at every translated location. A certified return at
one location therefore certifies the translated return at every other
location, provided the decoder treats the locations alike and the audited
error family is transformed by the same action. The symmetry-transformed
family $U_{g}\mathcal{F}_{s}U_{g}^{\dagger}$ is a classical relabelling of
operator descriptions, not an operation applied to hardware: for a lattice
translation, $U_{g}E_{s,a}U_{g}^{\dagger}$ is the same Pauli error written
on the translated qubits, nothing is executed on the device, and no new
error is created. The only risk is bookkeeping, namely a decoder that is not
equivariant or a declared family at $g\cdot s$ that differs from the
transformed one; this is why Eq.~\eqref{eq:sec27-decoder-equivariance} and
the family clause of Proposition~\ref{prop:sec27-reconstruction} are stated
as hypotheses. What is saved is the repeated independent certification and
storage of identical local physics; each orbit member still carries its own
symmetry-transformed certificate, obtained from the representative rather
than re-derived, and the net saving must be weighed against the cost of
verifying Eqs.~\eqref{eq:sec27-character-covariance} and~%
\eqref{eq:sec27-decoder-equivariance}. What is not saved is any syndrome
address: each translated address remains a separate check outcome with its
own block, its own certificate, and, by Proposition~\ref%
{prop:sec27-stability}, its own exact darkness.

An optional character basis organises each reconstructed orbit. With $%
\widehat{G}$ the Pontryagin dual, we use
\begin{equation}
P_{\chi }=\frac{1}{\lvert G\rvert }\sum_{g\in G}\overline{\chi (g)}%
\,U_{g},\qquad P_{\chi }P_{\chi ^{\prime }}=\delta _{\chi ,\chi ^{\prime
}}P_{\chi },\qquad \sum_{\chi \in \widehat{G}}P_{\chi }=I.
\label{eq:sec27-character-projector}
\end{equation}%
This is character-method operator bookkeeping\footnote{%
The character projectors $P_{\chi }$ of Eq.~%
\eqref{eq:sec27-character-projector} are central idempotents of the group
algebra; their role as commuting conserved charges resolving multi-sector
states is standard and implemented in the BPS literatures~\cite%
{Kemp:2019log,Berenstein:2022srd}. The present situation is a decoder-facing
operator layer, and the dynamics of each can be different.}. On an orbit on
which $G$ acts freely, the normalised combinations
\begin{equation}
B_{O,\chi }=\frac{1}{\sqrt{\lvert G\rvert }}\sum_{g\in G}\overline{\chi (g)}%
\,U_{g}B_{s_{O}}U_{g}^{\dagger },\qquad \chi \in \widehat{G},
\label{eq:sec27-character-basis}
\end{equation}%
form a mutually orthogonal basis of the orbit span in the Hilbert--Schmidt
inner product, each member having the common squared norm $\lVert B_{O,\chi
}\rVert _{\mathrm{HS}}^{2}=2^{k}$; multiplication by $2^{-k/2}$ would make
the basis orthonormal. This holds because the $\lvert G\rvert $ blocks $%
U_{g}B_{s_{O}}U_{g}^{\dagger }$ are mutually orthogonal with equal squared
norm $2^{k}$ by Lemma~\ref{lem:sec27-HS}, and the character table of a
finite Abelian group is unitary after division by $\sqrt{\lvert G\rvert }$.
The change of basis is invertible and preserves the rank $\lvert O\rvert
=\lvert G\rvert $. If the action on an orbit is not free, the sum over the
isotropy subgroup $G_{s_{O}}$ annihilates every character that is nontrivial
on $G_{s_{O}}$, so exactly $\lvert O\rvert =\lvert G\rvert /\lvert
G_{s_{O}}\rvert $ of the combinations in Eq.~\eqref{eq:sec27-character-basis}
are nonzero, namely those with $\chi $ trivial on $G_{s_{O}}$, each then
having squared norm $2^{k}\lvert G_{s_{O}}\rvert $, and they still span the
orbit space. The equation serves only as representation bookkeeping
compatible with the character conventions used elsewhere in the paper; it
compresses nothing and supplies no certification criterion. The $P_{\chi }$
resolve the group representation, not individual syndrome addresses or
logical classes, and they need not commute with each $\widetilde{P}_{s}$. A
character label is therefore neither a same-fibre certificate nor a licence
to truncate a nonzero direction. The convention is retained only for
compatibility with character bookkeeping elsewhere in the draft; it neither
deforms $Q$ nor supplies a new certification criterion.

\paragraph{Ledger counts and coverage scope.}

The ledger carries two integers with different meanings. The certified count
$N_{\mathrm{cert}}=\lvert\mathcal{S}_{\mathrm{impl}}\rvert$ of Eq.~%
\eqref{eq:sec27-certified-rank} is the number of independent certified
return actions and is fixed by the window alone. With the disjoint shells of
Eq.~\eqref{eq:sec27-higher-shells},
\begin{equation}
N_{\mathrm{cert}}=N_{w1}+\sum_{w=2}^{w_{\max}}\bigl\lvert\mathcal{S}_{%
\mathrm{cert}}^{(w)}\bigr\rvert .  \label{eq:sec27-general-count}
\end{equation}
The storage count $N_{\mathrm{store}}$ of Eq.~\eqref{eq:sec27-storage-count}
is the number of explicitly stored blocks and depends on the verified
symmetry; it equals $N_{\mathrm{cert}}$ whenever no symmetry passes the
verification of Eqs.~\eqref{eq:sec27-character-covariance} and~%
\eqref{eq:sec27-decoder-equivariance}. Neither integer is a threshold, a
logical-error-rate estimate, or a completion of higher-weight coverage, and $%
N_{\mathrm{store}}$ counts representative recovery descriptions rather than
bytes, latency, total certification cost, or logical-error performance,
since the group action and certificate metadata must also be recorded. The
count governs this optional certification ledger and lookup-style
pre-decoders; it is not a memory or runtime law for a message-passing base
decoder, which stores no syndrome table.

For the Steane $[[7,1,3]]$ control code of Subsec.~\ref%
{subsec:steane-static-audit}, all twenty-one weight-one syndromes are
distinct, so $N_{w1}=N_{\mathrm{cert}}=21$~\cite{Steane1996}. This equality
is a coverage count, not a symmetry reduction. For the Gross convention,
\begin{equation}
G_{\mathrm{gross}}=\mathbb{Z}_{12}\times \mathbb{Z}_{6},\qquad \lvert G_{%
\mathrm{gross}}\rvert =72,\qquad N_{w1}\leq 3n=432.
\label{eq:sec27-gross-count}
\end{equation}%
The group order alone fixes neither the orbit inventory nor the storage
saving. Three conditions must be verified on the frozen presentation before
a smaller count is quoted. First, the bound in Eq.~%
\eqref{eq:sec27-gross-count} must be saturated, which holds exactly when no
nonidentity element of $N(\mathcal{S})$ has weight at most two. The distance
$d=12$ excludes such elements outside $\mathcal{S}$~\cite{Bravyi2024}, but
it does not by itself exclude a weight-two product of stabilizer generators,
so saturation is checked by direct enumeration of the $3n$ weight-one
syndromes on the frozen presentation. Second, the translation group must act
freely on the realized weight-one addresses; it acts freely on the two qubit
blocks and permutes the checks accordingly, so freeness on addresses follows
once saturation holds. Third, the base decoder must satisfy Eq.~%
\eqref{eq:sec27-decoder-equivariance}. Under these three conditions, and
only then, the weight-one counts would be
\begin{equation}
N_{\mathrm{cert}}=432,\qquad N_{\mathrm{store}}=432/72=6,
\label{eq:sec27-gross-store}
\end{equation}%
one representative for each of the three Pauli species on each of the two
qubit blocks, with the remaining $426$ blocks generated by Eq.~%
\eqref{eq:sec27-orbit-reconstruction}. No such orbit audit is reported in
this paper; Eq.~\eqref{eq:sec27-gross-store} records what a verified audit
would certify. If decoder equivariance fails, the raw ledger of all $N_{w1}$
blocks is retained. If equivariance holds but the action is not free, the
actual orbits, possibly of unequal size, are counted instead of dividing by $%
72$. If saturation fails, $N_{\mathrm{cert}}=N_{w1}<432$ and the orbit
inventory must be recomputed. In every case the weight-one shell is only the
first audited shell for this code, and higher shells must be certified and
counted separately by Eq.~\eqref{eq:sec27-higher-shells}.

The retained ledger records, for each stored representative, the source
address, the correction coset, the certified fibre verdict, the physical
support, the orbit it generates, and the reconstructing group elements. Sec.~%
\ref{sec:software-supercharge-cohomology} retains the exact finite-field
residual classifier, and Sec.~\ref{sec:bb-circuit-reranking} implements the
detector fibres and circuit-level logical action.

\section{Software Supercharge and Decoder Certification}

\label{sec:software-supercharge-cohomology}

\renewcommand{\theequation}{4.\arabic{equation}} \setcounter{equation}{0} %
\renewcommand{\thethm}{4.\arabic{thm}} \setcounter{thm}{0} %
\renewcommand{\theprop}{4.\arabic{prop}} \setcounter{prop}{0} %
\renewcommand{\theprotocol}{4.\arabic{protocol}} \setcounter{protocol}{0} %
\renewcommand{\thecoro}{4.\arabic{coro}} \setcounter{coro}{0}

Section~\ref{sec:syndrome-fibres-KL-window} completed with a committed
correction $C_{s}$, a true error $E_{\mathrm{true}}$, and the residual $%
R_{s}=C_{s}E_{\mathrm{true}}$. The remaining question is physical: after
every visible check has become quiet, has the encoded state returned to the
same logical fibre? This section gives an exact finite-field answer. We
first separate the physical supercharge $Q$ from its software shadow $\tilde{%
Q}^{\mathrm{full}}$. We then complete the latter over both CSS sectors. The
completed middle cohomology is the full logical-Pauli address. It yields an
accept--reject certificate for the syndrome-silent failure probability $p_{%
\mathrm{fs}}$ and a computable ambiguity flag for a finite candidate list.

The CSS quotient identities used below are standard and are quoted rather
than claimed~\cite{Calderbank1996,Gottesman1997}. The methodological
contribution is their typed use in the decoder architecture: binarization
supplies the syndrome arrow, stabilizer completion supplies the boundary
arrow, and the two conjugate sectors close the one-sector false-accept
channel. The resulting absolute class is exact in a labelled audit, while
its relative form remains computable online.

\subsection{Typed demotion and stabilizer completion}

\label{ssec:scope-Q-vs-Qsw} \label{ssec:typed-demotion-ranks-weights}

The CSS matrices and single-sector complex are fixed in those of Eqs.~%
\eqref{eq:sec3-css-complex}--\eqref{eq:sec3-Qsw} in Subsec.~\ref%
{subsec:software-demotion}. Here $C_{2}$ labels $Z$-type generators, $C_{1}=%
\mathbb{F}_{2}^{n}$ data supports, and $C_{0}$ the visible $X$-check
responses. Only their typed maps are needed here: The differential acts by
\begin{equation}
\tilde{Q}^{\overline{Z}}\big|_{C_{2}}=H_{Z}^{\mathsf{T}},\qquad \tilde{Q}^{%
\overline{Z}}\big|_{C_{1}}=H_{X},\qquad \tilde{Q}^{\overline{Z}}\big|%
_{C_{0}}=0,\qquad (\tilde{Q}^{\overline{Z}})^{2}=0.  \label{eq:sw-Qsw-def}
\end{equation}

The physical and software gradings answer different questions. The
zero-degree harmonic sector of $Q$ is the protected code space,
\begin{equation}
H^{0}(Q)\cong C_{\mathrm{QEC}},\qquad \dim _{\mathbb{C}}H^{0}(Q)=2^{k}.
\label{eq:sw-H0Q}
\end{equation}
The distinction is essential because a finite-field incidence complex
carries neither the amplitudes nor the inner product of the physical theory.

Write $W=(w_{X},w_{Z})$ for the phase-suppressed symplectic image of a Pauli
operator. Its response to the $X$-type checks is
\begin{equation}
\sigma _{X}(W)=H_{X}w_{Z}^{\mathsf{T}}.  \label{eq:tdrw-binarization}
\end{equation}%
Binarization therefore supplies only the forward syndrome map $D_{1}=H_{X}$.
A stabilizer is invisible to every syndrome measurement, so the equivalence
between two silent residuals that differ by a stabilizer must be adjoined
from the stabilizer group. The staging is
\begin{equation}
Q\ \xrightarrow{\text{binarization restricted to $Z$-type supports}}\ D_{1}\ %
\xrightarrow{\text{stabilizer completion}}\ (D_{0},D_{1}),\qquad
D_{0}=H_{Z}^{\mathsf{T}}.  \label{eq:tdrw-staging}
\end{equation}%
This separation has a direct physical meaning. The outgoing arrow records
what the checks can see. The incoming arrow records what the encoded state
cannot feel. Without the second arrow, one can test syndrome closure but
cannot certify logical triviality.

The degree labels must remain explicit. On ${\tilde C}=C_{2}\oplus
C_{1}\oplus C_{0}$, write
\begin{equation}
v=%
\begin{pmatrix}
z \\
u \\
x%
\end{pmatrix}%
,\qquad z\in C_{2},\quad u\in C_{1},\quad x\in C_{0}.
\label{eq:tdrw-graded-domain}
\end{equation}%
In the ordered degree basis,
\begin{equation}
\tilde{Q}^{\overline{Z}}=%
\begin{pmatrix}
0 & 0 & 0 \\
H_{Z}^{\mathsf{T}} & 0 & 0 \\
0 & H_{X} & 0%
\end{pmatrix}%
.  \label{eq:tdrw-blockmatrix}
\end{equation}%
Only the middle summand is reached by one arrow and tested by the other.
Consequently its cohomology is
\begin{equation}
H_{\overline{Z}}^{1}=\frac{\ker H_{X}}{\func{im}H_{Z}^{\mathsf{T}}}.
\label{eq:tdrw-middle}
\end{equation}%
Flattening the three summands is harmless for storage, but a generic change
of basis on the flattened space mixes generator, qubit, and syndrome labels.
It is then no longer a map of the complex.

The demotion retains incidence, nilpotency, and the residual class. It
discards complex amplitudes, Pauli phases, ancilla-register occupation. Thus
$Q$ determines whether the ancilla register is bright, whereas $\tilde{Q}^{%
\overline{Z}}$ asks whether a residual that is already syndrome-silent has
changed the logical fibre.

The layers share generating ranks, not vector spaces. With $r_{X}=\func{rank}%
H_{X}$ and $r_{Z}=\func{rank}H_{Z}$,
\begin{equation}
r=r_{X}+r_{Z}=n-k,\qquad m_{X}+m_{Z}\geq r.  \label{eq:tdrw-ranks}
\end{equation}%
Equality holds only for an independent generating presentation. In
particular,
\begin{equation}
\dim _{\mathbb{C}}\mathcal{H}_{\mathrm{data}}=2^{n},\qquad \dim _{\mathbb{C}}%
\mathcal{H}_{\mathrm{anc}}=2^{r},\qquad \dim _{\mathbb{F}_{2}}C_{1}=n.
\label{eq:tdrw-dims}
\end{equation}%
Physical darkness is an intersection of check constraints. The software
logical address below is a direct sum of two conjugate coordinates.

\subsection{Two-sector completion and merged cohomology dictionary}

\label{subsec:full-logical-address}

A one-sector classifier is exact on its declared sector but incomplete for a
general Pauli residual. An $X$-type check responds to $Z$-type support,
while a $Z$-type check responds to $X$-type support. With a Pauli support
written as $(x,z)$, the crossed response is
\begin{equation}
s_{Z}=H_{Z}x^{\mathsf{T}},\qquad s_{X}=H_{X}z^{\mathsf{T}}.
\label{eq:fla-orthogonality}
\end{equation}%
The conjugate CSS identity is
\begin{equation}
H_{Z}H_{X}^{\mathsf{T}}=0.  \label{eq:fla-css-orth-conjugate}
\end{equation}%
It therefore classifies $Z$-type supports,
\begin{equation}
H_{\overline{Z}}^{1}=\frac{\ker H_{X}}{\func{im}H_{Z}^{\mathsf{T}}},\qquad
\dim _{\mathbb{F}_{2}}H_{\overline{Z}}^{1}=k,  \label{eq:fla-H1Z}
\end{equation}%
while exchanging the check families gives
\begin{equation}
H_{\overline{X}}^{1}=\frac{\ker H_{Z}}{\func{im}H_{X}^{\mathsf{T}}},\qquad
\dim _{\mathbb{F}_{2}}H_{\overline{X}}^{1}=k.  \label{eq:fla-H1X}
\end{equation}%
The subscript identifies the residual species; the responding check family
is the conjugate one.

The two conjugate differentials have the degree-resolved blocks
\begin{equation}
\tilde{Q}^{\overline{Z}}=%
\begin{pmatrix}
0 & 0 & 0 \\
H_{Z}^{\mathsf{T}} & 0 & 0 \\
0 & H_{X} & 0%
\end{pmatrix}%
,\qquad \tilde{Q}^{\overline{X}}=%
\begin{pmatrix}
0 & 0 & 0 \\
H_{X}^{\mathsf{T}} & 0 & 0 \\
0 & H_{Z} & 0%
\end{pmatrix}%
.  \label{eq:fla-sector-differentials}
\end{equation}%
Each square vanishes by CSS orthogonality over $\mathbb{F}_{2}$; this binary
nilpotency is not an operator identity inherited from $Q$. The completed
software differential is the direct sum
\begin{equation}
\tilde{Q}^{\mathrm{full}}=\tilde{Q}^{\overline{X}}\oplus \tilde{Q}^{%
\overline{Z}},\qquad \left( \tilde{Q}^{\mathrm{full}}\right) ^{2}=0.
\label{eq:fla-Qfull}
\end{equation}%
At the middle degree, ordered as $(x,z)$, its incoming and outgoing maps are
\begin{equation}
D_{0}=%
\begin{pmatrix}
H_{X}^{\mathsf{T}} & 0 \\
0 & H_{Z}^{\mathsf{T}}%
\end{pmatrix}%
,  \label{eq:fla-D0}
\end{equation}%
\begin{equation}
D_{1}=%
\begin{pmatrix}
H_{Z} & 0 \\
0 & H_{X}%
\end{pmatrix}%
,\qquad D_{1}D_{0}=0.  \label{eq:fla-D0D1}
\end{equation}%
Equivalently,
\begin{equation}
D_{0}(u_{X},u_{Z})=\left( H_{X}^{\mathsf{T}}u_{X},H_{Z}^{\mathsf{T}%
}u_{Z}\right) ,  \label{eq:fla-D0-action}
\end{equation}%
\begin{equation}
D_{1}(x,z)=\left( H_{Z}x^{\mathsf{T}},H_{X}z^{\mathsf{T}}\right) .
\label{eq:fla-D-actions}
\end{equation}%
The staging is
\begin{equation}
Q\ \xrightarrow{\text{binarization}}\ D_{1}\ \xrightarrow{\text{stabilizer
completion}}\ (D_{0},D_{1})~\xrightarrow{\text{assemble}}~\tilde{Q}^{\mathrm{%
full}}.
\end{equation}%
There is no cross term in the differential. Correlations between $X$- and $Z$%
-type faults belong to the noise prior and to subsequent reranking, not to
the definition of the exact class.

\begin{prop}[Two-sector completeness]
\label{prop:fla-completeness} Let $\mathcal{S}$ be an abelian CSS stabilizer
group with check matrices $H_X,H_Z$. Then
\begin{equation}
H^{1}_{\mathrm{full}} :=H^1\!\left(\tilde{Q}^{\mathrm{full}}\right) \cong
H^{1}_{\overline{X}}\oplus H^{1}_{\overline{Z}} \cong N(\mathcal{S})/%
\mathcal{S},  \label{eq:proof-split}
\end{equation}
and
\begin{equation}
\dim_{\mathbb{F}_{2}}H^{1}_{\mathrm{full}}=2k, \qquad \left|H^{1}_{\mathrm{%
full}}\right|=4^k.  \label{eq:fla-full-dimension}
\end{equation}
\end{prop}

\noindent To verify the statement, suppress phases and write a Pauli as $%
P(x,z)=X^{x}Z^{z}$. Membership in the normalizer is equivalent to
\begin{equation}
H_{Z}x^{\mathsf{T}}=0,\qquad H_{X}z^{\mathsf{T}}=0.
\label{eq:fla-normalizer-test}
\end{equation}%
Membership in the stabilizer means
\begin{equation}
x\in \func{im}H_{X}^{\mathsf{T}},\qquad z\in \func{im}H_{Z}^{\mathsf{T}}.
\label{eq:fla-stabilizer-test}
\end{equation}%
Quotienting sector by sector yields Eq.~\eqref{eq:proof-split}. This
identification is standard stabilizer formalism~\cite{Gottesman1997}; its
use here shows why a one-sector acceptance test cannot certify a general
Pauli residual. The omitted coordinate is a conjugate logical direction, not
a finer reading of the same coordinate.

The count $4^{k}$ refers to logical Pauli operators, while $2^{k}$ is the
dimension of the protected state space. These counts agree with $\dim _{%
\mathbb{C}}\func{End}(\mathcal{K})=2^{2k}=4^{k}$.

\begin{coro}[Manifest quotient]
\label{cor:fla-quotient} At the middle degree, define
\begin{equation}
Z^{1}=\ker \!\left( \tilde{Q}^{\mathrm{full}}\big|_{C_{1}^{\mathrm{full}%
}}\right) ,\qquad B^{1}=\func{im}\!\left( \tilde{Q}^{\mathrm{full}}\big|%
_{C_{2}^{\mathrm{full}}}\right) .  \label{eq:fla-cycles-boundaries}
\end{equation}%
Then
\begin{equation}
H_{\mathrm{full}}^{1}=Z^{1}/B^{1}=\frac{\ker D_{1}}{\func{im}D_{0}}.
\label{eq:manifest-quotient}
\end{equation}%
A closed residual has zero class if and only if it is exact.
\end{coro}

For $R_s=C_sE_{\mathrm{true}}$ with binary support $u_s=(x_s,z_s)$, the full
silent logical obstruction is
\begin{equation}
D_1u_s=0, \qquad [u_s]\neq 0\ \text{in }H^{1}_{\mathrm{full}}.
\label{eq:obstruction}
\end{equation}
Writing $R_s^{\mathrm{full}}=(x_s,z_s)$, exact acceptance is
\begin{equation}
[R_s^{\mathrm{full}}]=0 \quad\Longleftrightarrow\quad x_s^{\mathsf{T}}\in%
\func{im}H_X^{\mathsf{T}} \ \text{and}\ z_s^{\mathsf{T}}\in\func{im}H_Z^{%
\mathsf{T}} \quad\Longleftrightarrow\quad R_s\in\mathcal{S},
\label{eq:fla-residual-pair}
\end{equation}
up to the suppressed Pauli phase. If $p_{\mathrm{fs}}^X$ and $p_{\mathrm{fs}%
}^Z$ denote the two sector failure probabilities, then
\begin{equation}
p_{\mathrm{fs}} \leq p_{\mathrm{fs}}^X+p_{\mathrm{fs}}^Z.
\label{eq:fla-correlation-safe}
\end{equation}
This is a union bound and assumes no statistical independence between fault
species.

The absolute class is available when $E_{\mathrm{true}}$ is declared, as in
an injected-fault audit. It is not available on live hardware. For two
record-compatible candidates $C_{s,a}$ and $C_{s,b}$, however, the relative
class
\begin{equation}
\left[C_{s,a}C_{s,b}^{-1}\right] \in H^{1}_{\mathrm{full}}
\label{eq:fla-relative-class}
\end{equation}
is computable because the unknown error cancels. It partitions a candidate
list into logical fibres but does not identify which fibre contains the
truth.

The one-sector cohomology dictionary is now recorded in its proper role: it
is standard bookkeeping needed by later audits.
\begin{equation}
H^{0}(\tilde{Q}^{\overline{Z}})=\func{coker}H_{X},\qquad H^{1}(\tilde{Q}^{%
\overline{Z}})=\frac{\ker H_{X}}{\func{im}H_{Z}^{\mathsf{T}}},\qquad H^{2}(%
\tilde{Q}^{\overline{Z}})=\ker H_{Z}^{\mathsf{T}}.
\label{eq:sw-cohomology-groups}
\end{equation}%
Rank--nullity gives
\begin{equation}
\dim _{\mathbb{F}_{2}}H^{0}(\tilde{Q}^{\overline{Z}})=m_{X}-\func{rank}H_{X},
\label{eq:sw-H0-dim}
\end{equation}%
\begin{equation}
\dim _{\mathbb{F}_{2}}H^{2}(\tilde{Q}^{\overline{Z}})=m_{Z}-\func{rank}H_{Z},
\label{eq:sw-outer-dims}
\end{equation}%
\begin{equation}
\dim _{\mathbb{F}_{2}}H^{1}(\tilde{Q}^{\overline{Z}})=n-\func{rank}H_{X}-%
\func{rank}H_{Z}=k.  \label{eq:sw-H1-dim}
\end{equation}%
The outer groups record presentation redundancy, not encoded qubits. The
middle group is intrinsic to the logical quotient.

For the standard full-rank Steane $[[7,1,3]]$ presentation,
\begin{equation}
\func{rank}H_{X}=\func{rank}H_{Z}=3,\qquad H^{0}(\tilde{Q}^{\overline{Z}%
})=H^{2}(\tilde{Q}^{\overline{Z}})=0,  \label{eq:sw-steane-values}
\end{equation}%
and one sector has dimension one while the completed address has dimension
two. For the gross $[[144,12,12]]$ bivariate-bicycle presentation, $%
m_{X}=m_{Z}=72$ and both ranks are $66$, so
\begin{equation}
\dim _{\mathbb{F}_{2}}H^{0}(\tilde{Q}^{\overline{Z}})=\dim _{\mathbb{F}%
_{2}}H^{2}(\tilde{Q}^{\overline{Z}})=6,\qquad \dim _{\mathbb{F}_{2}}H^{1}(%
\tilde{Q}^{\overline{Z}})=12.  \label{eq:sw-bb-values}
\end{equation}%
Hence
\begin{equation}
\left\vert H^{1}(\tilde{Q}^{\overline{Z}})\right\vert =2^{12},\qquad
\left\vert N(\mathcal{S})/\mathcal{S}\right\vert =2^{24}.
\label{eq:sw-bb-cardinalities}
\end{equation}%
The redundant checks are properties of the chosen presentation; they are not
additional logical qubits~.

The projected-return block $B_s$ tests the route back into the protected
space. The residual class $[R_s]$ tests the destination after a correction
has been committed. Returning to the code space need not mean returning to
the correct logical fibre~\cite{KnillLaflamme1997}. The two diagnostics are
therefore complementary, not interchangeable.

For a non-CSS abelian stabilizer code, the two-sector splitting is absent.
The classifier is then the undecomposed symplectic quotient $N(\mathcal{S})/%
\mathcal{S}$, still of binary dimension $2k$. What is special to CSS codes
is the decomposition into the two conjugate summands, not the quotient
itself.

\paragraph{Non-CSS symplectic completion.}

The physical supercharge of Eq.~\eqref{eq:sg-Q-def} uses only the
commutation relations $[S_{a},S_{b}]=0$ of Eq.~\eqref{eq:sg-stab-algebra},
whereas its software shadow has so far been written for
Calderbank--Shor--Steane (CSS) presentations, in which every generator is of
pure $X$ or pure $Z$ type~\cite{Calderbank1996,Steane1996}. The purpose of
this paragraph is to close that gap. We show that the completed address $%
H^{1}_{\mathrm{full}}=H^{1}_{\overline X}\oplus H^{1}_{\overline Z}$ of
Proposition~\ref{prop:fla-completeness} is the block-diagonal special case
of one symplectic complex that exists for every abelian Pauli stabilizer
group. The good effect is twofold. The scope claimed in the abstract becomes
exact for the software layer, and the tier dictionary of Subsec.~\ref%
{ssec:tier-taxonomy-certificate} together with the online relative class of
Eq.~\eqref{eq:fla-relative-class} acquires a form that never mentions a
species label. The algebra is the standard binary symplectic description of
stabilizer codes; what is specific to this paper is its placement in the
typed demotion of Subsec.~\ref{ssec:typed-demotion-ranks-weights}.

A Pauli operator modulo phase is a binary row $u=(x\,|\,z)\in \mathbb{F}%
_{2}^{2n}$. Stack the $m$ listed generators into one check matrix $%
H_{S}=[H_{S}^{X}\;\big|\;H_{S}^{Z}]\in \mathbb{F}_{2}^{\,m\times 2n}$,$~$and
introduce the symplectic form,
\begin{equation}
\Lambda =%
\begin{pmatrix}
0 & I_{n} \\
-I_{n} & 0%
\end{pmatrix}%
.  \label{eq:ncss-symplectic-data}
\end{equation}%
Row $a$ of $H_{S}$ records the $X$- and $Z$-support of $S_{a}$. Two Pauli
operators commute if and only if their symplectic product $u\Lambda v^{%
\mathsf{T}}$ vanishes over $\mathbb{F}_{2}$~\cite{NielsenChuang2010}, so the
abelian property of $S$ is one matrix identity,
\begin{equation}
H_{S}\,\Lambda \,H_{S}^{\mathsf{T}}=0\qquad \text{over }\mathbb{F}_{2}.
\label{eq:ncss-self-orthogonality}
\end{equation}%
Demotion now proceeds exactly as in Eq.~\eqref{eq:tdrw-staging}, without a
species split. The graded commutator of Eq.~\eqref{eq:sg-error-to-syndrome}
lights ancilla-register mode $a$ precisely when $S_{a}$ anticommutes with
the residual. Binarization keeps only the list of lit modes and discards
amplitudes, Pauli phases, and register occupation. That list is the linear
map $u\mapsto H_{S}\Lambda u^{\mathsf{T}}\in \mathbb{F}_{2}^{m}$, the binary
shadow of the forward oracle of Subsec.~\ref{subsec:forward-oracle}. A
stabilizer is invisible to every check, so this map alone cannot tell a
harmless dressing from a logical operator, and stabilizer completion adjoins
the incoming arrow from the declared generators. On the graded spaces $C_{2}=%
\mathbb{F}_{2}^{m}$, $C_{1}=\mathbb{F}_{2}^{2n}$, $C_{0}=\mathbb{F}_{2}^{m}$%
,
\begin{equation}
D_{0}=H_{S}^{\mathsf{T}}:\;C_{2}\rightarrow C_{1},\qquad D_{1}=H_{S}\Lambda
:\;C_{1}\rightarrow C_{0},\qquad D_{1}D_{0}=H_{S}\Lambda H_{S}^{\mathsf{T}%
}=0.  \label{eq:ncss-complex}
\end{equation}%
Nilpotency is thus not an extra assumption; it is Eq.~%
\eqref{eq:ncss-self-orthogonality}, from check commutativity. The physical
square $Q^{2}=0$ of Eq.~\eqref{eq:sg-Qsq-zero} cancels because violation
projectors commute while register creators anticommute; the software square
cancels because the listed checks commute. In the degree ordering $%
(C_{2},C_{1},C_{0})$ of Eq.~\eqref{eq:tdrw-graded-domain} the two arrows
assemble into one strictly triangular software differential, the shape of
Eq.~\eqref{eq:tdrw-blockmatrix} with the middle space doubled,
\begin{equation}
\tilde{Q}^{\mathrm{sp}}=%
\begin{pmatrix}
0 & 0 & 0 \\
H_{S}^{\mathsf{T}} & 0 & 0 \\
0 & H_{S}\Lambda & 0%
\end{pmatrix}%
,\qquad (\tilde{Q}^{\mathrm{sp}})^{2}=0.  \label{eq:ncss-block-matrix}
\end{equation}%
Only the corner block survives squaring, and it is Eq.~%
\eqref{eq:ncss-self-orthogonality}; every other composite vanishes by strict
triangularity. The three binary degrees are the generator, support, and
syndrome labels of the complex. As in Subsec.~\ref%
{ssec:typed-demotion-ranks-weights}, $D_{1}$ is a shadow of $Q$, $D_{0}$ is
adjoined data, and the physical layer and the binary layer share generating
rank, not vector spaces, as in Eq.~\eqref{eq:tdrw-ranks}.

\begin{prop}[Symplectic software supercharge]
\label{prop:ncss-symplectic} Let $\mathcal{S}=\langle S_{1},\dots
,S_{r}\rangle \subset \mathcal{P}_{n}$ be any abelian Pauli stabilizer group
with check matrix $H_{S}$ of rank $r$, so that $k=n-r$. Then Eq.~%
\eqref{eq:ncss-complex} is a complex, and its middle cohomology is the full
logical Pauli group,
\begin{equation}
H^{1}\bigl(\tilde{Q}^{\mathrm{sp}}\bigr)=\frac{\ker \!\bigl(H_{S}\Lambda %
\bigr)}{\func{im}\!\bigl(H_{S}^{\mathsf{T}}\bigr)}\;\cong \;\frac{N(\mathcal{%
S})}{\mathcal{S}}\;\cong \;\mathbb{F}_{2}^{2k},\qquad \dim _{\mathbb{F}%
_{2}}H^{1}\bigl(Q_{\mathrm{sw}}^{\mathrm{sp}}\bigr)=2k.
\label{eq:ncss-middle-cohomology}
\end{equation}%
The outer groups $H^{2}=\ker H_{S}^{\mathsf{T}}$ and $H^{0}=\mathbb{F}%
_{2}^{m}/\func{im}(H_{S}\Lambda )$ record presentation redundancy, as in Eq.~%
\eqref{eq:sw-H0-dim},\eqref{eq:sw-outer-dims}.
\end{prop}

\noindent {\textit{Proof}}. The numerator is the set of supports commuting
with every generator, which is $N(\mathcal{S})$ modulo phase. The
denominator is the row space of $H_{S}$, which is $S$ modulo phase. The
quotient is $N(\mathcal{S})/\mathcal{S}$ by definition~\cite{Gottesman1997}.
Since $\Lambda $ is invertible, $\func{rank}(H_{S}\Lambda )=r$, so $\dim
\ker D_{1}=2n-r$. Subtracting $\dim \func{im}D_{0}=r$ gives $2n-2r=2k$. The
outer dimensions follow from rank--nullity in $\mathbb{F}_{2}^{m}$. {\hfill $%
\square $}\newline

The raw kernel $\ker D_{1}$ has dimension $2k+r$; the extra $r$ directions
are the stabilizers themselves, which is why the boundary arrow $D_{0}$ must
be adjoined. The three tiers of Subsec.~\ref{ssec:tier-taxonomy-certificate}
now read without a sector label: a residual support $u$ is Tier-1 if $%
H_{S}\Lambda u^{\mathsf{T}}\neq 0$, Tier-2 if it is syndrome-silent and lies
in $\func{im}H_{S}^{\mathsf{T}}$, and Tier-3 if it is syndrome-silent with $%
[u]\neq 0$ in $H^{1}(\tilde{Q}^{\mathrm{sp}})$. The relative class of two
record-compatible candidates, Eq.~\eqref{eq:fla-relative-class}, is computed
by the same quotient, because the unknown fault still cancels over $\mathbb{F%
}_{2}$.

The CSS two-sector address is recovered without taking a limit. Order the $X$%
-type checks first. Then $H_{S}$ is block diagonal and
\begin{equation}
H_{S}=%
\begin{pmatrix}
H_{X} & 0 \\
0 & H_{Z}%
\end{pmatrix}%
,\qquad H_{S}\Lambda
\begin{pmatrix}
x \\
z%
\end{pmatrix}%
=%
\begin{pmatrix}
H_{X}z \\
H_{Z}x%
\end{pmatrix}%
,  \label{eq:ncss-css-block}
\end{equation}%
and here the equations and matrix multiplications are over $\mathbb{F}_{2}$.
The symplectic form crosses the species: $X$-type checks read $Z$-type
support and conversely. This is the crossed response of Eq.~%
\eqref{eq:fla-orthogonality}, and it shows that the one-sector response $%
\sigma _{X}(W)=H_{X}w_{Z}^{\mathsf{T}}$ of Eq.~\eqref{eq:tdrw-binarization}
is the restriction of $H_{S}\Lambda $ to the selected sector. Keeping the
middle coordinates ordered as $(x\,|\,z)$ and ordering the outgoing check
coordinates as $(Z,X)$, the pair $(D_{0},D_{1})$ of Eq.~%
\eqref{eq:ncss-complex} becomes exactly the block-diagonal pair of Eqs.~%
\eqref{eq:fla-D0}--\eqref{eq:fla-D0D1}, and Eq.~%
\eqref{eq:ncss-self-orthogonality} becomes the two CSS identities $%
H_{X}H_{Z}^{\mathsf{T}}=0$ and $H_{Z}H_{X}^{\mathsf{T}}=0$. The middle
cohomology splits accordingly,
\begin{equation}
H^{1}\bigl(\tilde{Q}^{\mathrm{sp}}\bigr)\;\cong \;\frac{\ker H_{Z}}{\func{im}%
H_{X}^{\mathsf{T}}}\;\oplus \;\frac{\ker H_{X}}{\func{im}H_{Z}^{\mathsf{T}}}%
\;=\;H_{\overline{X}}^{1}\oplus H_{\overline{Z}}^{1}=H_{\mathrm{full}}^{1},
\label{eq:ncss-css-split}
\end{equation}%
with dimension $k+k=2k$ on both sides, in agreement with Eq.~%
\eqref{eq:proof-split}. CSS is therefore a block-diagonal specialization
reached by a discrete choice of presentation. No parameter tends anywhere
over $\mathbb{F}_{2}$, and the excluded wall $t_{X}\rightarrow 0$ or $%
t_{Z}\rightarrow 0$ of Sec.~\ref{sec:weighted-lift} deletes a differential
rather than approaching the mixed case. In the generic non-CSS cases, $H_{%
\mathrm{full}}^{1}$ becomes $H^{1}(\tilde{Q}^{\mathrm{sp}})$.

The quotient is also frame invariant. A local Clifford unitary acts on
supports by a binary symplectic matrix $M$ with $M\Lambda M^{\mathsf{T}%
}=\Lambda $, under which $H_{S}\mapsto H_{S}M$ and Eq.~%
\eqref{eq:ncss-self-orthogonality} is preserved. The triple $(\mathrm{id},M^{%
\mathsf{T}},\mathrm{id})$ intertwines old and new arrows, since $%
(H_{S}M\Lambda )M^{\mathsf{T}}=H_{S}\Lambda $, so it is an isomorphism of
complexes and $H^{1}(\tilde{Q}^{\mathrm{sp}})$ is unchanged. The $X$/$Z$
split is therefore a property of the frame, not of the code. The XZZX
surface code is the standard witness: Hadamard gates on alternating qubits
map it to the CSS surface code, while the logical quotient is untouched~\cite%
{Ataides:2021bdg}. The certificate depends on the code and the committed
residual, never on the frame in which the checks are written.

A genuinely non-CSS check is the perfect $[[5,1,3]]$ code with cyclic
generators $XZZXI$, $IXZZX$, $XIXZZ$, $ZXIXZ$, each mixing both species.
Here $n=5$, $m=r=4$, $k=1$, and
\begin{equation}
\dim \ker \bigl(H_{S}\Lambda \bigr)=10-4=6,\qquad \dim H^{1}\bigl(Q_{\mathrm{%
sw}}^{\mathrm{sp}}\bigr)=6-4=2=2k,  \label{eq:ncss-five-qubit}
\end{equation}%
two independent logical directions, $\overline{X}$ and $\overline{Z}$, hence
four classes; the four removed directions are stabilizer boundaries. The
Steane code of Subsec.~\ref{subsec:steane-static-audit} and the gross code
of Sec.~\ref{sec:bb-circuit-reranking} are CSS, so Proposition~\ref%
{prop:ncss-symplectic} adds coverage there but no new numbers.

Proposition~\ref{prop:ncss-symplectic} is a statement about exact binary
classification and inherits no benchmark or lower $p_{\mathrm{fs}}$. Most of
constructions in this paper remain CSS-specific: the check-species weights $%
t_{X},t_{Z}$ and the invariance proof of Proposition~\ref%
{prop:weight-invariance}, which need separate blocks that a mixed row does
not have. CSS therefore remains the working dictionary for weighting and for
the audits of Secs.~\ref{sec:bb-circuit-reranking} and~\ref%
{sec:weighted-lift}; the symplectic completion fixes the exact address for
every abelian code and is used below only through the sector-free tier tests
just stated.

\subsection{Tier taxonomy, $p_{\mathrm{fs}}$, and the residual certificate}

\label{ssec:tier-taxonomy-certificate}

The purpose of this subsection is to turn the completed two-sector
cohomology $H_{\mathrm{full}}^{1}=H^{1}(\tilde{Q}^{\mathrm{full}})$ of
Subsec.~\ref{subsec:full-logical-address} into an accept--reject certificate
for a committed recovery. Its good effect is to separate syndrome clearance
from preservation of the encoded logical state across both CSS sectors,
supplying the Tier-2 and Tier-3 definitions used in Subsecs.~\ref%
{subsec:ambiguity-flag-rate} and~\ref{subsec:circuit-static-bridge}~\cite%
{KnillLaflamme1997,Gottesman1997,Terhal2015}.

For the selected correction $C_{s}$ and unobserved fault $E_{\mathrm{true}}$%
, write the residual $R_{s}=C_{s}E_{\mathrm{true}}$ of Eq.~%
\eqref{eq:sec3-residual} via its phase-suppressed binary support $%
u_{s}=(x_{s},z_{s})\in C_{1}^{\mathrm{full}}$. By Eqs.~\eqref{eq:fla-D0}--%
\eqref{eq:fla-D-actions}, $D_{1}u_{s}=(H_{Z}x_{s}^{\mathsf{T}},H_{X}z_{s}^{%
\mathsf{T}})$ is the full syndrome and $\func{im}D_{0}$ is the
stabilizer-boundary subspace. The class $[u_{s}]\in H_{\mathrm{full}}^{1}$
is defined when $D_{1}u_{s}=0$. The silent-failure event $F$ of Subsec.~\ref%
{subsec:fs-target} and its probability $p_{\mathrm{fs}}$ over the fault
ensemble are then
\begin{equation}
F=\bigl\{R_{s}\in N(\mathcal{S})\setminus \mathcal{S}\bigr\}=\bigl\{%
D_{1}u_{s}=0,\ [u_{s}]\neq 0\ \text{in }H_{\mathrm{full}}^{1}\bigr\},\qquad
p_{\mathrm{fs}}=\func{Prob}(F).  \label{eq:sw-three-way}
\end{equation}%
Here $F$ is a physical event, whereas $p_{\mathrm{fs}}$ is its probability: $%
D_{1}u_{s}=0$ means every check is quiet, while $[u_{s}]\neq 0$ means a
nontrivial logical Pauli has acted on $\mathcal{K}$~\cite%
{Dennis2002,Terhal2015}.

A Tier-1 residual has $D_{1} u_{s} \neq 0$ and is excluded whenever $C_{s}$
is syndrome-matched ($\sigma(C_{s}) = \sigma(E_{\mathrm{true}}) = s
\Rightarrow \sigma(R_{s}) = 0$). Every closed residual ($D_{1} u_{s} = 0$)
falls into one of two silent tiers:
\begin{equation}
D_{1} u_{s} = 0, \quad u_{s} \in \func{im} D_{0} \; \Longleftrightarrow \;
[u_{s}] = 0 \ \text{in } H^{1}_{\mathrm{full}} \qquad \text{(Tier-2,
stabilizer boundary)},  \label{eq:sw-accept-cond}
\end{equation}
\begin{equation}
D_{1} u_{s} = 0, \quad u_{s} \notin \func{im} D_{0} \; \Longleftrightarrow
\; [u_{s}] \neq 0 \ \text{in } H^{1}_{\mathrm{full}} \qquad \text{(Tier-3,
syndrome-silent logical)}.  \label{eq:sw-accept-concl}
\end{equation}
By Corollary~\ref{cor:fla-quotient}, $[u_{s}] = 0$ means $u_{s} = D_{0} w$
for some generator vector $w \in C_{2}^{\mathrm{full}}$, so $R_{s} \in
\mathcal{S}$ acts as the identity on $\mathcal{K}$ up to phase;
contrapositively, $[u_{s}] \neq 0$ isolates the Tier-3 logical obstruction~%
\cite{Calderbank1996,Steane1996,Gottesman1997}. Both sectors are essential
because a purely $X$-type logical fault ($z_{s} = 0$, $H_{Z} x_{s}^{\mathsf{T%
}} = 0$, $x_{s}^{\mathsf{T}} \notin \func{im} H_{X}^{\mathsf{T}}$) is
invisible to $H_{X}$ alone; full acceptance requires Eq.~%
\eqref{eq:fla-residual-pair} and obeys Eq.~\eqref{eq:fla-correlation-safe},
while non-CSS codes use $H^{1}(\tilde{Q}^{\mathrm{sp}})$ of Eq.~%
\eqref{eq:ncss-middle-cohomology}.

This absolute verdict requires $E_{\mathrm{true}}$ and is an offline audit
tool: a known class $[u_{s}]$ determines the inverse logical class, whereas
online check records cannot reveal a silent fault. Online, only the
candidate-relative class of Eq.~\eqref{eq:fla-relative-class} is computable
to flag candidate disagreement (Subsec.~\ref{subsec:ambiguity-flag-rate}),
and soft scores in Secs.~\ref{sec:bb-circuit-reranking} and~\ref%
{sec:weighted-lift} rank candidates only after this exact quotient test~\cite%
{Panteleev2021,Bravyi2024}.


\subsection{Ambiguity flag and unflagged silent mass}

\label{subsec:ambiguity-flag-rate}

The absolute certificate requires $E_{\mathrm{true}}$, which a running
memory does not observe. The live decoder does observe a syndrome record $s$
and a finite candidate list $\{C_{s,j}\}_{j=1}^{M_{s}}$. The relative return
of two candidates is
\begin{equation}
\Delta_{jj^{\prime }} = C_{s,j}\big(C_{s,j^{\prime }}\big)^{-1}, \qquad \big[%
\Delta_{jj^{\prime }}\big] \in H^{1}_{\mathrm{full}}.
\label{eq:relativeclass}
\end{equation}
This is not the truth-referenced residual. The unknown error cancels
exactly,
\begin{equation}
R_{s,j}R_{s,j^{\prime }}^{-1} = C_{s,j}E_{\mathrm{true}} E_{\mathrm{true}%
}^{-1} \big(C_{s,j^{\prime }}\big)^{-1} = \Delta_{jj^{\prime }}.
\label{eq:relative-cancellation}
\end{equation}
Two candidates that differ by a stabilizer tell the same logical story. Two
candidates that differ by a nontrivial class tell distinct logical stories
that the stabilizer measurements cannot reconcile. This motivates the
per-record indicator
\begin{equation}
I_{\mathrm{amb}}(s) = \mathbf{1}\Big\{ \exists\, j\neq j^{\prime }: \big[%
\Delta_{jj^{\prime }}\big]\neq 0 \text{ in }H^{1}_{\mathrm{full}} \Big\}.
\label{eq:ambflag}
\end{equation}
Since both candidates reproduce the same record, their relative return is
automatically closed. The decision thus reduces to the exactness test
\begin{equation}
\big[\Delta_{jj^{\prime }}\big]=0 \Longleftrightarrow \delta u_{jj^{\prime
}}\in\func{im} D_{0},  \label{eq:exactcriterion}
\end{equation}
where $\delta u_{jj^{\prime }}$ is the binary support of $\Delta_{jj^{\prime
}}$. It is one finite-field linear solve and requires neither a noise model
nor a confidence score.

For a frozen list-generation rule, the ambiguity rate and its empirical
estimator over $N$ independent records are
\begin{equation}
p_{\mathrm{amb}}=\sum_{s}\Pr (s)\,I_{\mathrm{amb}}(s),\qquad \hat{p}_{%
\mathrm{amb}}=\frac{1}{N}\sum_{i=1}^{N}I_{\mathrm{amb}}(s_{i}),\qquad
\mathbb{E}\big[\hat{p}_{\mathrm{amb}}\big]=p_{\mathrm{amb}}.
\label{eq:ambrate}
\end{equation}%
Unlike $p_{\mathrm{fs}}$, this rate can be measured on unlabelled live
records. It is nevertheless not a logical error rate. Let $F$ denote the
Tier-3 failure event. Splitting it by the flag gives an identity valid even
when one flag value never occurs:
\begin{equation}
p_{\mathrm{fs}}=\Pr (F\ \text{and}\ I_{\mathrm{amb}}=1)+p_{\mathrm{miss}%
},\qquad p_{\mathrm{miss}}:=\Pr (F\ \text{and}\ I_{\mathrm{amb}}=0).
\label{eq:bayessilent}
\end{equation}%
When $1-p_{\mathrm{amb}}>0$, the missed joint mass also has the conditional
form
\begin{equation}
p_{\mathrm{miss}}=(1-p_{\mathrm{amb}})\Pr (F\mid I_{\mathrm{amb}}=0).
\label{eq:missbridge}
\end{equation}%
The joint definition remains meaningful if this conditional probability is
undefined. Since the flagged contribution is at most $p_{\mathrm{amb}}$, $p_{%
\mathrm{fs}}\leq p_{\mathrm{amb}}+p_{\mathrm{miss}}$. No independence
assumption enters these identities. A singleton list is never ambiguous but
can still be wrong, so $p_{\mathrm{miss}}$ is the exact blind mass of a
triage policy driven only by the flag. Every reported $p_{\mathrm{amb}}$
should therefore be accompanied by an injected-fault estimate of $p_{\mathrm{%
miss}}$ on the same frozen decoder stack. Circuit-level detector histories
are treated in Sec.~\ref{sec:bb-circuit-reranking}. The static quotient
developed here does not, by itself, prove the circuit-to-static
identification.

\subsection{Steane $[[7,1,3]]$: exact static audit of the CSS decoder
certificate}

\label{subsec:steane-static-audit}

The purpose of this subsection is to run the certificate of Secs.~\ref%
{sec:syndrome-fibres-KL-window} and \ref{sec:software-supercharge-cohomology}
on one code that can be checked by hand. Its good effect is twofold. The
return block $B_{s}$ of Eq.~(\ref{eq:projected-return-block}) and the class $%
[u_{s}]\in H_{\mathrm{full}}^{1}$ of Eq.~(\ref{eq:manifest-quotient}) are
seen to agree on every branch of an enumerated window, and the online
ambiguity flag of Subsec.~\ref{subsec:ambiguity-flag-rate} is seen to fire
on a genuine logical disagreement.

\paragraph{Code data.}

The Steane code \cite{Calderbank1996,Steane1996} has $n=7$, $r=6$, $k=1$, $%
t=1$, and the full-rank presentation
\begin{equation}
H_X=H_Z=H=%
\begin{pmatrix}
0 & 0 & 0 & 1 & 1 & 1 & 1 \\
0 & 1 & 1 & 0 & 0 & 1 & 1 \\
1 & 0 & 1 & 0 & 1 & 0 & 1%
\end{pmatrix}%
, \qquad H_XH_Z^{\mathsf{T}}=0\ \text{over }\mathbb{F}_2 .
\label{eq:steane-merged-params}
\end{equation}
Write $e_i$ for the $i$-th unit column of $\mathbb{F}_2^{7}$ and $h_i=He_i$
for the $i$-th column of $H$; the seven columns are the seven nonzero
vectors of $\mathbb{F}_2^{3}$. Pauli supports $u=(x,z)$ are binary rows, as
in Eq.~(\ref{eq:fla-D-actions}). By Eq.~(\ref{eq:fla-full-dimension}), $%
\dim_{\mathbb{F}_2}H^{1}_{\mathrm{full}}=2k=2$, one coordinate each for $%
\bar X$ and $\bar Z$, while the protected manifold $\mathcal{K}$ of Eq.~(\ref%
{eq:sg-K-equals-CQEC}) has $\dim_{\mathbb{C}}\mathcal{K}=2^{k}=2$. The first
number counts silent logical directions, the second counts encoded states.

\paragraph{Declared window.}

Since $t=1$, the audited catalogue is $\mathcal{E}_{w=1}=\{X_i,Y_i,Z_i%
\}_{i=1}^{7}$, and the declared family at each record is $\mathcal{F}_s=%
\mathcal{E}_{w=1}\cap\sigma^{-1}(s)$, a restriction of Eq.~(\ref%
{eq:audited-family}). In the order $(s_Z,s_X)$ of Eq.~(\ref%
{eq:fla-orthogonality}),
\begin{equation}
\sigma(X_i)=(h_i,0),\qquad \sigma(Z_i)=(0,h_i),\qquad \sigma(Y_i)=(h_i,h_i).
\label{eq:steane-merged-window}
\end{equation}
Distinct columns give twenty-one distinct words, so every $\mathcal{F}_s$ is
a singleton, the minimum-weight decoder returns $E_s=E$ with $C_s=E^{-1}$,
and the bound of Eq.~(\ref{eq:sec27-weight-one-count}) is saturated,
\begin{equation}
\mathcal{S}_{\mathrm{impl}}=\sigma(\mathcal{E}_{w=1})\subseteq\mathcal{S}_{%
\mathrm{faith}},\qquad N_{w=1}=3n=21=N_{\mathrm{cert}} .
\label{eq:steane-merged-counts}
\end{equation}
Physically, every single-qubit fault lights its own pattern of
ancilla-register modes, so no two declared faults compete for one record.

\paragraph{Acceptance.}

With $R_s(E)=C_sE=I$, the scalar-return test of Eq.~(\ref%
{eq:all-branch-scalar-return}) and the Tier-2 test of Eq.~(\ref%
{eq:sw-accept-cond}) pass together,
\begin{equation}
P_{\mathcal{K}}\,C_sE\,P_{\mathcal{K}}=P_{\mathcal{K}},\qquad B_s=P_{%
\mathcal{K}}C_s\widetilde{P}_s\neq 0,\qquad u_s=(x_s,z_s)=0,\qquad
[u_s]=0\in H^{1}_{\mathrm{full}} ,  \label{eq:steane-residual}
\end{equation}
so $p_{\mathrm{fs}}|_{\mathcal{E}_{w=1}}=0$ for any fault ensemble supported
on the catalogue \cite{KnillLaflamme1997}. The $Y_i$ branch lights both
check species, which is why the two-sector class of Eq.~(\ref{eq:proof-split}%
) is the object quoted.

\paragraph{Online relative residual class.}

Now let a supplied candidate list at $s=\sigma (X_{1})$ contain the decoder
output $C_{s}$ and a second record-compatible correction $C_{s}^{\prime
}=C_{s}X_{1}X_{2}X_{3}$, which is not a minimum-weight choice. Because $%
h_{1}+h_{2}+h_{3}=0$, both clear the syndrome and both have nonzero return
blocks, so arrival in $\mathcal{K}$ cannot tell them apart. In the
convention of Eq.~(\ref{eq:relativeclass}), their relative residual operator
and its binary support are
\begin{equation}
\Delta _{s}=C_{s}^{\prime }C_{s}^{-1}=X_{1}X_{2}X_{3},\qquad u_{\Delta }=%
\big((e_{1}+e_{2}+e_{3})^{\mathsf{T}},\,0\big).
\label{eq:steane-relative-operator}
\end{equation}%
The support has weight three, whereas every nonzero element of the row space
of $H$ has weight four, hence
\begin{equation}
D_{1}u_{\Delta }=0,\qquad \lbrack u_{\Delta }]\neq 0\in H_{\mathrm{full}%
}^{1},\qquad I_{\mathrm{amb}}(s)=1.  \label{eq:steane-rejection}
\end{equation}%
The operator $X_{1}X_{2}X_{3}$ represents $\bar{X}$: the two candidates
differ by a logical Pauli, so applying one in place of the other flips the
encoded qubit while every register mode stays dark, the Tier-3 class of Eq.~(%
\ref{eq:sw-accept-concl}). The unknown fault cancelled in Eq.~(\ref%
{eq:relative-cancellation}), so the flag is computed online; it reports
disagreement, not which candidate is right. Equations~(\ref%
{eq:steane-residual})--(\ref{eq:steane-rejection}) are exact
finite-catalogue statements, not a threshold or a noise-averaged rate; the
circuit-level audit is the task of Sec.~\ref{sec:bb-circuit-reranking}.

\section{Circuit-Level Logical-Fibre Audit and Reranking for qLDPC Codes: A
Bivariate-Bicycle Example}

\label{sec:bb-circuit-reranking}

\renewcommand{\theequation}{5.\arabic{equation}} \setcounter{equation}{0} %
\renewcommand{\thethm}{5.\arabic{thm}} \setcounter{thm}{0} %
\renewcommand{\theprop}{5.\arabic{prop}} \setcounter{prop}{0} %
\renewcommand{\theprotocol}{5.\arabic{protocol}} \setcounter{protocol}{0}

This section carries the logical-fibre audit to circuit-level decoding of
the Gross $[[144,12,12]]$ bivariate-bicycle code~\cite{Bravyi2024}. The
operational question is whether detector-compatible fault histories act
identically on the encoded information. The clean protected manifold,
projected-return blocks, certified faithful window, and completed two-sector
residual classifier are inherited from Secs.~\ref{sec:susy-stabilizer}, \ref%
{sec:syndrome-fibres-KL-window}, and \ref%
{sec:software-supercharge-cohomology}.

For a frozen extraction circuit, fault basis, detector convention, and
logical Pauli frame, the detector and logical-action maps give an exact
relative classification. Candidates reproducing the same detector record lie
in the same logical fibre precisely when their relative history has
vanishing logical action. The unknown true history cancels from this
comparison, but logical equivalence does not establish that either candidate
is correct. The layer therefore audits and, where appropriate, reranks lists
supplied by a base decoder; it does not replace syndrome inference or
provide an unconditional online recovery certificate.

We first specify the code and circuit conventions, then develop detector
fibres and audit the static--circuit bridge on a predeclared finite domain.
Candidate reranking, finite-weight enumeration, and matched evaluation
follow this classification. Logical-error reduction, compression, and
latency advantages remain conditional on the extraction schedule, noise
model, candidate-list coverage, and decoding configuration; none follows
from logical classification alone.

\subsection{Bivariate-bicycle conventions and the Gross benchmark}

\label{ssec:bb-family} \label{ssec:bb-scope} \label{ssec:bb-symmetry}

The purpose of this subsection is to declare the code, extraction circuit,
and decoder conventions before any circuit-level audit is attempted. Its
good effect is that every exact statement later in this section refers to
one fixed object. The bivariate-bicycle family uses two qubit blocks indexed
by a finite Abelian translation group,
\begin{equation}
G_{\ell ,m}=\mathbb{Z}_{\ell }\times \mathbb{Z}_{m},\qquad \lvert G_{\ell
,m}\rvert =\ell m.  \label{eq:bb-family-group}
\end{equation}%
Its binary group ring is $R_{\ell ,m}=\mathbb{F}_{2}[x,y]/(x^{\ell
}-1,y^{m}-1)$. Consequently the number of data qubits is
\begin{equation}
n_{\ell ,m}=2\lvert G_{\ell ,m}\rvert =2\ell m.  \label{eq:bb-family-n}
\end{equation}%
The CSS check matrices $H_{X}$ and $H_{Z}$ are assembled from circulant
blocks and satisfy
\begin{equation}
H_{X}H_{Z}^{\mathsf{T}}=0\qquad \text{over }\mathbb{F}_{2}.
\label{eq:bb-family-orth}
\end{equation}%
These code-family identities are standard inputs, not new circuit-level
conclusions. The orthogonality makes the static quotient used below well
defined. It says nothing by itself about how a fault propagates through a
chosen syndrome-extraction schedule.

The Gross instance of Ref.~\cite{Bravyi2024} has
\begin{equation}
(\ell,m)_{\mathrm{gross}}=(12,6), \qquad G_{\mathrm{gross}}=\mathbb{Z}%
_{12}\times\mathbb{Z}_{6}, \qquad \lvert G_{\mathrm{gross}}\rvert=72,
\label{eq:bb-group}
\end{equation}
\begin{equation}
[[n,k,d]]=[[144,12,12]], \qquad n=2\lvert G_{\mathrm{gross}}\rvert=144.
\label{eq:bb-code-params}
\end{equation}
The number $144$ counts data qubits, not translation sectors. The clean
protected manifold $\mathcal{K}$ has dimension $2^k=4096$; a complete
logical Pauli address instead contains $2k=24$ binary coordinates. The two
cyclic factors are $12$ and $6$. Replacing them by two copies of $6$, or by
the distinct square instance $G_{12,12}$, would change the code under
discussion. These distinctions fix the scope of later counts without
repeating the static construction of Secs.~\ref{sec:susy-stabilizer}--\ref%
{sec:software-supercharge-cohomology}.

All subsequent statements refer to one frozen operational stack. Declare the
number and ordering of extraction rounds, the preparation and measurement
bases of the ancilla register that implements the checks, every
Clifford-gate orientation, the elementary fault basis of cardinality $N_f$,
and the detector convention. Declare also the ordered logical Pauli
generators, final Pauli frame, correction rule, candidate-list generator,
list cap, and latency allowance. The benchmark candidate rule can be belief
propagation with ordered-statistics post-processing~\cite{Panteleev2021};
the audit does not require this specific engine. A new schedule, frame, or
candidate rule requires a new audit: the same physical fault can acquire a
different detector shadow or final logical action.

The group $G_{\mathrm{gross}}$ can reorganize qubit and check labels. At
circuit level it can support optional offline orbit or character bookkeeping
only when the extraction circuit, detector map, logical action, and relevant
noise weights respect the corresponding action. The character convention
remains the one fixed in Subsec.~\ref{subsec:compress-Bs}; the operational
invariance test belongs to Eq.~\eqref{eq:bb-enum-orbit-invariance} below. In
particular, translation symmetry neither identifies two logical fibres nor
promotes an unaudited history into a certified one. Schedule-dependent
circuit constructions illustrate why this qualification matters~\cite%
{Shaw2025}.

Once the stack is fixed, detector compatibility and relative logical action
are exact finite-field statements. A lower logical-error rate, smaller audit
ledger, or real-time advantage is not implied. Those are conditional claims
for the enumeration and matched ablation in Subsecs.~\ref%
{ssec:bb-enumeration} and~\ref{ssec:bb-metrics}.

\subsection{Circuit detectors, fault histories, and the exact logical-fibre
criterion}

\label{ssec:bb-fibre-test}

A static syndrome labels errors at one instant. A noisy extraction circuit
instead yields a sequence of measurement comparisons. The purpose of this
subsection is to ask the exact question needed by the bridge: do two
explanations of the same circuit record act identically on the encoded
information? It does not select the correct explanation.

For a fixed Clifford circuit and elementary Pauli-fault basis, let $f\in%
\mathbb{F}_2^{N_f}$ denote a spacetime fault history. The observed detector
parities obey
\begin{equation}
d=Df\in\mathbb{F}_2^{N_d}, \qquad D\in\mathbb{F}_2^{N_d\times N_f}.
\label{eq:bb-detector-matrix}
\end{equation}
A coordinate of $f$ records an elementary fault mechanism; a coordinate of $%
d $ records a parity of measurement outcomes. This detector description of
repeated extraction is standard~\cite{Dennis2002}. Thus $D$ retains the
visible shadow of a history, not its full physical effect. For one observed
record, the compatible histories form an affine fibre,
\begin{equation}
F_d =\left\{f\in\mathbb{F}_2^{N_f}:Df=d\right\}.  \label{eq:bb-fibre}
\end{equation}
A base decoder proposes one or more candidates in $F_d$. Detector
compatibility alone cannot certify that any candidate was the true history.

The missing address is obtained by propagating the elementary faults in the
declared logical frame. Write the corresponding linear map as
\begin{equation}
A\in \mathbb{F}_{2}^{2k\times N_{f}},\qquad f\longmapsto Af\in \mathbb{F}%
_{2}^{2k}.  \label{eq:bb-logical-action}
\end{equation}%
The label $Af$ records the induced logical Pauli coordinates modulo
stabilizers~\cite{Gottesman1997}. Both conjugate CSS species are retained,
as required by the completed static address of Subsec.~\ref%
{subsec:full-logical-address}:
\begin{equation}
\dim _{\mathbb{F}_{2}}H^{1}\!\left( \tilde{Q}^{\mathrm{full}}\right) =2k.
\label{eq:bb-width-match}
\end{equation}%
This dimension is the width of a logical label, not the dimension $2^{k}$ of
the protected state space. Neither the existence of a detector map nor the
resulting logical-coset bookkeeping is claimed as a new theorem.

Take $f_a,f_b\in F_d$ and set
\begin{equation}
\Delta f=f_a+f_b.  \label{eq:bb-delta}
\end{equation}
Binary addition gives at once
\begin{equation}
D\,\Delta f=Df_a+Df_b=d+d=0.  \label{eq:relative-history}
\end{equation}
The unknown true fault cancels from the comparison of the two candidates;
writing $f_a=f_{\mathrm{true}}+e_a$ and $f_b=f_{\mathrm{true}}+e_b$ leaves $%
\Delta f=e_a+e_b$, so no online oracle is required to compute their relative
address. Linearity of $A$ gives
\begin{equation}
A\,\Delta f=Af_a+Af_b,
\end{equation}
and therefore the exact same-fibre criterion is
\begin{equation}
A\,\Delta f=0 \ \Longleftrightarrow\ Af_a=Af_b.  \label{eq:bb-fibre-test}
\end{equation}

\begin{prop}
\label{prop:bb-fibre-classification} For a fixed detector matrix $D$,
logical-action map $A$, and logical frame, the condition $A(f_a+f_b)=0$ is
an equivalence relation on $F_d$. Its classes are the level sets of $%
f\mapsto Af$.
\end{prop}

Indeed, equality of the binary labels is reflexive, symmetric, and
transitive. This familiar fact is retained because the audit needs an
unambiguous meaning of ``same logical fibre.'' Physically, the hardware can
see that two histories cast the same detector shadow while remaining blind
to an encoded Pauli displacement between them. The dangerous relative event
is
\begin{equation}
D\,\Delta f=0, \qquad A\,\Delta f\neq 0.  \label{eq:bb-silent-wrong-fibre}
\end{equation}
It describes rival logical accounts of one detector record. Without the true
history, it does not prove that a particular observed shot failed. A
posterior score can choose between these accounts, but cannot make the two
classes equivalent. The next subsection checks a different issue: whether
this circuit-level address agrees with the static cohomological address
computed after propagation and residualization.

\subsection{Audited static--circuit bridge and low-weight certification}

\label{subsec:circuit-static-bridge} \label{subsec:bb-bridge}

\paragraph{Purpose and two address spaces.}

The kernel of $D$ contains detector-silent spacetime histories. The $H^{1}\!(%
\tilde{Q}^{\mathrm{full}})$ contains logical classes of closed end-of-round
data residuals. They inhabit different vector spaces. The purpose of the
bridge is to compare two separately constructed full logical addresses on a
declared finite domain. Agreement makes the circuit-to-static interface
usable there. Disagreement supplies an explicit counterexample rather than a
third kind of logical state. Throughout, the sets denoted by $\Omega $ are
finite sets of fault histories.

The fixed syndrome-extraction schedule \cite{Bravyi2024},\cite{Shaw2025},%
\cite{Gidney2021} and correction convention define a proposed
residualization on detector-null relative histories,
\begin{equation}
R_{\mathrm{circ}}:\ker D\longrightarrow \mathbb{F}_{2}^{n}\oplus \mathbb{F}%
_{2}^{n},  \label{eq:rcirc-def}
\end{equation}%
\begin{equation}
R_{\mathrm{circ}}(\Delta f)=\left( r_{X}(\Delta f),r_{Z}(\Delta f)\right) .
\label{eq:rcirc-value}
\end{equation}%
The two components are the final $X$- and $Z$-support vectors under the
fixed propagation and relative-correction rule. No linearity of $R_{\mathrm{%
circ}}$ is assumed. Use $R_{\mathrm{circ}}(\Delta f)$ only when every
admitted candidate pair with $\Delta f=f_{a}+f_{b}$ gives the same final
supports; selecting one pair does not establish this independence. If the
supports differ, retain the history and the pairwise outputs, record the
unsatisfied prerequisite, and keep the history in the parent coverage counts.

The residual must also be closed before it has a static cohomology class. On
the audited domain defined below, verify
\begin{equation}
H_{Z}r_{X}(\Delta f)=0,\qquad H_{X}r_{Z}(\Delta f)=0.
\label{eq:rcirc-closure}
\end{equation}%
Failure of either equation is an interface failure to be reported, not a
nonzero element of the static quotient. On closed inputs the completed
two-sector address inherited from Sec.~\ref%
{sec:software-supercharge-cohomology} is
\begin{equation}
H^{1}(\tilde{Q}^{\mathrm{full}})=\frac{\ker H_{Z}}{\func{im}H_{X}^{\mathsf{T}%
}}\oplus \frac{\ker H_{X}}{\func{im}H_{Z}^{\mathsf{T}}}.
\label{eq:completed-address}
\end{equation}%
It has binary dimension $2k$. In the first summand a closed $X$-support
vector is identified modulo an $X$-stabilizer boundary; the second does the
same for $Z$ support. Thus the quotient forgets harmless local dressing but
retains a protected logical displacement. With $\pi $ denoting quotient
projection, the static address induced by the propagated circuit residual is
\begin{equation}
L_{\mathrm{circ}}=\pi \circ R_{\mathrm{circ}},\qquad L_{\mathrm{circ}%
}(\Delta f)=\left( [r_{X}(\Delta f)],[r_{Z}(\Delta f)]\right) .
\label{eq:lcirc-def}
\end{equation}%
Here $L_{\mathrm{circ}}$ is defined only for histories whose residual is
well defined and closed. In particular, no general map from $\ker D$ into
the quotient has been silently assumed. The completed address, combining
both CSS sectors, must be compared with the $2k$ coordinates of $A$.

\paragraph{Predeclared candidate domain.}

Let $\mathcal{D}_{\mathrm{audit}}\subseteq \mathbb{F}_{2}^{N_{d}}$ be the
declared detector-record domain. Its end-of-round static sources must be
identified through the frozen interface with the certified faithful window $%
\mathcal{S}_{\mathrm{faith}}$ of Subsec.~\ref%
{subsec:certified-faithful-window}; this identification is an input to be
checked, not an automatic consequence of Eq.~\eqref{eq:bb-family-orth}. For
each $d\in \mathcal{D}_{\mathrm{audit}}$,$~$let $\mathcal{L}%
_{d}=\{f_{1},\ldots ,f_{M}\}\subseteq \mathcal{F}_{d}~$denote the finite
candidate list returned by the base decoder, held fixed throughout the
audit. The parent set of relative histories is
\begin{equation}
\Omega _{D,2}^{\mathrm{cand}}=\{\Delta f=f_{a}+f_{b}:d\in \mathcal{D}_{%
\mathrm{audit}},~~f_{a},f_{b}\in \mathcal{L}_{d},~Df_{a}=Df_{b}=d,\func{wt}%
(\Delta f)\leq 2\}.  \label{eq:pre-audit-domain}
\end{equation}%
The subscript $D$ in $\Omega _{D,2}^{\mathrm{cand}}~$refers to detector
compatibility; subscript $2$ in $\Omega _{D,2}^{\mathrm{cand}}~$is a cutoff
on the weight of the relative history. The parent set is fixed before either
logical verdict is examined. It is generally a small subset of $\ker D$
because it also depends on the declared records and candidate lists.

Weight one checks the response to an individual fault, including an
exceptional zero column of $D$. Weight two is the first possible shell in
which separately detector-visible fault contributions cancel. It does not
guarantee that any resulting relative history has nontrivial logical action.
Record the shell counts
\begin{equation}
N_{w}^{\mathrm{cand}}=\left\vert \left\{ \Delta f\in \Omega _{D,2}^{\mathrm{%
cand}}:\func{wt}(\Delta f)=w\right\} \right\vert ,\qquad w\in \{0,1,2\}.
\label{eq:candidate-counts}
\end{equation}%
$N_{w}^{\mathrm{cand}}~$counts distinct admitted relative histories at
weight $w$. The code distance $d=12$ supplies no substitute for
circuit-level shell coverage.

\paragraph{The full-address test.}

Choose the ordered logical Pauli generators that were used to construct $A$.
They induce the symplectic frame identification
\begin{equation}
\mathbb{F}_{2}^{2k}\cong H^{1}(\tilde{Q}^{\mathrm{full}}).
\label{eq:frame-identification}
\end{equation}%
For a closed residual, the static half vanishes precisely when both
boundary-membership statements hold:
\begin{equation}
r_{X}(\Delta f)\in \func{im}H_{X}^{\mathsf{T}},\qquad r_{Z}(\Delta f)\in
\func{im}H_{Z}^{\mathsf{T}}.  \label{eq:boundary-membership}
\end{equation}%
This follows from the definition of each quotient in Eq.~%
\eqref{eq:completed-address}.

Once the domain and the identification of Eq.~\eqref{eq:frame-identification}
have been fixed, set
\begin{equation}
m:\Omega _{D,2}^{\mathrm{cand}}\longrightarrow \{0,1\},\qquad m(\Delta f)=0\
\Longleftrightarrow \ A\,\Delta f=L_{\mathrm{circ}}(\Delta f).
\label{eq:bridge-address}
\end{equation}%
The bit is defined on the whole parent set only after the well-definedness
and closure checks above pass throughout that set. A failure of those
prerequisites remains a reported interface counterexample and prevents a
claim that the parent set has been fully classified. The bridge-consistent
subset is then
\begin{equation}
\Omega _{D,2}^{\mathrm{cert}}=\left\{ \Delta f\in \Omega _{D,2}^{\mathrm{cand%
}}:m(\Delta f)=0\right\} .  \label{eq:certified-subset}
\end{equation}%
The superscript $\mathrm{cert}$ certifies agreement of two addresses on a
declared history. It does not identify the true fault or guarantee that a
committed decoder representative is correct.

Passing the full comparison implies the weaker condition
\begin{equation}
A\,\Delta f=0\quad \Longleftrightarrow \quad L_{\mathrm{circ}}(\Delta f)=0.
\label{eq:bridge-biconditional}
\end{equation}%
The full-address equality\ $A\,\Delta f=L_{\mathrm{circ}}(\Delta f)~$under
the frame identification \eqref{eq:frame-identification}, is strictly
stronger than a comparison of zero versus nonzero: it can detect a frame
permutation that maps one nontrivial logical class to another.

\paragraph{Gross-code finite audit.}

The natural gross check presentation has $m_X=m_Z=72$ and $\func{rank}H_X=%
\func{rank}H_Z=66$~\cite{Bravyi2024}. For each conjugate sector,
\begin{equation}
n-\func{rank}H_X-\func{rank}H_Z =144-66-66=12=k.  \label{eq:gross-dimension}
\end{equation}
This is a static dimension count, not the schedule-dependent number $N_f$ of
elementary fault coordinates fixed by the extraction circuit~\cite%
{Shaw2025,Gidney2021}.

The audit compares two logical addresses of one relative history $\Delta
f=f_{a}+f_{b}$, built by independent routes. The circuit route propagates
faults through the fixed schedule and reads $A\,\Delta f\in \mathbb{F}%
_{2}^{2k}$ of Subsec.~\ref{ssec:bb-fibre-test}. The static route
residualizes $\Delta f$ into $R_{\mathrm{circ}}(\Delta f)=(r_{X}(\Delta
f),r_{Z}(\Delta f))$ and reads $L_{\mathrm{circ}}(\Delta f)\in H^{1}\!\left(
\tilde{Q}^{\mathrm{full}}\right) $ of Subsec.~\ref%
{subsec:full-logical-address}. Agreement, history by history on a declared
domain, is what permits the static classifier of Sec.~\ref%
{sec:software-supercharge-cohomology} at circuit level. The protocol is a
reproducible procedure.

\begin{protocol}[Gross-code static--circuit bridge]
\label{protocol:gross-bridge copy(1)} Fix the syndrome-extraction schedule,
detector and Pauli-frame conventions, elementary fault basis, correction
rule, decoder list rule, $\mathcal{D}_{\mathrm{audit}}$, and every list $%
\mathcal{L}_{d}$. For each distinct $\Delta f\in \Omega _{D,2}^{\mathrm{cand}%
}$, check $D\,\Delta f=0$ and compute $A\,\Delta f$. Construct the final
supports independently of $A$. Use $R_{\mathrm{circ}}(\Delta f)$ only when
every admitted pair representing that same $\Delta f$ gives the same
supports. That common output is what makes the residual well defined. If the
supports disagree, or if either closure condition in Eq.~%
\eqref{eq:rcirc-closure} does not hold, retain the history and its outputs
as an unsatisfied prerequisite. Do not assign $m(\Delta f)$ to that history,
and do not remove it from the parent coverage counts. For a well-defined
closed residual, solve
\begin{equation}
r_{X}(\Delta f)=H_{X}^{\mathsf{T}}u_{X},\qquad r_{Z}(\Delta f)=H_{Z}^{%
\mathsf{T}}u_{Z},
\end{equation}%
over $\mathbb{F}_{2}$, with $u_{X}\in \mathbb{F}_{2}^{m_{X}}$ and $u_{Z}\in
\mathbb{F}_{2}^{m_{Z}}$. Solving these systems shows only whether $L_{%
\mathrm{circ}}(\Delta f)=0$. To obtain the full class, compute the $2k$
commutation bits of the residual Pauli with the fixed conjugate logical
generators. Its logical $X$ coordinates are read by the logical $Z$
generators, and conversely \cite{Gottesman1997}. These bits are unchanged by
stabilizer multiplication, so they give $L_{\mathrm{circ}}(\Delta f)$ in the
frame of Eq.~\eqref{eq:frame-identification}. Compare this complete label
with $A\,\Delta f$, record $m(\Delta f)$, and report the outcomes below.
\end{protocol}

The parent set is fixed before either address is inspected, and $A$ is not
used to construct the residual supports. The shared frame names the
coordinates. It does not impose agreement between them. An unsatisfied
prerequisite means that the circuit-to-static conversion is not yet well
defined, or that the residual is not $\tilde{Q}^{\mathrm{full}}$-closed.
Closure means $H_{Z}r_{X}(\Delta f)=0$ and $H_{X}r_{Z}(\Delta f)=0$. It is
not the stronger statement that the class is exact. A mismatch instead means
that two well-defined addresses disagree. Neither statement by itself
identifies the unknown fault. Stabilizer-circuit simulation can supply the
propagation data \cite{Gidney2021}, while repeated extraction explains why
one detector record can have several histories \cite{Dennis2002}.

\paragraph{The ledger.}

The ledger separates logical ambiguity from a discrepancy in the
circuit-to-static comparison. Provided every prerequisite passes throughout $%
\Omega _{D,2}^{\mathrm{cand}}$, that parent set partitions into
\begin{equation}
\Omega _{D,2}^{\mathrm{T2}}=\left\{ \Delta f\in \Omega _{D,2}^{\mathrm{cand}%
}:A\,\Delta f=0,\ m(\Delta f)=0\right\} ,  \label{eq:ledger-tier2}
\end{equation}%
\begin{equation}
\Omega _{D,2}^{\mathrm{T3}}=\left\{ \Delta f\in \Omega _{D,2}^{\mathrm{cand}%
}:A\,\Delta f\neq 0,\ m(\Delta f)=0\right\} ,  \label{eq:ledger-tier3}
\end{equation}%
\begin{equation}
\Omega _{D,2}^{\mathrm{disc}}=\left\{ \Delta f\in \Omega _{D,2}^{\mathrm{cand%
}}:m(\Delta f)=1\right\} .  \label{eq:ledger-disc}
\end{equation}%
Thus
\begin{equation}
\Omega _{D,2}^{\mathrm{cand}}=\Omega _{D,2}^{\mathrm{T2}}\sqcup \Omega
_{D,2}^{\mathrm{T3}}\sqcup \Omega _{D,2}^{\mathrm{disc}},\qquad \Omega
_{D,2}^{\mathrm{cert}}=\Omega _{D,2}^{\mathrm{T2}}\sqcup \Omega _{D,2}^{%
\mathrm{T3}}.  \label{eq:ledger-partition}
\end{equation}%
If a prerequisite is not satisfied, the history remains outside this
partition and is reported separately. It is not assigned an artificial
mismatch bit.

A Tier-2 difference is a stabilizer boundary. It may have nonzero data
support, but it has no logical action. A Tier-3 difference has one
nontrivial logical action, identified by both constructions. The candidates
are then logically inequivalent, although the comparison does not reveal
which candidate matches the unknown fault. This is the circuit-level form of
the silent logical event classified in Subsec.~\ref%
{subsec:full-logical-address}. A discrepant history instead exposes an
inconsistency in the comparison. It is not a third logical outcome. This
distinction tells Subsec.~\ref{ssec:bb-reranking} which classes may be
merged and which records require repair.

For each weight $w\in\{0,1,2\}$, report $N_w^{\mathrm{cand}}$, the
unsatisfied-prerequisite count, and $N_w^{\mathrm{T2}}$, $N_w^{\mathrm{T3}}$%
, and $N_w^{\mathrm{disc}}$. Count each relative history once. When all
prerequisites pass, the three class counts sum to $N_w^{\mathrm{cand}}$.
Otherwise the unsatisfied-prerequisite count supplies the remaining
histories. Pair multiplicities, if reported, are separate quantities. Retain
one witness for each nonempty Tier-3 or discrepant class,
\begin{equation}
\left( \Delta f,\ \func{wt}(\Delta f),\ D\,\Delta f,\ A\,\Delta f,\ R_{%
\mathrm{circ}}(\Delta f),\ L_{\mathrm{circ}}(\Delta f),\ m(\Delta f) \right).
\label{eq:witness-tuple}
\end{equation}
For an unsatisfied prerequisite, retain the unsatisfied check and the
available pairwise outputs. Undefined entries must not be replaced by zero.

A shell with $N_{w}^{\mathrm{cand}}=0$ provides no tested evidence at that
weight. When all prerequisites pass throughout the parent set, one may also
report
\begin{equation}
t_{\mathrm{audit}}^{\mathrm{eff}}=\max \left\{ t\in \{0,1,2\}:N_{w}^{\mathrm{%
disc}}=0\text{ for every }0\leq w\leq t\right\} .  \label{eq:audit-radius}
\end{equation}%
Leave this radius undefined if a prerequisite is not satisfied, if both
nonzero shells are empty, or if the maximization set is empty. Always report
the shell coverage with it. The radius summarizes address agreement only. A
nonzero Tier-3 count does not lower it, because the two addresses still
agree. It is neither a correction radius nor a bound on the logical error.
Those questions belong to the separate evaluation in Subsec.~\ref%
{ssec:bb-enumeration}. Higher-weight histories, unlisted candidates, and a
changed extraction schedule remain outside this audit \cite{Shaw2025}.

\paragraph{Decoder-facing reading of the ledger.}

The audited bridge carries one instruction for a deployed stack. Heavy
decoding effort is then spent only where logical routing is genuinely
ambiguous, at no cost in physical qubits and with no change of decoding
algorithm~\cite{Bravyi2024,Hillmann2024}.

The same detector record can admit two explanations, and one of them may
quietly rotate the encoded qubit. For two candidates $f_{a}$ and $f_{b}$
already on the live list, write $\Delta f=f_{a}+f_{b}$, $D\Delta f=0$. The
second equality says that the hardware does not distinguish them. The map $A$
tests whether they act in the same way on the encoded state. The bit $%
m(\Delta f)$ tests whether the circuit-level logical-address verdict agrees
with the independently constructed static verdict; it does not compare two
numerical counts. The purpose of the three rules below is to spend extra
decoding effort only when that quiet rotation is possible. Merging a
logically trivial pair is standard degenerate decoding. The new content is
the mismatch flag $m(\Delta f)$ and the discrepant class.

If
\begin{equation}
A\Delta f=0,\qquad m(\Delta f)=0,
\end{equation}%
the pair belongs to $\Omega _{D,2}^{\mathrm{T2}}$. The two explanations then
differ only by a stabilizer, which acts as the identity on the code space,
so they represent the same logical outcome. The decoder may merge them,
retain one representative, update the Pauli frame, and stop. No further
search is needed.

If
\begin{equation}
A\Delta f\neq 0,\qquad m(\Delta f)=0,
\end{equation}%
the pair belongs to $\Omega _{D,2}^{\mathrm{T3}}$. The record still looks
clean, yet the two explanations would send the memory to different logical
states. This is the silent event counted by $p_{\mathrm{fs}}$ in Eq.~%
\eqref{eq:sw-three-way}. The pair must not be merged, the ordinary winner
must not be treated as certified, and the Pauli frame must not be updated
yet. The shot is escalated inside the allowed time and then committed in
Subsec.~\ref{ssec:bb-reranking}; if the time expires, the shot is marked
unsafe rather than guessed. The layer still cannot see which logical state
is true, because a silent residual commutes with every check.

If
\begin{equation}
m(\Delta f)=1,
\end{equation}%
the pair belongs to $\Omega _{D,2}^{\mathrm{disc}}$. The two address
constructions disagree. The base decoder keeps running. Repair the schedule
or frame and repeat the audit. This is a bookkeeping fault, not a third
logical state. Visible errors never enter this list. A pair outside the
audited domain is not merged. How the extra work is done, how often each
class occurs, and whether it helps are specified in Subsecs.~\ref%
{ssec:bb-reranking}, \ref{ssec:bb-enumeration}, and \ref{ssec:bb-metrics}.

Three boundaries close the subsection. The bridge is exact only on $\Omega
_{D,2}^{\mathrm{cert}}$ and conditional on the frozen stack, so a change of
schedule, frame, sector, or decoder convention requires a fresh audit~\cite%
{Shaw2025}. Nothing here replaces the base decoder. Stabilizer-circuit
simulation and symbolic fault-tolerance verification provide complementary
tools for systematically analyzing fault-tolerant QEC circuit
implementations. Descriptions of circuit-level fault correction in QEC
circuits \cite{Gidney2021}, and formal verification of QEC programs and
circuit-level fault tolerance e.g. \cite{Chen:2025jfp},\cite{Fang:2023lcf},
are available; what is new here is the typed two-sector residualization into
$H^{1}(\tilde{Q}^{\mathrm{full}})$, the full-address comparison of Eq.~%
\eqref{eq:bridge-address} against the independently precomputed $A$, and the
falsifiable ledger that separates a Tier-3 event from a genuine interface
discrepancy. The order of logic therefore are as follows. Static residuals
are classified in Subsecs.~\ref{subsec:full-logical-address} and \ref%
{ssec:tier-taxonomy-certificate}; detector fibres are classified by the pair
$(D,A)$ in Subsec.~\ref{ssec:bb-fibre-test}; their compatibility is verified
here through $R_{\mathrm{circ}}$; only then may Subsec.~\ref%
{ssec:bb-reranking} consume the label $A$, Subsec.~\ref{ssec:bb-enumeration}
attach probabilities to the ledger. With the bridge in place, the classifier
$H^{1}\!\left( \tilde{Q}^{\mathrm{full}}\right) $, the projected-return
block $B_{s}$ of Subsec.~\ref{subsec:projected-return-faithfulness}, the
certified window $\mathcal{S}_{\mathrm{faith}}$, and the circuit-level map $%
A $ become four readings of one same-fibre separation rather than four
separate formal devices. Certification precedes use.

\subsection{Posterior reranking pipeline and supercharge features}

\label{ssec:bb-reranking}

Subsections~\ref{ssec:bb-scope}--\ref{subsec:circuit-static-bridge} fix the
extraction stack, detector map $D$, logical-action map $A$, and
circuit--static bridge. The task here is different: choose one explanation
of a detector record without confusing a plausible history with a certified
logical address. The map $A$ classifies relative logical actions exactly. A
score can only rank candidates supplied by a decoder.

For an observed detector word $d$, let the frozen base decoder supply a
bounded list, for example by belief propagation with ordered-statistics
post-processing~\cite{Panteleev2021}, which gives
\begin{equation}
\mathcal{L}_{d}=\{f_{1},\ldots ,f_{M}\}\subseteq \mathcal{F}_{d},\qquad
Df_{a}=d.  \label{eq:bb-candidate-list}
\end{equation}%
The full $2k$-component label $\ell =Af$ partitions the returned list:
\begin{equation}
\mathcal{L}_{d,\ell }=\{f\in \mathcal{L}_{d}:Af=\ell \},\qquad \mathcal{L}%
_{d}=\bigsqcup_{\ell \in \mathbb{F}_{2}^{2k}}\mathcal{L}_{d,\ell }.
\label{eq:bb-class-partition}
\end{equation}%
If $f_{a},f_{b}\in \mathcal{F}_{d}$, then $D(f_{a}+f_{b})=d+d=0$.
Nevertheless, $A(f_{a}+f_{b})$ can be nonzero: two histories may leave the
same detector shadow and move the encoded state to different logical fibres.
Equation~\eqref{eq:bb-fibre-test} makes membership in one $\mathcal{L}%
_{d,\ell }$ an exact test of relative logical equivalence. No score may
merge distinct classes. The partition says nothing about histories omitted
by $\mathcal{L}_{d}$ and cannot identify the unknown true history.

The order of operations matters:
\begin{equation}
d\longmapsto \mathcal{L}_{d}\xrightarrow{\ A\ }\{\mathcal{L}_{d,\ell
}\}_{\ell }\longmapsto \widehat{\ell }(d)\longmapsto \text{committed
representative and Pauli frame}.  \label{eq:bb-pipeline}
\end{equation}%
The arrow labelled by $A$ is exact; the later choices are list-restricted
and model-dependent. Classifying first prevents a high posterior weight from
being misread as evidence of logical equivalence.

Fix $\lambda _{1}$, $\lambda _{2}$, and the tie rule before comparing
decoder arms. With the calibrated features of Sec.~\ref{sec:weighted-lift},
keep the single objective
\begin{equation}
J(f)=-\log P_{\mathrm{base}}(f\mid d)+\lambda _{1}W_{\mathrm{syn}%
}(f)+\lambda _{2}\kappa (f).  \label{eq:bb-objective}
\end{equation}%
The three terms are declared once and frozen. The first, $P_{\mathrm{base}%
}(f\mid d)$, is the weight that the frozen base decoder assigns to the
listed candidate $f$ given the record $d$. Any normalisation that depends
only on $d$ shifts $J$ by a constant and cancels in both selection rules
below. The second, $W_{\mathrm{syn}}(f)$, is the calibrated syndrome cost of
Eq.~\eqref{eq:Wsyn}, evaluated on the end-of-round data support of $f$ under
the residualization convention of Subsec.~\ref{subsec:circuit-static-bridge}%
. The third, $\kappa (f)$, is an optional auxiliary cost term on the
candidate, set to zero unless declared. It must not read the label $Af$, so
that the arm $\mathrm{ord}$ receives no logical information through $J$. The
weights $\lambda _{1},\lambda _{2}\geq 0$ are fixed before any arm is
compared.

For the reranking arm $X=\mathrm{ord}$, with its declared feature set, the
per-class-minimum policy is
\begin{equation}
\widehat{\ell }(d)=\limfunc{arg}_{\ell :\,\mathcal{L}_{d,\ell }\neq
\varnothing }\Bigl(\min_{f\in \mathcal{L}_{d,\ell }}J(f)\Bigr),\qquad
\widehat{f}_{X}(d)\in \mathcal{L}_{d,\widehat{\ell }(d)}.
\label{eq:bb-selection-rule}
\end{equation}%
The committed representative $\widehat{f}_{X}(d)$ minimises $J$ inside the
selected class, with the fixed tie rule deciding equal scores. The
minimisation in Eq. \eqref{eq:bb-selection-rule} is itself standard. The
distinctive content here is its use after the exact subfibre partition in
Eq. \eqref{eq:bb-class-partition}, together with audit-triggered escalation
before a correction is committed. With ties fixed, this commitment coincides
with the global minimum of $J$ over $\mathcal{L}_{d}$, so $\mathrm{ord}$
needs no logical label. The label-aware arms $\mathrm{conv}$ and $\mathrm{%
soft}$ instead select by the list-restricted class mass, $\widehat{\ell }%
(d)=\arg\min_{\ell }\bigl(-\log \sum_{f\in \mathcal{L}_{d,\ell }}\mathrm{e}%
^{-J(f)}\bigr)$, then minimise $J$ inside $\mathcal{L}_{d,\widehat{\ell }%
(d)} $, with the same tie rule and fallback. The minimum favours a class's
best listed history; the mass includes its other listed explanations.
Neither rule estimates the weight of omitted histories without an additional
assumption. The score $J$ ranks plausibility, whereas $Af$ records logical
action. The posterior weight, $W_{\mathrm{syn}}$, and $\kappa $ are not
certificates; a character-sector label can reorganise only already certified
data under Subsec.~\ref{subsec:compress-Bs}.

The auditor compares candidate addresses before the selector commits a
frame. A listed relative history in $\Omega_{D,2}^{\mathrm{T3}}$ of Eq.~%
\eqref{eq:ledger-tier3} exposes rival logical actions. A pair in $%
\Omega_{D,2}^{\mathrm{disc}}$ of Eq.~\eqref{eq:ledger-disc} instead exposes
disagreement between the circuit and static address constructions, not a
third logical outcome. The declared policy may enlarge the list or invoke a
slower solver within its latency allowance. For a discrepant pair,
certification-dependent routing is suspended and the counterexample retained
for repair. If a deadline expires, a declared fallback still supplies the
representative required for rate evaluation; an unsafe flag records that
this choice was not certified. Neither escalation nor a favourable $J$
reveals the true fault. These tier and discrepancy statements apply only to
the predeclared candidate-pair domain $\Omega_{D,2}^{\mathrm{cand}}$, not to
the whole detector fibre. When all competitive candidates share one address,
the layer makes no additional logical-routing decision.

For the resource-matched test of Subsec.~\ref{ssec:bb-metrics}, write
\begin{equation}
X\in \{\mathrm{base},\mathrm{ord},\mathrm{conv},\mathrm{soft}\},\qquad
\widehat{f}_{X}(d)\in \mathcal{L}_{d}.  \label{eq:bb-reranking-stacks}
\end{equation}%
The index $X$ on decoder quantities names an arm, not an $X$-type Pauli.
Only $\mathrm{ord}$ follows Eq.~\eqref{eq:bb-selection-rule}; $\mathrm{conv}$
and $\mathrm{soft}$ share the class-mass rule stated above, and the base arm
retains its own declared policy. Subsection~\ref{ssec:bb-enumeration}
evaluates these committed choices against injected histories. Their relative
logical labels can be read online, but their truth-referenced failure
indicator is an offline observable defined there.

\subsection{Finite-weight circuit enumeration and the logical-error envelope}

\label{ssec:bb-enumeration}

For a circuit fault history $f$ with record $Df$, let $\widehat{f}_{X}(Df)$
be the representative committed by arm $X$. Its policy, including fallback
on an escalated record, must be defined on every history used for an
unconditional error rate. Compare the true and committed histories through
the detector-null decoder residual
\begin{equation}
r_{\mathrm{dec}}^{X}(f)=f+\widehat{f}_{X}(Df),\qquad Dr_{\mathrm{dec}%
}^{X}(f)=0.  \label{eq:bb-enum-residual}
\end{equation}%
The symbol $r_{\mathrm{dec}}^{X}$ is not the $X$-support component $r_{X}$
of $R_{\mathrm{circ}}$ in Subsec.~\ref{subsec:circuit-static-bridge}. Define
the wrong-fibre indicator, with values in $\{0,1\}$, by
\begin{equation}
\Theta _{X}(f)=1\quad \Longleftrightarrow \quad Ar_{\mathrm{dec}}^{X}(f)\neq
0;~~\qquad \Theta _{X}(f)=0\ \text{otherwise}.  \label{eq:bb-enum-failure}
\end{equation}%
The detector record has been matched, yet $\Theta _{X}(f)=1$ says that the
encoded information reaches a different logical fibre. This is the circuit
form of the gap between detectability and correctability in the
Knill--Laflamme criterion~\cite{KnillLaflamme1997,Gottesman1997}. This
truth-referenced verdict is an offline evaluator, not an online certificate.
Only if this particular residual also belongs to $\Omega _{D,2}^{\mathrm{cert%
}}$ and satisfies the bridge prerequisites does its static cohomology
address agree with the circuit label. Certification of other candidate-pair
differences does not prove that condition for every committed residual.

Count wrong-fibre histories at each fault weight:
\begin{equation}
N_{w}^{X} =\bigl|\{f\in\mathbb{F}_{2}^{N_f}: \func{wt}(f)=w,\ \Theta_X(f)=1\}%
\bigr|.  \label{eq:bb-enum-Nw}
\end{equation}
The integers depend on the frozen circuit, frame, fault alphabet, and
decoder policy, not on the code alone. Under a declared model of $N_f$
independent binary fault locations of common rate $p$, each weight-$w$
history has probability $p^w(1-p)^{N_f-w}$~\cite{Dennis2002}. Because the
histories are disjoint, the exact logical-error rate is
\begin{equation}
p_L^X(p) =\sum_{w=0}^{N_f}N_w^X p^w(1-p)^{N_f-w}.
\label{eq:bb-pL-enumeration}
\end{equation}
The Bernoulli sum is standard; its input here is the decoder-specific test $%
\Theta_X$. A different fault law changes the probabilities of histories, not
the logical predicate for the fixed circuit and frame.

Suppose every shell through $w_{\star }$ is exhaustively counted. Its
contribution is
\begin{equation}
p_{L}^{X,(\leq w_{\star })}(p)=\sum_{w=0}^{w_{\star
}}N_{w}^{X}p^{w}(1-p)^{N_{f}-w}.  \label{eq:bb-pL-truncated}
\end{equation}%
Each omitted term is nonnegative, and $N_{w}^{X}\leq \binom{N_{f}}{w}$. The
two facts give the conservative envelope
\begin{equation}
p_{L}^{X,(\leq w_{\star })}(p)\leq p_{L}^{X}(p)\leq p_{L}^{X,(\leq w_{\star
})}(p)+\Pr \!\left[ \func{Bin}(N_{f},p)>w_{\star }\right] ,\qquad 0\leq
p\leq 1.  \label{eq:bb-pL-two-sided-bracket}
\end{equation}%
Here $\func{Bin}(N_{f},p)$ denotes the binomial distribution for $N_{f}$
independent trials with error probability $p$. The upper side treats every
uncounted history as fatal. This bound records the coverage of a finite
enumeration; it does not establish a threshold.

A stronger small-$p$ statement requires a separate all-histories premise.
Define a circuit-correction radius $t_{\mathrm{eff}}^{X}$ only after
verifying
\begin{equation}
\func{wt}(f)\leq t_{\mathrm{eff}}^{X}\quad \Longrightarrow \quad \Theta
_{X}(f)=0\quad \text{for every such }f\in \mathbb{F}_{2}^{N_{f}}.
\label{eq:bb-enum-teff-def}
\end{equation}%
It follows that $N_{w}^{X}=0$ for $w\leq t_{\mathrm{eff}}^{X}$ and
\begin{equation}
p_{L}^{X}(p)\leq \Pr \!\left[ \func{Bin}(N_{f},p)\geq t_{\mathrm{eff}}^{X}+1%
\right] =O\!\left( p^{t_{\mathrm{eff}}^{X}+1}\right) \quad (p\longrightarrow
0).  \label{eq:bb-enum-teff-bound}
\end{equation}%
The bridge radius $t_{\mathrm{audit}}^{\mathrm{eff}}$ of Eq.~%
\eqref{eq:audit-radius} measures agreement of two address constructions on
an audited candidate-pair domain; $t_{\mathrm{eff}}^{X}$ measures successful
recovery by a declared decoder arm. They may coincide, but the former alone
does not imply the latter. Neither radius is set by the block-code distance $%
d=12$ without a circuit audit \cite{Bravyi2024,Shaw2025}.

The Gross translation symmetry can reduce offline enumeration only if the
frozen circuit passes a schedule-dependent test~\cite{Shaw2025}. A
sufficient condition for the convention used here is
\begin{equation}
D(gf)=g\cdot Df,\qquad A(gf)=Af,\qquad g\in G_{\mathrm{gross}},
\label{eq:bb-enum-orbit-invariance}
\end{equation}%
where $g\cdot $ denotes the induced permutation of detector coordinates;
strict invariance, $D(gf)=Df$, is the special case of a trivial action.
Orbit-weighted counts additionally require $\func{wt}(gf)=\func{wt}(f)$ and $%
\Theta _{X}(gf)=\Theta _{X}(f)$ for the decoder arm being evaluated. The
latter must be checked for the committed policy; detector and logical-map
invariance alone do not license orbit counting. Character bookkeeping then
follows Subsec.~\ref{subsec:compress-Bs}, without enlarging $\Omega _{D,2}^{%
\mathrm{cert}}$ or altering the exact test~\eqref{eq:bb-enum-failure}. The
resulting $N_{w}^{\mathrm{base}}$, $N_{w}^{\mathrm{ord}}$, $N_{w}^{\mathrm{%
conv}}$, and $N_{w}^{\mathrm{soft}}$ are passed to the matched ablation of
Subsec.~\ref{ssec:bb-metrics}.

\subsection{Matched ablation, mechanism attribution, and the reporting ledger%
}

\label{ssec:bb-metrics}

\paragraph{Matched configurations.}

The counts of Subsection~\ref{ssec:bb-enumeration} tell us how often a
decoder commits the wrong logical fibre. They do not tell us why a second
decoder behaves differently. The present test separates the value of
logical-class information from the value of independently auditing it, then
prices any extra search. We compare four configurations of the same frozen
decoding stack:
\begin{equation}
X\in\{\mathrm{base},\mathrm{ord},\mathrm{conv},\mathrm{soft}\}.
\label{eq:bb-metrics-arms}
\end{equation}
The base configuration retains its declared correction rule. The ordinary
configuration adds generic posterior reranking but receives no logical
labels; such post-processing is already used in quantum low-density
parity-check decoding~\cite{Panteleev2021,Tsubouchi:2025hgy}. The
conventional control receives the full label $Af$ of each listed candidate,
computed by ordinary symplectic stabilizer bookkeeping~\cite{Gottesman1997}.
The informed configuration receives the same labels and selector, together
with the independent circuit--static audit of Subsection~\ref%
{subsec:circuit-static-bridge}. A label specifies the logical action of a
candidate. It does not tell the running decoder which candidate was the true
fault history.

All configurations share one extraction circuit, noise ensemble, Pauli
frame, fault histories, candidate generator, lists, score $J$, random seeds,
tie rule, stopping rule, fallback, and hardware. The reranking
configurations have equal list caps and the same allowed deadline:
\begin{equation}
M_{\mathrm{ord}}=M_{\mathrm{conv}}=M_{\mathrm{soft}}, \qquad T_{\mathrm{%
budget}}^{\mathrm{ord}} =T_{\mathrm{budget}}^{\mathrm{conv}} =T_{\mathrm{%
budget}}^{\mathrm{soft}} =T_{\mathrm{budget}}.
\label{eq:bb-metrics-matched-control}
\end{equation}
Here $T_{\mathrm{budget}}$ is the allowance used in Eq.~%
\eqref{eq:cal-latency-observables}. These are equal allowances, not equal
measured runtimes or a claim that no shot misses the deadline. The common
offline failure test is $\Theta_X(f)$ of Eq.~\eqref{eq:bb-enum-failure}.
Only the candidate labels, never the unobserved true $Af$, can enter an
online selection rule.

A necessary null check takes one sentence. Minimizing the same score $J$
first over each class of Eq.~\eqref{eq:bb-class-partition} and then over
classes returns the same global minimizer as ordinary reranking, provided
ties agree. Grouping alone cannot change the commitment. For the
escalation-off experiment, both label-aware configurations must use the same
predeclared, decision-changing variant of Subsection~\ref{ssec:bb-reranking}%
. The variant is an input to the test, not an original decoding theorem.
Degeneracy-aware OSD is a related, distinct finite-list post-processing
method~\cite{Kung:2024adosd}. Optional deep searches stay off during this
comparison, so extra time is not credited to logical information.

\paragraph{Logical attribution and audited domain.}

Write $\mathrm{soft,off}$ for the informed configuration with optional
deeper search disabled. Its logical-failure rate $p_L^{\mathrm{soft,off}}(p)$
and fault-shell counts $N_w^{\mathrm{soft,off}}$ are evaluated by the same
predicate and enumeration rule as the $\mathrm{soft}$ configuration of
Subsection~\ref{ssec:bb-enumeration}. The argument $p$ is the physical fault
rate, and $w$ counts elementary fault locations. Define
\begin{equation}
\Delta p_L^{\mathrm{full}}(p) =p_L^{\mathrm{base}}(p)-p_L^{\mathrm{soft,off}%
}(p), \qquad \Delta p_L^{\mathrm{attr}}(p) =p_L^{\mathrm{ord}}(p)-p_L^{%
\mathrm{soft,off}}(p).  \label{eq:bb-metrics-gains}
\end{equation}
The corresponding shell difference is
\begin{equation}
\Delta N_w^{\mathrm{attr}} =N_w^{\mathrm{ord}}-N_w^{\mathrm{soft,off}}.
\label{eq:bb-metrics-shell-gain}
\end{equation}
Adding and subtracting the ordinary failure rate yields
\begin{equation}
\Delta p_L^{\mathrm{full}}(p) =\bigl[p_L^{\mathrm{base}}(p)-p_L^{\mathrm{ord}%
}(p)\bigr] +\Delta p_L^{\mathrm{attr}}(p).
\label{eq:bb-metrics-decomposition}
\end{equation}
The bracket measures ordinary reranking. The remaining term measures the
added class-sensitive decision, not a uniquely supercharge-based way to
calculate its labels. When every shell through $w_{\star}$ has been
exhaustively counted under the independent binary-fault model,
\begin{equation}
\Delta p_L^{\mathrm{attr},(\leq w_{\star})}(p) =\sum_{w=0}^{w_{\star}}
\Delta N_w^{\mathrm{attr}}p^w(1-p)^{N_f-w}.
\label{eq:bb-metrics-Delta-pL-trunc}
\end{equation}
Here $N_f$ is the number of fault locations. The differences $\Delta N_w^{%
\mathrm{attr}}$ can have either sign. Thus this partial sum is not generally
a bound on the full gain. Its first nonzero term marks a net policy
difference, not the onset of all logical failures.

On identical injected fault histories the estimate remains paired:
\begin{equation}
\widehat{\Delta p}_L^{\mathrm{attr}}(p) =\frac{1}{N_{\mathrm{shots}}}
\sum_{j=1}^{N_{\mathrm{shots}}} \left[ \Theta_{\mathrm{ord}}(f_j) -\Theta_{%
\mathrm{soft,off}}(f_j) \right].  \label{eq:bb-metrics-paired-estimator}
\end{equation}
The extension of the index $X$ to $\mathrm{soft,off}$ only names a specified
correction policy; $\Theta_X$ is unchanged. Report the failure count in each
configuration, both directions of discordant pairs, and a paired interval
with its confidence level, independent sampling unit, and stopping rule
fixed before inspecting the results.

For one detector word $d$, let $\delta f=\widehat{f}_{\mathrm{ord}}(d)+%
\widehat{f}_{\mathrm{soft,off}}(d)$ be the difference of the committed
histories over $\mathbb{F}_{2}$. A truth-labelled benefit witness obeys
\begin{equation}
D\delta f=0,\qquad A\delta f\neq 0,\qquad \lbrack R_{\mathrm{circ}}(\delta
f)]\neq 0\quad \text{in }H^{1}(\tilde{Q}^{\mathrm{full}}).
\label{eq:bb-metrics-wrong-fibre-witness}
\end{equation}%
The last clause applies only when $\delta f\in \Omega _{D,2}^{\mathrm{cert}}$
of Eq.~\eqref{eq:certified-subset} and the bridge's residualization,
closure, and frame tests pass. The same injected history must give $\Theta _{%
\mathrm{ord}}(f)=1$ and $\Theta _{\mathrm{soft,off}}(f)=0$. The detector
sees one record although the two choices act differently on the encoded
state. The failure bits, not the relative label, identify the correct
choice. Outside the audited domain the $D,A$ part remains exact, but no
static cohomology agreement is asserted. A conventional--informed benefit
instead retains the \emph{specific listed candidate pair} that triggered a
changed safeguard; it need not be the difference of the final two committed
corrections.

The predeclared paired confidence interval tests the attribution gain. A
lower endpoint above zero supports a positive gain; an interval crossing
zero is inconclusive. An upper endpoint at or below zero rules out a
strictly positive gain at the stated confidence level for this stack. None
of these outcomes changes the exact same-fibre criterion of Eq.~%
\eqref{eq:bb-fibre-test}.

To distinguish a logical-label benefit from an implementation benefit, keep
optional deeper search off and write
\begin{equation}
\Delta p_L^{\mathrm{info}}(p) =p_L^{\mathrm{ord}}(p)-p_L^{\mathrm{conv,off}%
}(p), \qquad \Delta p_L^{\mathrm{impl}}(p) =p_L^{\mathrm{conv,off}}(p)-p_L^{%
\mathrm{soft,off}}(p).  \label{eq:bb-metrics-info-impl}
\end{equation}
The first comparison asks what is gained by seeing a logical label. The
second asks whether the independently audited realization changes an outcome
relative to conventional bookkeeping with the same label. By simple
subtraction, $\Delta p_L^{\mathrm{attr}} =\Delta p_L^{\mathrm{info}}+\Delta
p_L^{\mathrm{impl}}$. This equality is accounting, not a new decoding
theorem. The signed implementation gain is estimated on the same injected
histories,
\begin{equation}
\widehat{\Delta p}_L^{\mathrm{impl}}(p) =\frac{1}{N_{\mathrm{shots}}}
\sum_{j=1}^{N_{\mathrm{shots}}} \bigl[\Theta_{\mathrm{conv,off}}(f_j)
-\Theta_{\mathrm{soft,off}}(f_j)\bigr].
\label{eq:bb-metrics-impl-paired-estimator}
\end{equation}
Report both directions of discordance and a predeclared paired confidence
interval. Only a positive lower endpoint supports reduced logical failure in
this matched implementation comparison. If the two label-aware
configurations receive the same labels, list, selector, ties, safeguards,
and fallback, they commit the same correction on every record. The
independent audit does not make a shared exact label more accurate by itself.

An audit fraction requires an explicit population. Before sampling, fix
which pairs from each returned list are submitted to the bridge. For a
sampled record $d$, put $e(d)=1$ when that pair set is nonempty and its
residualization, closure, and frame prerequisites all pass; otherwise put $%
e(d)=0$. On eligible records put $q(d)=1$ when at least one submitted pair
has the mismatch bit $m=1$ of Eq.~\eqref{eq:bridge-address}, and put $q(d)=0$
when none does. Set $q(d)=0$ on ineligible records. Then, on the same
noise-weighted record ensemble used for $p_L^X(p)$, define
\begin{equation}
\eta_{\mathrm{cert}} =\Pr[e(d)=1,\ q(d)=0], \qquad r_{\mathrm{disc}} =\Pr[%
e(d)=1,\ q(d)=1].  \label{eq:bb-metrics-audit-rates}
\end{equation}
The quantity $\eta_{\mathrm{cert}}$ is the fraction of records for which all
\emph{submitted} pairwise addresses agree, not the fraction of correctly
recovered records. The quantity $r_{\mathrm{disc}}$ is the fraction
containing an audited interface disagreement, not a logical failure rate.
They sum to $\Pr[e(d)=1]$, which need not be one. Report eligible,
ineligible, and failed-prerequisite counts separately. No absence of
discrepancies on an empty audited population is evidence of correctness;
candidates or faults omitted from the domain remain uncertified.

There is a conditional limit on the implementation comparison. Suppose $%
\mathrm{conv,off}$ and $\mathrm{soft,off}$ commit identical corrections
whenever $q(d)=0$, including ineligible records, and use the same record
distribution in $r_{\mathrm{disc}}$ and $\Delta p_{L}^{\mathrm{impl}}$.
Their paired failure bits may then differ only when $q(d)=1$. Thus, for a
sampled history $f$ producing record $d$, the exact expectation identity
under the stated matching condition is
\begin{equation}
\Delta p_{L}^{\mathrm{impl}}(p)=\mathbb{E}\!\left[ \bigl(\Theta _{\mathrm{%
conv,off}}(f)-\Theta _{\mathrm{soft,off}}(f)\bigr)q(d)\right] .
\end{equation}%
As each failure bit lies in $\{0,1\}$,
\begin{equation}
|\Delta p_{L}^{\mathrm{impl}}(p)|\leq \Pr [q(d)=1]=r_{\mathrm{disc}}.
\label{eq:bb-metrics-impl-bound}
\end{equation}%
This bound follows by restricting the expectation of the signed bit
difference to the discrepancy records. It does not predict a positive gain.
If only the informed configuration changes fallback on a bridge mismatch, a
measured difference tests that \emph{fallback policy}, not a more accurate
supercharge label. A claim about implementation advantage requires
comparison against a conventional audit with matched safeguards, accepted
accuracy, and coverage, including online time and storage.

\paragraph{Conditional triage cost.}

Turning on optional deeper decoding is a different experiment. Its logical
yield is
\begin{equation}
\Delta p_{L}^{\mathrm{triage}}(p)=p_{L}^{\mathrm{soft,off}}(p)-p_{L}^{%
\mathrm{soft,on}}(p).  \label{eq:bb-metrics-triage-gain}
\end{equation}%
It measures additional computation, not a new logical label, and is not part
of $\Delta p_{L}^{\mathrm{impl}}$. Let $I_{\mathrm{trig}}$ be the trigger
indicator of the predeclared \emph{union} of the events that actually invoke
a deep procedure: ambiguity, and any implemented nonconvergence or
bridge-discrepancy trigger. An overlap is charged once. For a conditional
critical path, let $T_{\mathrm{fast}}$ include production of the initial
candidate list, $T_{\mathrm{cert}}$ include the online audit and trigger
decision, and $T_{\mathrm{esc}}$ be the measured \emph{additional} time
after escalation. Since the last term vanishes when $I_{\mathrm{trig}}=0$,
averaging $T=T_{\mathrm{fast}}+T_{\mathrm{cert}}+I_{\mathrm{trig}}T_{\mathrm{%
esc}}$ gives
\begin{equation}
\mathbb{E}[T]=\mathbb{E}[T_{\mathrm{fast}}+T_{\mathrm{cert}}]+\Pr [I_{%
\mathrm{trig}}=1]\mathbb{E}[T_{\mathrm{esc}}\mid I_{\mathrm{trig}}=1].
\label{eq:bb-metrics-conditional-cost}
\end{equation}%
No independence between record difficulty and escalation time is needed.
Define the final term as zero when the trigger never fires. Count dispatch,
transfer, and queueing on the measured path. Deduct fast-stage work from a
deep-path total only if it is reused; a restart is not a continuation. An
always-on parallel engine is not described by this gated law.

Compare with the same declared deep procedure applied to every record at its
measured logical accuracy. Report the trigger rate and a confidence gate
tuned to that rate, along with $q_{0.99}$, $P_{\mathrm{miss}}$, and
throughput under the $T_{\mathrm{budget}}$ of Eq.~%
\eqref{eq:cal-latency-observables}. Mean time alone does not settle a
deadline claim~\cite{Maurer:2025tbw}. The $R_{\mathrm{useful}}$ of Eq.~%
\eqref{eq:cal-useful-escalation} is conditioned on the distinct weight-sweep
flag $\Phi $ of Definition~\ref{def:sweep-trigger}; an $I_{\mathrm{trig}}$%
-conditioned yield must be reported under its own name. The scheduling
probability $P_{\mathrm{miss}}$ and the unflagged silent-failure mass $p_{%
\mathrm{miss}}$ of Eq.~\eqref{eq:missbridge} count different events.

A positive $\Delta p_{L}^{\mathrm{info}}$ with paired evidence supports
logical-information usefulness on the frozen stack. Evidence of
supercharge-specific operational usefulness instead requires an
output-matched conventional audit with a measured advantage in accepted
accuracy, coverage, or resources.

\section{Weighted Software Lift $\tilde{Q}^{\mathrm{full}}(t_{X},t_{Z}) $:
Two--Sector Invariance, Bias Calibration, and Decoder Diagnostics}

\renewcommand{\theequation}{6.\arabic{equation}} \setcounter{equation}{0} %
\renewcommand{\thethm}{6.\arabic{thm}} \setcounter{thm}{0} %
\renewcommand{\theprop}{6.\arabic{prop}} \setcounter{prop}{0} %
\renewcommand{\thelemm}{6.\arabic{lemm}} \setcounter{lemm}{0}

\label{sec:weighted-lift}

Everything this section consumes has been fixed already and is recalled here
once: the completed finite--field classifier $\tilde{Q}^{\mathrm{full}}=%
\tilde{Q}^{\overline{X}}\oplus \tilde{Q}^{\overline{Z}}$ of Eq.~%
\eqref{eq:fla-Qfull}, its middle cohomology $H_{\mathrm{full}}^{1}:=H^{1}(%
\tilde{Q}^{\mathrm{full}}) $ of binary dimension $2k$, the crossed
check--family response, and the circuit--level logical--action map $A$ of
Subsection~\ref{ssec:bb-fibre-test}, all fixed in Subsection~\ref%
{subsec:full-logical-address}.

Hardware noise, may have biases. Circuit--level faults can be strongly
biased between the two Pauli species, and deployed quantum low--density
parity--check decoders already exploit continuous channel information when
they order the candidate explanations of a single detector record \cite%
{Hillmann2024,Bravyi2024}. The purpose of admitting that information is
operational rather than foundational, and the good effect is concrete: a
bias--aware ordering concentrates expensive decoding effort on the few
detector records whose logical routing is genuinely fragile, which is the
allocation discipline that real--time latency budgets reward \cite%
{Maurer:2025tbw}. The architectural order of the paper is preserved without
exception: the exact two--sector verdict first, calibrated weighting second,
escalation third. Subsection~\ref{subsec:invariance} constructs the weighted
lift and proves that the completed logical address is invariant under it,
and Subsection~\ref{subsec:calibration} calibrates the weights, exposes the
single soft feature a deployed stack consumes, states the one new
diagnostic, and declares the deployment observables that Subsection~\ref%
{ssec:bb-metrics} reports.

\subsection{Weighted software lift and logical-fibre invariance}

\label{subsec:invariance}

\paragraph{Purpose and scope.}

Hardware noise is not binary. Circuit-level faults in a superconducting
quantum memory are frequently biased between the two Pauli species, and a
decoding stack that is forbidden to price $X$-type and $Z$-type residual
support differently is throwing away channel information that practical
quantum low-density parity-check decoders already exploit when they order
candidate explanations of one detector record \cite{Hillmann2024,Bravyi2024}%
. The purpose of this subsection is to admit exactly that information, and
to prove in advance that admitting it is safe. What must be guaranteed
before any such pricing is trusted is that a continuous weight can neither
create nor destroy a logical class, because the logical action rather than
the physical error count is the quantity a quantum-memory analysis must
ultimately control \cite{Terhal2015,KnillLaflamme1997}. Proposition~\ref%
{prop:weight-invariance} below is that guarantee, and it is the only theorem
this subsection needs.

\paragraph{Check-family assignment of the two amplitudes.}

Two positive amplitudes $t_{X}>0$ and $t_{Z}>0$ are introduced, and their
assignment is fixed before any matrix is displayed. Each amplitude follows
its own check family and never the direction of the arrow on which that
family happens to sit,
\begin{equation}
t_{X}~\;\mathrm{on}~~H_{X},\;H_{X}^{\mathsf{T}},~\qquad t_{Z}~\;\mathrm{on}%
~~H_{Z},\;H_{Z}^{\mathsf{T}}.  \label{eq:weighted-placement}
\end{equation}%
The physical reason is the crossed response fixed in Subsec.~\ref%
{subsec:full-logical-address}: an $X$-type check is violated by $Z$-type
support and never by $X$-type support, so the penalty that prices a given
Pauli species must remain attached to the check family able to see that
species, in whichever degree the corresponding matrix appears \cite%
{Gottesman1997}. The named order $(t_{X},t_{Z})$ is used throughout and
matches the rate pair $(p_{X},p_{Z})$ consumed in Subsec.~\ref%
{subsec:calibration}. Under Eq.~\eqref{eq:weighted-placement} the two sector
differentials read
\begin{equation}
\tilde{Q}^{\overline{Z}}(t_{X},t_{Z})=%
\begin{pmatrix}
0 & 0 & 0 \\
t_{Z}H_{Z}^{\mathsf{T}} & 0 & 0 \\
0 & t_{X}H_{X} & 0%
\end{pmatrix}%
,  \label{eq:weighted-sector-Zbar}
\end{equation}%
\begin{equation}
\tilde{Q}^{\overline{X}}(t_{X},t_{Z})=%
\begin{pmatrix}
0 & 0 & 0 \\
t_{X}H_{X}^{\mathsf{T}} & 0 & 0 \\
0 & t_{Z}H_{Z} & 0%
\end{pmatrix}%
,  \label{eq:weighted-sector-Xbar}
\end{equation}%
and the completed weighted classifier is their direct sum,
\begin{equation}
\tilde{Q}^{\mathrm{full}}(t_{X},t_{Z})=\tilde{Q}^{\overline{X}%
}(t_{X},t_{Z})\;\oplus \;\tilde{Q}^{\overline{Z}}(t_{X},t_{Z}),\qquad \tilde{%
Q}^{\mathrm{full}}(1,1)=\tilde{Q}^{\mathrm{full}}.  \label{eq:weighted-full}
\end{equation}%
The undeformed point is thus exactly the certified object of Eq.~%
\eqref{eq:fla-Qfull}, which is what makes this a reweighting rather than a
redefinition. The square of each block reduces to one surviving composite,
\begin{equation}
\left. \left( \tilde{Q}^{\overline{Z}}(t_{X},t_{Z})\right) ^{2}\right\vert _{%
\text{degree two}}=t_{X}t_{Z}\,H_{X}H_{Z}^{\mathsf{T}}=0,\qquad \left.
\left( \tilde{Q}^{\overline{X}}(t_{X},t_{Z})\right) ^{2}\right\vert _{\text{%
degree two}}=t_{X}t_{Z}\,H_{Z}H_{X}^{\mathsf{T}}=0,
\label{eq:weighted-nilpotency}
\end{equation}%
every other composition vanishing by strict triangularity, so that $\bigl(%
\tilde{Q}^{\mathrm{full}}(t_{X},t_{Z})\bigr)^{2}=0$. The common prefactor $%
t_{X}t_{Z}$ already signals that the assignment respects the exchange $%
H_{X}\leftrightarrow H_{Z}$ together with $t_{X}\leftrightarrow t_{Z}$.

\begin{lemm}[Check-family placement of the amplitudes]
\label{lem:weight-placement} Assigning one amplitude to every incoming arrow
and the other to every outgoing arrow, in place of Eq.~%
\eqref{eq:weighted-placement}, agrees with Eq.~\eqref{eq:weighted-placement}
in the barred-$Z$ block and disagrees with it in the conjugate barred-$X$
block whenever $t_{X}\neq t_{Z}$. Only the check-family assignment induces
one common quadratic form on the middle degree of both sectors.
\end{lemm}

\noindent {\textit{Proof}}. In the barred-$Z$ block of Eq.~%
\eqref{eq:weighted-sector-Zbar} the outgoing map contributes $%
t_{X}^{2}H_{X}^{\mathsf{T}}H_{X}$ and the incoming map contributes $%
t_{Z}^{2}H_{Z}^{\mathsf{T}}H_{Z}$, which is Eq.~%
\eqref{eq:weighted-middle-operator} below. In the conjugate block of Eq.~%
\eqref{eq:weighted-sector-Xbar} the two check families exchange arrow role
but not species, so the same sum recurs. An arrow-position assignment
delivers instead $t_{Z}^{2}H_{X}^{\mathsf{T}}H_{X}+t_{X}^{2}H_{Z}^{\mathsf{T}%
}H_{Z}$ in the conjugate block, which differs whenever $t_{X}\neq t_{Z}$ and
contradicts the sector-exchange symmetry. {\hfill $\square $}\newline

The useful effect of Lemma~\ref{lem:weight-placement} is that it removes a
label-reversal that is easy to write, dimensionally consistent, and
invisible to every binary test: it forbids pricing abundant $Z$-type support
by the check family that cannot detect it.

\paragraph{Invariance of the logical-fibre address.}

\begin{prop}[Weight-independence of the full logical-fibre address]
\label{prop:weight-invariance} Assume CSS orthogonality condition for the
chosen representatives. Then for all $t_{X}>0$ and $t_{Z}>0$,
\begin{equation}
H^{1}\!\left( \tilde{Q}^{\mathrm{full}}(t_{X},t_{Z})\right) \;\cong \;H_{%
\overline{X}}^{1}\oplus H_{\overline{Z}}^{1}\;=\;H_{\mathrm{full}}^{1}.
\label{eq:weighted-invariance}
\end{equation}%
In particular the residual pair and acceptance rule of Eq.~%
\eqref{eq:fla-residual-pair}, the quotient presentation of Eq.~%
\eqref{eq:manifest-quotient}, the target event of Subsec.~\ref%
{subsec:fs-target}, and the circuit-level same-fibre criterion of Eq.~%
\eqref{eq:bb-fibre-test} are unchanged by the weighting.
\end{prop}

\noindent {\textit{Proof}}. A nonzero scalar is invertible, so $%
\ker(tL)=\ker L$ and $\func{im}(tL)=\func{im}L$ for every linear map $L$.
Applying this to both arrows of each sector leaves the middle kernels and
images of Eqs.~\eqref{eq:weighted-sector-Zbar} and %
\eqref{eq:weighted-sector-Xbar} untouched, hence also their quotients;
block-diagonal additivity of middle cohomology then gives Eq.~%
\eqref{eq:weighted-invariance}. {\hfill $\square $}\newline

\paragraph{Shared middle-space scoring geometry and scope}

The amplitudes reach a decoder not through the differential but through a
quadratic object on the middle degree, and it is this object that a deployed
engine evaluates. With respect to a declared inner product there, the object
is not postulated but computed. Declare on the auxiliary $\mathbb{K}$-valued
chain groups the Euclidean or Hermitian inner product, with respect to which
the dagger denotes the adjoint of the lifted maps alone, and restrict the
anticommutator of a weighted sector differential with that adjoint to the
middle degree. Every cross term vanishes by the strict triangularity of Eqs.~%
\eqref{eq:weighted-sector-Zbar} and \eqref{eq:weighted-sector-Xbar}, so in
the barred-$Z$ sector only the outgoing composite $t_{X}^{2}H_{X}^{\mathsf{T}%
}H_{X}$ and the incoming composite $t_{Z}^{2}H_{Z}^{\mathsf{T}}H_{Z}$
survive,
\begin{equation}
\left. \left\{ \tilde{Q}^{\overline{Z}}(t_{X},t_{Z}),\left( \tilde{Q}^{%
\overline{Z}}(t_{X},t_{Z})\right) ^{\dagger }\right\} \right\vert _{\text{%
degree one}}=t_{Z}^{2}\,H_{Z}^{\mathsf{T}}H_{Z}+t_{X}^{2}\,H_{X}^{\mathsf{T}%
}H_{X},  \label{eq:weighted-middle-anticommutator}
\end{equation}%
equivalently, for a middle-degree residual coordinate $u$, the induced
quadratic form is the total squared check response of that residual,
\begin{equation}
\left\langle u,\left( t_{Z}^{2}H_{Z}^{\mathsf{T}}H_{Z}+t_{X}^{2}H_{X}^{%
\mathsf{T}}H_{X}\right) u\right\rangle =\left\Vert t_{Z}H_{Z}u\right\Vert
^{2}+\left\Vert t_{X}H_{X}u\right\Vert ^{2}.
\end{equation}%
The barred-$X$ sector returns the identical expression, since its two arrows
exchange role but not check family. We therefore define
\begin{equation}
\tilde{H}(t_{X},t_{Z})=t_{Z}^{2}\,H_{Z}^{\mathsf{T}}H_{Z}+t_{X}^{2}\,H_{X}^{%
\mathsf{T}}H_{X},  \label{eq:weighted-middle-operator}
\end{equation}%
and on the completed classifier the induced object is the direct sum $\tilde{%
H}(t_{X},t_{Z})\oplus \tilde{H}(t_{X},t_{Z})$, which doubles the
multiplicity of each eigenvalue and not the eigenvalues themselves. By Lemma~%
\ref{lem:weight-placement} both conjugate sectors induce this same form,
which is precisely the consistency a two-sector address demands. Two
conventions follow. Since the amplitudes enter Eq.~%
\eqref{eq:weighted-middle-operator} quadratically while multiplying the
differential linearly, a preference for linear penalties $w_{X}$ and $w_{Z}$
per violated check is accommodated by $t_{X}=\sqrt{w_{X}}$ and $t_{Z}=\sqrt{%
w_{Z}}$. Accordingly this subsection fixes the weighted lift, the admissible
placement of the amplitudes, the invariance of the completed logical
address, and the single operator through which weighting acts, and it
selects no value of $t_{X}/t_{Z}$. Fixing that ratio from a declared
circuit-level noise bias, the resulting candidate score, the treatment of
calibration uncertainty, and the escalation rule are the business of Subsec.~%
\ref{subsec:calibration}. Downstream, the certified compression of Subsec.~%
\ref{subsec:compress-Bs} may reorganise already certified blocks and the
audited circuit analysis of Sec.~\ref{sec:bb-circuit-reranking} may consume
the resulting feature, but neither may enlarge $\mathcal{S}_{\mathrm{faith}}$%
, and the protected-fibre count of Sec.~\ref{sec:witten-static-decoder} is
independent of every choice of amplitudes.

\subsection{Noise calibration, soft score, and ambiguity escalation}

\label{subsec:calibration}

\paragraph{Purpose and inherited input.}

The weighted lift carries two free positive numbers, and a construction with
free parameters is of no use to a running decoder until they are tied to
something measurable. The purpose of this subsection is to tie them to a
declared circuit--level bias, to expose exactly one real--valued feature
built from them, and to turn the residual uncertainty in that tie into an
auditable alarm rather than a hidden tuning knob. The invariance of
Proposition~\ref{prop:weight-invariance}, the check--family placement of
Lemma~\ref{lem:weight-placement}, the middle--space operator $\tilde{H}%
(t_{X},t_{Z})$ of Eq.~\eqref{eq:weighted-middle-operator}, the candidate
list and its exact partition in Eq.~\eqref{eq:bb-class-partition}, the
penalised objective of Eq.~\eqref{eq:bb-objective}, and the
audit--controlled escalation and reporting policy of Subsections~\ref%
{ssec:bb-reranking} and \ref{ssec:bb-metrics} are analysed in the previous
subsections.

\paragraph{Declared rates and the crossed response.}

Fix two declared rates $p_{X},p_{Z}\in(0,\tfrac{1}{2})$ for the $X$--type
and $Z$--type components of the residual channel on the frozen extraction
stack; the restriction to rates below one half is what makes every log--odds
below strictly positive, so that asserting a fault always costs rather than
pays. They may be read as data--qubit rates per extraction round or as
effective residual rates after the declared circuit reduction, but the two
readings must not be mixed, since preparation, measurement, idle and
ancilla--register faults propagate through the schedule and make the
effective rates a property of the stack rather than of the code \cite%
{Bravyi2024}. The response orientation is the crossing already fixed in
Subsection~\ref{subsec:full-logical-address}: the $H_{X}$ sector responds to
$Z$--type support and the $H_{Z}$ sector to $X$--type support. The placement
of the weights therefore follows the responding check family, with $t_{X}$
on $H_{X}$ and $H_{X}^{\mathsf{T}}$ and $t_{Z}$ on $H_{Z}$ and $H_{Z}^{%
\mathsf{T}}$, as proved once in Lemma~\ref{lem:weight-placement} and
recorded in Eq.~\eqref{eq:weighted-placement}. This is the single point at
which a plausible--looking formula can be silently wrong, because a rule
that tracks the matrix letter and forgets the crossing prices each Pauli
species by the check family that cannot see it.

\paragraph{Derivation of the calibration.}

We derive the weights rather than announce them, since a reader must be able
to check the ratio line by line. Under the declared independent biased Pauli
model the prior of a residual pair $(x,z)$ factorises over the $n$ data
qubits,
\begin{equation}
\Pr
(x,z)\;=\;\prod_{i=1}^{n}\,p_{X}^{\,x_{i}}\,(1-p_{X})^{\,1-x_{i}}\;p_{Z}^{%
\,z_{i}}\,(1-p_{Z})^{\,1-z_{i}}.  \label{eq:cal-prior}
\end{equation}%
For one Bernoulli bit $e\in \{0,1\}$ of rate $p$ the negative logarithm of
its probability splits into a candidate--dependent and a
candidate--independent piece,
\begin{equation}
-\log \bigl[\,p^{\,e}\,(1-p)^{\,1-e}\,\bigr]\;=\;e\,\log \frac{1-p}{p}%
\;+\;\log \frac{1}{1-p},  \label{eq:cal-bernoulli}
\end{equation}%
and the second piece is the same for every candidate, so it cancels in every
comparison made inside one detector fibre. Each asserted unit of support
therefore costs exactly the log--odds of its own species, and summing Eq.~%
\eqref{eq:cal-bernoulli} over qubits and over both species gives
\begin{equation}
-\log \Pr (x,z)\;=\;|x|\,\log \frac{1-p_{X}}{p_{X}}\;+\;|z|\,\log \frac{%
1-p_{Z}}{p_{Z}}\;+\;\mathrm{constant},  \label{eq:cal-nll}
\end{equation}%
with $|x|$ and $|z|$ the Hamming weights of the two conjugate supports.
Equations \eqref{eq:cal-prior}--\eqref{eq:cal-nll} contain the whole
probabilistic input, and the only remaining step is bookkeeping: by the
crossing recalled above, the sector built from $H_{X}$ is the one that
prices $|z|$, and the sector built from $H_{Z}$ is the one that prices $|x|$%
. Reading the two coefficients of Eq.~\eqref{eq:cal-nll} off in that order
gives the response--sector costs
\begin{equation}
\lambda _{X}^{\mathrm{resp}}\;=\;\log \frac{1-p_{Z}}{p_{Z}},\qquad \lambda
_{Z}^{\mathrm{resp}}\;=\;\log \frac{1-p_{X}}{p_{X}},
\label{eq:cal-response-costs}
\end{equation}%
so that the reversal of subscripts is derived and not stipulated. Matching
these costs to the two sector coefficients of $\tilde{H}(t_{X},t_{Z})$,
which is quadratic in the residual, fixes one ratio and one gauge,
\begin{equation}
\rho ^{2}\;\equiv \;\frac{t_{X}^{2}}{t_{Z}^{2}}\;=\;\frac{\lambda _{X}^{%
\mathrm{resp}}}{\lambda _{Z}^{\mathrm{resp}}},\qquad t_{Z}=1,\qquad
t_{X}\;=\;\left( \frac{\log \bigl[(1-p_{Z})/p_{Z}\bigr]}{\log \bigl[%
(1-p_{X})/p_{X}\bigr]}\right) ^{1/2}.  \label{eq:cal-ratio-gauge}
\end{equation}%
The gauge is legitimate because the common rescaling multiplies every score
by one positive factor and cancels in any comparison inside one fibre, so
only the single number $\rho =t_{X}/t_{Z}$ carries essential meaning. The
matching defines a noise-informed heuristic and not an identity of
likelihoods.

\paragraph{Orientation checks and one excluded shortcut.}

Two limits confirm that Eq.~\eqref{eq:cal-ratio-gauge} points the right way.
At $p_{X}=p_{Z}$ one recovers $\rho =1$, so the overlay adds nothing to the
exact verdict, whereas for $p_{Z}\gg p_{X}$ one has $\lambda _{X}^{\mathrm{%
resp}}\ll \lambda _{Z}^{\mathrm{resp}}$ and hence $\rho \ll 1$: the term
pricing the abundant $Z$--type support is cheap and the term pricing the
rare $X$--type support is expensive, so the ranking is reluctant to
postulate rare faults and tolerant of common ones, and an implementation
showing the opposite ordering has dropped the crossing somewhere in its
pipeline. The admissible domain is the open quadrant, since a vanishing
weight deletes a sector differential and changes the complex rather than
representing the bias.

\paragraph{The soft score.}

For a candidate representative $E$ returned by the frozen base decoder
inside a detector fibre, in the residualization convention declared in
Subsection~\ref{subsec:circuit-static-bridge}, the calibrated syndrome cost
is the quadratic form of Subsection~\ref{subsec:invariance} evaluated on
that candidate,
\begin{equation}
W_{\mathrm{syn}}(E;t_{X},t_{Z})\;=\;\bigl\langle E,\,\tilde{H}%
(t_{X},t_{Z})\,E\bigr\rangle\;=\;t_{Z}^{2}\,\bigl\Vert H_{Z}E\bigr\Vert%
^{2}\;+\;t_{X}^{2}\,\bigl\Vert H_{X}E\bigr\Vert^{2},  \label{eq:Wsyn}
\end{equation}%
with the inner product, the row normalisations and any circuit--response
multiplicities declared once and held fixed in every matched comparison.
This is the only quantity the present subsection contributes to the deployed
pipeline, and it enters solely as the calibrated feature of the objective of
Eq.~\eqref{eq:bb-objective}. Its status is fixed in one sentence: it is
heuristic, it is evaluated on candidates whose exact labels $A$ have already
been computed by Eq.~\eqref{eq:bb-logical-action}, and it may reorder
candidates while retaining those labels but may never merge, override or
certify one. One structural property must be stated, namely that the score
orders explicit representatives whose exact labels travel with them, and
that any statement intended to be class level is formed from pairwise
differences, whose logical content is decided by $A$.

\paragraph{The one new diagnostic.}

A declared bias is never known exactly, so the honest input is an interval.
Because only $\rho$ carries physics, the sweep is one dimensional.

\begin{definition}[Weight--sweep ambiguity trigger]
\label{def:sweep-trigger} Let $\mathcal{U}\subset(0,\infty)$ be a declared
compact interval of admissible ratios containing the calibrated value of Eq.~%
\eqref{eq:cal-ratio-gauge}. For each $\rho\in\mathcal{U}$, and under a fixed
deterministic tie rule, let $f_{\star}(\rho)$ be the member of the
base--decoder list minimising the score of Eq.~\eqref{eq:Wsyn} at $t_{Z}=1$,
$t_{X}=\rho$, and let $[\,f_{\star}(\rho)\,]$ denote its exact logical class
under $A$ or, after residualization on the certified domain, in $H^{1}_{%
\mathrm{full}}$. The detector record $d$ is called weight--sweep unstable
when
\begin{equation}
[\,f_{\star}(\rho)\,] \ \text{is not constant for}\ \rho\in\mathcal{U},
\label{eq:cal-sweep-flag}
\end{equation}
and the corresponding per--shot flag is written $\Phi(d)=1$, with $\Phi(d)=0$
otherwise.
\end{definition}

The physical content of Eq.~\eqref{eq:cal-sweep-flag} is sharper than a
sensitivity analysis, and this is why the trigger is worth carrying. Bias
uncertainty is operationally irrelevant while it only permutes
representatives inside one class, since those candidates are interchangeable
by Eq.~\eqref{eq:bb-class-partition}; it becomes decisive exactly when it
moves the selected class, because two accounts of one record that the
extraction hardware cannot distinguish are then being separated by a number
the experimenter does not know precisely. Instability is an alarm and
stability is not a certificate. On a flagged record, and equally on a
discrepant member of Eq.~\eqref{eq:ledger-disc}, the shot is escalated
before any representative is committed, under the audit--controlled policy
of Subsection~\ref{ssec:bb-reranking} and within the declared latency
allowance: enlarge the candidate list, deepen the ordered--statistics
support, restart belief propagation with fresh seeds, or defer to the
offline solver \cite{Maurer:2025tbw}. The regime in which this pays is the
working regime of a high--rate memory, where one bounded list routinely
holds near--degenerate detector--compatible candidates drawn from distinct
exact subfibres \cite{Bravyi2024,Hillmann2024}.

\paragraph{Deployment observables and cost.}

This subsection serves also the description of the three engineering
quantities that Subsection~\ref{ssec:bb-metrics} reports. For a declared
per--shot latency budget $T_{\mathrm{budget}}$ and a run of $N_{\mathrm{shots%
}}$ shots with critical--path latencies $T_{j}$ and flag values $\Phi _{j}$
from Definition~\ref{def:sweep-trigger}, define the tail latency, the
missed--deadline probability, and the useful--escalation rate by
\begin{equation}
q_{0.99}=\inf \bigl\{\,T:\ \#\{\,j:T_{j}\leq T\,\}\ \geq \ 0.99\,N_{\mathrm{%
shots}}\,\bigr\},\qquad P_{\mathrm{miss}}=\frac{\#\{\,j:\ T_{j}>T_{\mathrm{%
budget}}\,\}}{N_{\mathrm{shots}}},  \label{eq:cal-latency-observables}
\end{equation}%
\begin{equation}
R_{\mathrm{useful}}=\frac{\#\{\,j:\ \Phi _{j}=1\ \text{and escalation
changes the committed logical class}\,\}}{\#\{\,j:\ \Phi _{j}=1\,\}}.
\label{eq:cal-useful-escalation}
\end{equation}%
These three are engineering quantities attached to a frozen stack and never
logical invariants; their purpose is to expose the trade the overlay
actually makes, since a flag rate that is large while $R_{\mathrm{useful}}$
is small means effort is being spent without changing any logical outcome,
and a small $R_{\mathrm{useful}}$ together with a non--vanishing $P_{\mathrm{%
miss}}$ is direct evidence against deploying the diagnostic on that platform
\cite{Maurer:2025tbw}. The price is modest by construction. Evaluating Eq.~%
\eqref{eq:Wsyn} costs two sparse matrix--vector products per candidate, with
$H_{X}$ and $H_{Z}$ and the row normalisations preconditioned once offline
for the frozen schedule, and a sweep over a declared grid in $\mathcal{U}$
multiplies only that cost, leaving the base decoding algorithm, the
candidate generator and the physical qubit count unchanged \cite%
{Hillmann2024}. The overlay is therefore software on the existing stack
rather than an addition to the device.

\section{Witten Index as Cohomological Count of Protected Logical Fibres}

\label{sec:witten-static-decoder}\label{sec:witten-static}

\renewcommand{\theequation}{7.\arabic{equation}} \setcounter{equation}{0} %
\renewcommand{\thethm}{7.\arabic{thm}} \setcounter{thm}{0} %
\renewcommand{\theprop}{7.\arabic{prop}} \setcounter{prop}{0}

Every certification statement made so far in this paper is a statement about
a route. The projected-return block of Sec.~\ref%
{sec:syndrome-fibres-KL-window} tests whether a decoder-selected correction
carries a syndrome sector back into the clean protected manifold $\mathcal{K}
$, the certified faithful window $\mathcal{S}_{\mathrm{faith}}$ records
where that test has been discharged, and the residual classifier of Sec.~\ref%
{sec:software-supercharge-cohomology} names the obstruction where it has
not. What none of these constructions fixes is the size of the destination.
The purpose of the present section is to supply that single missing datum
and to prove that it is stable. The index constructed here counts how many
protected logical fibres exist, whereas the projected-return blocks and the
Knill--Laflamme Gram condition decide which fibre a given recovery actually
reaches~\cite{KnillLaflamme1997,NielsenChuang2010}. Because the arena is
pinned before any weighting, ranking or symmetry labelling is applied, Secs.~%
\ref{sec:weighted-lift} and \ref{subsec:compress-Bs} may proceed without
ever restating it.

The treatment is static and kinematic. We use only the nilpotent supercharge
$Q$ of Sec.~\ref{sec:susy-stabilizer}, the fermion-parity grading $(-1)^{F}$
of Eq.~\eqref{eq:sg-nu}, the non-negative operator $H_{\mathrm{SUSY}} = \{ Q
, Q^{\dagger} \}$ of Eq.~\eqref{eq:sg-HSUSY-clean}, and the clean protected
projector $P_{\mathcal{K}}$ of Eq.~\eqref{eq:sg-PK}. No new operator is
introduced and nothing established earlier is re-derived. The object we
build is the stabilizer-code counterpart of the graded trace by which
supersymmetric quantum mechanics counts zero-energy states~\cite{Witten1982}%
, read here as a cohomological audit of the landing arena of a decoder.

\subsection{Projected static index: definition and evaluation}

\label{subsec:witten-definition}\label{subsec:witten-purpose}\label%
{subsec:witten-hodge}\label{subsec:witten-eval}

The graded trace that the certification layer needs is the one taken against
the clean protected projector, so we define

\begin{equation}
I_{W}\;\equiv \;\mathrm{Tr}_{\mathcal{H}}\left[ \,(-1)^{F}\,P_{\mathcal{K}%
}\,e^{-\beta H_{\mathrm{SUSY}}}\right] ,\qquad P_{\mathcal{K}}=P_{\mathcal{C}%
_{\mathrm{QEC}}}\otimes P_{0,\mathrm{anc}}.  \label{eq:witten-IW-PK}
\end{equation}

Both tensor factors are load-bearing, for the reason already fixed in
Subsec.~\ref{subsec:sg-hodge}: the harmonic kernel taken before the
degree-zero cut is the raw cylinder $\mathcal{K}_{\mathrm{raw}}$ of Eq.~%
\eqref{eq:sg-Kraw}, which carries every possible register record above each
encoded state and is therefore not the manifold on which a certified
recovery lands. The first factor demands that no stabilizer check is
violated, the second that no stale flag survives in the ancilla register,
and together they select exactly the target that the projected-return blocks
of Subsec.~\ref{subsec:projected-return-faithfulness} are required to hit.
The index is consequently a purely kinematic integer, with no regulator
hidden inside it.\footnote{%
Because $H_{\mathrm{SUSY}}P_{\mathcal{K}}=0$ by the harmonic identification
of Subsec.~\ref{subsec:sg-hodge}, one has $\mathrm{Tr}_{\mathcal{H}%
}[(-1)^{F}P_{\mathcal{K}}e^{-\beta H_{\mathrm{SUSY}}}]=I_{W}$ for every $%
\beta \geq 0$, which is protected and independent of $\beta ~$\cite%
{Witten1982}; the regulated form is recorded only for contact with the
standard supersymmetric presentation~\cite{Witten1982}. One may introduce
extra fugacities that conjugate to extra quantum numbers or charges, c.f.
\cite{Kinney:2005ej}, in various different contexts.}

The physical mechanism that turns Eq.~\eqref{eq:witten-IW-PK} into a count
of protected states, rather than an arbitrary partial trace, is the pairwise
cancellation of all syndrome-active directions, and one paragraph suffices
because the mechanism was established earlier. By the Hodge decomposition of
Eq.~\eqref{eq:sg-hodge}, a vector outside the harmonic sector lies in the
image of $Q$ or of $Q^{\dagger }$, and the supercharge maps it to a partner
of opposite fermion parity. The two members of such a pair share the same
strictly positive eigenvalue of $H_{\mathrm{SUSY}}$, which by Eq.~%
\eqref{eq:sg-syndrome-weight} is the syndrome weight; they contribute $+1$
and $-1$ and drop out. A state of $\mathcal{K}$ has no partner at all, since
$Q$ annihilates it because no check is violated and $Q^{\dagger }$ has no
lower degree into which to map. The census therefore returns the clean
harmonic representatives and

\begin{equation}
I_{W}=\dim \mathcal{K}=\dim H^{0}(Q)=2^{k},  \label{eq:witten-H0}
\end{equation}%
in agreement with the degree-zero identification of Subsec.~\ref%
{subsec:sg-hodge} and with the logical dimension of an $[[n,k,d]]$
stabilizer code~\cite{Gottesman1997,NielsenChuang2010}. In plain words, this
is why the index is useful to a decoder at all: directions that announce
themselves through a syndrome enter the census with cancelling signs and
disappear, while the silent protected directions in which the encoded
information resides are counted exactly once. That is index
mechanism~transcribed into the language of stabilizer checks and
ancilla-register bookkeeping, and it is the same alternating bookkeeping
that governs degreewise cohomology. One reading must be kept straight, since
it is the commonest source of confusion: $I_{W}$ counts protected states, so
it is the dimension of $\mathcal{K}$ and not the cardinality of the
logical-operator quotient, which is $4^{k}$ for the same code.

Two anchors and one cross-layer check complete the evaluation, and neither
requires a calculation here. On the Steane $[[7,1,3]]$ code, the
hand-verified control of Subsec.~\ref{subsec:steane-static-audit}, the index
takes the value $I_{W}=2$ already recorded in Sec. 4; on the gross $%
[[144,12,12]]$ bivariate-bicycle code of Ref.~\cite{Bravyi2024} it takes the
value $I_{W}=2^{12}=4096$, the arena size presupposed throughout the
circuit-level audit of Sec.~\ref{sec:bb-circuit-reranking}. Independently,
the finite-field layer reproduces the same integer, since the middle
cohomology of the selected Calderbank--Shor--Steane sector has binary
dimension $k$, so that $\log _{2}I_{W}=\frac{1}{2}\dim _{\mathbb{F}%
_{2}}H^{1}(\tilde{Q}^{\mathrm{full}})=k$, one half of the full
logical-address width $2k$ fixed by the two-sector completion of Sec.~\ref%
{sec:software-supercharge-cohomology}. The purpose of this is diagnostic and
it is worth using as such: it is a falsifiable diagnostic test between the
physical and the finite-field descriptions.

\subsection{Deformation pinning of the protected-fibre count}

\label{subsec:witten-pinning}\label{subsec:witten-deformation}

An integer that depended on how the complex happens to be presented would be
useless as a certificate. The purpose of this subsection is therefore to
identify exactly what a deformation must preserve in order for $I_{W}$ to be
rigid, and the good effect is concrete: a reader may rely on the arena size
while the surrounding construction is reparametrised, reweighted or
relabelled, and no later layer can silently change the number of protected
fibres.

We use the additive presentation in which deformations of the supercharge
are introduced. Let $\lambda $ label a family

\begin{equation}
Q(\lambda )=Q+\Delta Q(\lambda ),\qquad \Delta Q(0)=0,
\label{eq:witten-family}
\end{equation}%
of operators on $\mathcal{H}$ that remain odd under the parity and nilpotent,

\begin{equation}
\left\{ (-1)^{F},Q(\lambda )\right\} =0,\qquad Q(\lambda )^{2}=0,
\label{eq:witten-family-conditions}
\end{equation}%
with associated non-negative operator $H_{\mathrm{SUSY}}(\lambda
)=\{Q(\lambda ),Q(\lambda )^{\dagger }\}$ and clean harmonic manifold

\begin{equation}
\mathcal{K}(\lambda )=\ker Q(\lambda )\cap \ker Q(\lambda )^{\dagger }\cap
\ker N_{\mathrm{anc}}.  \label{eq:witten-deformed-manifold}
\end{equation}%
No degree condition beyond Eq.~\eqref{eq:witten-family-conditions} is
imposed, and the omission is deliberate rather than an oversight: the
deformation of Sec.~\ref{subsec:compress-Bs} multiplies the data factor
without raising the ancilla degree by one, so imposing $[N_{\mathrm{anc}%
},Q(\lambda )]=Q(\lambda )$ here would place the principal application
outside its own family. One clause is needed in exchange. Degreewise
cohomological language, and in particular the reading of Eq.~%
\eqref{eq:witten-H0} through $H^{0}(Q)$, is used only for members that
preserve the ancilla grading; for a general member the statement below is
made through the clean harmonic manifold of Eq.~%
\eqref{eq:witten-deformed-manifold} alone, which is the object the census
actually reads. The one structural property that the census genuinely
requires is that the deformation leave that clean landing manifold where it
was.

\begin{prop}[Pinning of the projected index under manifold preservation]
\label{prop:witten-pinning} Let $Q(\lambda)$ satisfy Eqs.~%
\eqref{eq:witten-family} and \eqref{eq:witten-family-conditions}, and
suppose that the clean harmonic manifold of Eq.~%
\eqref{eq:witten-deformed-manifold} is preserved,

\begin{equation}
\mathcal{K}(\lambda )=\mathcal{K}\text{,}~\text{for every }\lambda \text{ in
the declared domain}.  \label{eq:witten-manifold-preservation}
\end{equation}%
Then the projected index of Eq.~\eqref{eq:witten-IW-PK} is constant along
the family,

\begin{equation}
I_{W}\!\left[ Q(\lambda) \right] = I_{W}\!\left[ Q(0) \right] = 2^{k} .
\label{eq:witten-pinning}
\end{equation}
\end{prop}

\noindent {\textit{Proof}}. By Eq.~\eqref{eq:witten-manifold-preservation}
the orthogonal projector onto the clean harmonic manifold of $Q(\lambda)$ is
the operator $P_{\mathcal{K}}$ of Eq.~\eqref{eq:sg-PK} itself, independently
of $\lambda$. The trace in Eq.~\eqref{eq:witten-IW-PK} is therefore
literally the same number at every $\lambda$, and Eq.~\eqref{eq:witten-H0}
evaluates it. For members that additionally preserve the grading the same
conclusion may be read cohomologically: directions of nonzero syndrome
energy come in partners of opposite parity and cancel, so the census returns
the dimension of the unpaired clean sector, which by hypothesis has not
moved. {\hfill $\square $}\newline

The content of Proposition~\ref{prop:witten-pinning} lies entirely in its
hypothesis, and that hypothesis must be established for each deformation
rather than assumed: nilpotency alone does not imply Eq.~%
\eqref{eq:witten-manifold-preservation}, and neither does the presence of a
spectral gap above the clean sector. The mechanism used in this paper is
direct verification of the darkness condition, in which one shows that a
state is annihilated by the deformed operator and its adjoint in the clean
degree-zero sector precisely when it was annihilated by the undeformed pair.
That is the route taken for the character-resolved deformation of Sec.~\ref%
{subsec:compress-Bs}, where nilpotency and harmonic-manifold preservation
are proved together, and it makes that deformation the primary verified
instance of the present proposition.

A second, purely optional mechanism is worth naming once, because it
explains why whole families can be pinned without a case-by-case check. If
there exist even invertible operators $V(\lambda )$, with $V(0)=\mathbf{1}$,
such that

\begin{equation}
Q+\Delta Q(\lambda )=V(\lambda )\,Q\,V(\lambda )^{-1},\qquad V(\lambda )\,%
\mathcal{K}=\mathcal{K},~~V(\lambda )^{\dagger }\,\mathcal{K}=\mathcal{K},
\label{eq:witten-similarity}
\end{equation}%
then Eq.~\eqref{eq:witten-manifold-preservation} follows at once, and the
associated cohomologies are isomorphic as well. Equation~%
\eqref{eq:witten-similarity} is a convenience and not a requirement, and
nothing in Proposition~\ref{prop:witten-pinning} presupposes it.

The check-species rescaling recorded in Sec.~\ref{sec:susy-stabilizer}
illustrates the above mechanisms. For a Calderbank--Shor--Steane
presentation~\cite{Calderbank1996,Steane1996} it reads

\begin{equation}
Q(t_{X},t_{Z})=Q+(t_{X}-1)\,Q_{X}+(t_{Z}-1)\,Q_{Z},\qquad t_{X}>0,\quad
t_{Z}>0,  \label{eq:witten-css-family}
\end{equation}%
which is additive in the sense of Eq.~\eqref{eq:witten-family} and nilpotent
for all positive amplitudes, because the cancellation behind Eq.~%
\eqref{eq:sg-Qsq-zero} plays the symmetry of the projector product against
the antisymmetry of the register creators and is therefore blind to the
coefficients. It preserves the clean manifold directly, since the amplitudes
enter the non-negative operator quadratically and so rescale only the
excitation penalty of a violated check of each species, never entering the
darkness condition. It is also of the form of Eq.~%
\eqref{eq:witten-similarity}, the conjugating operator being the even,
register rescaling that fixes the clean register vacuum.\footnote{%
The register map is stated here in the direction that conjugates the
undeformed supercharge into Eq.~\eqref{eq:witten-css-family} for the
convention of Eq.~\eqref{eq:sg-Q-def}; if the opposite particle-hole
convention is adopted, the map is replaced by its conjugate.} Either
observation gives

\begin{equation}
\mathcal{K}(t_{X},t_{Z}) = \mathcal{K} , \qquad I_{W}(t_{X},t_{Z}) = 2^{k} ,
\qquad t_{X},t_{Z} > 0 .  \label{eq:witten-css-pinning}
\end{equation}

The deformation-pinned index fixes the dimension of the clean protected
manifold, not the logical action of a selected recovery: $I_{W}=\dim
\mathcal{K}=2^{k}$. This is the useful distinction between a count of
protected logical states and a certificate of return to the intended logical
address. To see it directly, let $L$ be a nontrivial logical Pauli and set $%
U=L\otimes I_{\mathrm{anc}}$. Since $L$ normalizes the stabilizer group, $U$
preserves the clean projector $P_{\mathcal{K}}$ while acting nontrivially on
$\mathcal{K}$~\cite{Gottesman1997}. The projected graded trace is therefore
unchanged, although the encoded state need not be. A silent logical Pauli
changes the fibre while leaving $I_{W}$ unchanged, so routing stays with the
projected-return blocks $B_{s}$ on $S_{\mathrm{faith}}$. The stability of
index under these deformations protects the size of the landing space, not a
decoder's choice of destination within it~\cite{Witten1982}.

Faithful return must instead be established on the declared family of errors
by the projected-return blocks and the fibrewise Knill--Laflamme test of
Subsec.~\ref{subsec:certified-faithful-window}; their passing addresses form
$S_{\mathrm{faith}}$~\cite{KnillLaflamme1997}. The residual-class audit of
Sec.~\ref{sec:software-supercharge-cohomology} and the circuit-level
comparison of Subsec.~\ref{ssec:bb-fibre-test} address the corresponding
static and detector-history questions. In particular, $I_{W}$ alone does not
give a nontrivial bound on the syndrome-silent failure probability $p_{%
\mathrm{fs}}$ nor a code distance or decoding threshold. Those quantities
require error geometry, a noise model, and a specified recovery procedure.

\section{Discussion}

\label{sec:discussion}

This paper develops a diagnostic and audit layer for stablizer codes and
decoders. The supercharge $Q$ records violated checks, the projected-return
block $B_{s}$ tests the return route, the certified faithful window $%
\mathcal{S}_{\mathrm{faith}}$ records proved same-fibre recovery, and the
class in $H^{1}(\tilde{Q}^{\mathrm{full}})$ diagnoses the final residual.
The consequence is a certificate on a declared domain, rather than only a
posterior score. This distinction is useful for quantum computing because
candidate generation and candidate validation become separate, independently
testable services.

Supersymmetry plays a constructive role in that service. In the construction
of Sec.~\ref{sec:susy-stabilizer}, nilpotency prevents a check record from
being counted twice, fermionic grading gives each syndrome register a
definite degree, and $H_{\mathrm{SUSY}}$ converts syndrome weight into an
auxiliary energy. Bright syndrome-bearing directions pair, while the clean
degree-zero protected sector remains unpaired, following the familiar
zero-mode logic of supersymmetric quantum mechanics~\cite{Witten1982}. The
degree-zero cut also has a concrete operational purpose: it removes code
states accompanied by stale ancilla-register records. Thus $\mathcal{K}$
identifies the space a recovery must actually reach, and the code space is
not chosen by hand. From this single starting point the paper obtains the
return route $B_{s}$ (Sec.~\ref{sec:syndrome-fibres-KL-window}), the
residual class in $H^{1}(\tilde{Q}^{\mathrm{full}}) $ (Sec.~\ref%
{sec:software-supercharge-cohomology}), and the protected count $I_{W}$
(Sec.~\ref{sec:witten-static-decoder}). Each object inherits its meaning
from the same complex, so the pieces fit together without ad hoc matching.
Supersymmetry therefore supplies natural bookkeeping algebra for the
decoders.

The novelty for quantum error correction is the organization of destination,
route, warrant, and diagnosis into one certification language. The
destination is $\mathcal{K}$. The route is tested by $B_{s}$, built
internally from the graded commutator in Sec.~\ref%
{sec:syndrome-fibres-KL-window}. A nonzero block, however, only proves
arrival in the code space, not preservation of encoded information. The
branchwise same-fibre condition defining $\mathcal{S}_{\mathrm{faith}}$
supplies the missing warrant and reflects the state-independent structure of
the Knill--Laflamme criterion~\cite{KnillLaflamme1997}. The finite-field
diagnosis of Sec.~\ref{sec:software-supercharge-cohomology} then
distinguishes a stabilizer boundary from an undetectable logical class.
Keeping these roles separate is a genuine strength of the present method: it
permits syndrome clearance, logical faithfulness, and residual
classification to be tested without being collapsed into one opaque number.

The framework is also directly suited to existing quantum-computing stacks.
It covers an arbitrary abelian Pauli stabilizer code, adds no physical
qubits, fixes no base-decoding algorithm, and sits above engines that
already generate candidates. Because a live experiment does not reveal $E_{%
\mathrm{true}}$, the absolute residual $R_{s}=C_{s}E_{\mathrm{true}}$
belongs to a truth-labelled audit. Relative candidate verification remains
available online. In Sec.~\ref{sec:bb-circuit-reranking}, equal detector
shadows and equal logical actions are represented by different conditions:
the first only says that two circuit histories look alike to the extraction
hardware, whereas the second says that their difference does not change
encoded information. Only logical equality licenses merging. This lets a
decoder reserve expensive search for records in which sound alternatives
disagree logically, instead of imposing it everywhere.

The two worked settings show that the method is executable, not merely
formal. On the Steane $[[7,1,3]]$ code, Subsec.~\ref%
{subsec:steane-static-audit} checks the whole weight-one catalogue by hand,
and the operator reading agrees with the finite-field reading \cite%
{Steane1996}. That agreement is the right first credit: the dictionary can
be audited. On the gross $[[144,12,12]]$ bivariate-bicycle code, the same
logical question is posed for spacetime fault histories \cite{Bravyi2024}.
The static--circuit bridge is specified precisely enough to be run and
falsified. The evidence therefore sits at three distinct levels. The
ancilla-register construction, the identification of $\mathcal{K}$, the
sectorwise return identity, and the declared-family faithfulness test are
algebraic results valid for every abelian Pauli stabilizer code; the
completed two-sector residual classification holds in the CSS setting of
Sec.~\ref{sec:software-supercharge-cohomology}. The Steane calculation is a
completed finite-domain verification. The gross-code shell enumeration,
mismatch ledger, matched ablation, and latency study of Sec.~\ref%
{sec:bb-circuit-reranking} are specified experiments that have not yet been
performed. This paper therefore does not claim that the bridge has already
lowered the silent-failure probability, raised a threshold, or shortened
latency. The method is already strong enough to tell a later measurement
which quantity it is entitled to claim.

The supporting methods have a common purpose: they make an exact audit
easier to use without changing what it certifies. Positive weights $%
t_{X},t_{Z}$ of Sec.~\ref{sec:weighted-lift} let a noise model influence
candidate ranking while retaining the binary logical address. A
weight-sensitive change of preferred logical class can identify records
worth escalating, but the score cannot decide which class is true. Character
resolution with $\chi \in \widehat{G}$ in Subsec.~\ref{subsec:compress-Bs}
can organize already-certified return blocks when the required covariance is
verified. An exact Gram test can then expose redundant operator directions.
Neither symmetry nor linear dependence grants certification to an untested
error family, so neither can enlarge $\mathcal{S}_{\mathrm{faith}}$. The
order matters. First identify the protected manifold. Then certify the
route. Then classify the residual and audit the circuit interface. Only
after that should one weight, rerank, or compress. The standard
Calderbank--Shor--Steane cohomology is the software language for one sector
\cite{Calderbank1996}. We use it. We do not claim it.

The projected index supplies a final, more conceptual cross-check. It counts
the clean protected states: $I_{W}=\dim \mathcal{K}=2^{k}$. It does not
count the $4^{k}$ logical Pauli classes or choose the state reached by a
correction. A nontrivial logical Pauli can act within $\mathcal{K}$ without
changing its dimension. The index therefore fixes the landing arena, not the
address reached inside it, even though both arise from the graded
construction. The connection to supersymmetric zero-mode counting is useful
for understanding the protected sector \cite{Witten1982}. In the
deformations considered here, constancy of the projected count also requires
preservation of the clean harmonic manifold; nilpotency alone does not
supply that conclusion. The index cannot, by itself, bound $p_{\mathrm{fs}}$.

Four conclusions stand out. First, returning to the code space and returning
to the intended logical fibre are decided by different tests, and the second
test is exact on any declared error family (Subsec.~\ref%
{subsec:certified-faithful-window}). Second, a running decoder can detect
logical disagreement among its own candidates without knowing the true
error. When two residuals are multiplied, the unknown error cancels, and
only the product of the two corrections remains. A nontrivial relative class
therefore means that at most one candidate is right (Subsec.~\ref%
{subsec:ambiguity-flag-rate}). Third, the certificate survives two changes
of setting. It survives demotion to exact finite-field arithmetic, and it
carries over to spacetime fault histories, where the test $A(f_{a}+f_{b})=0$
of Eq.~\eqref{eq:bb-fibre-test} replaces the static class. Fourth, later
refinements, such as weighting, compression, and index counting, organize
certified data.

The layer strengthens existing decoders. Belief propagation with
ordered-statistics post-processing, localized-statistics decoding, real-time
message passing, and automorphism ensemble decoding, are fast engines \cite%
{Panteleev2021,Hillmann2024,Muller:2025rue,Maurer:2025tbw,Koutsioumpas2025},
which the audit layer of this paper is compatible with and strengthens. The
certificate adds one piece of information that these engines do not report:
whether their candidates describe one encoded event. Extra decoding effort
can then be spent only where candidates disagree logically. This complements
degeneracy-aware post-processing \cite{Tsubouchi:2025hgy}, and any
improvement in the base decoder passes through unchanged. The diagnostic and
audit layer is also compatible with and may strengthen other related aspects
\cite{Raveendran:2020hzu} and \cite{Bradshaw:2025huj}.

The framework opens several concrete directions. The first is to run the
gross-code bridge on a frozen circuit. A stabilizer-circuit simulator such
as Stim makes this practical \cite{Gidney2021}. The resulting ledger should
publish every populated fault shell, every static--circuit mismatch, and a
witness history for each harmful logical difference, so that the audit can
be reproduced and challenged. The second is a resource-matched control. The
proposed layer should be compared against an ordinary logical-label audit
that uses the same candidates, scores, fallback, and deadlines, and the
comparison should report tail latency, that is, high-percentile delay,
together with the average cost \cite{Maurer:2025tbw}. Only such a test can
measure the value of logical information directly. Because the operator
construction needs only an abelian Pauli stabilizer group, it extends
naturally to other code families of current interest, where the
corresponding audits remain to be carried out. The lightest silent class
suggests a route to bounds on $p_{\mathrm{fs}}$ once a noise model is fixed.
More broadly, the results suggest that supercharge and cohomological methods
can serve as a working language for fault tolerance. We could see, in one
structure, how quantum information are stored in the protected manifold, and
how it returns after the error correction, to the right logical states.

Several other related directions connect this algebraic framework to
physical implementations. A separate related direction is physical lattice
supersymmetry. While the supercharge here serves as an auxiliary diagnostic
layer, lattice models with $N$=2 supersymmetry \cite{Fendley:2003} realize
physical supersymmetric Hamiltonians governed by hard-core exclusion.
Programmable neutral-atom and Rydberg arrays, as well as digital
superconducting processors simulating two-dimensional lattice dynamics \cite%
{Alam:2025ofn,Bernien2017,Burgers2022,Browaeys2020}, may provide platforms
for exploring such constrained models and their quantum simulations. These
works may also motivate possible hardware-facing investigations, among other
aspects.

\section*{Acknowledgments}

We would like to thank B. Guo, Q. Li, and H. Xu for related discussions or
communications. The work was supported in part by National Key R\&D Program
of China grant 2020YFA0713000, by Grant No. TH-533310008, by National
high-level talents program, by Fundamental Research Funds for the Central
Universities of China, by Grant No. 4007012320, and by Grant No. 3207012204.













\end{document}